\documentstyle[12pt,subeqnarray, epsf,epsfig,verbatim, cite]{article}
\DeclareFontFamily{U}{msb}{} %
\DeclareFontShape{U}{msb}{m}{n}{ <5> <6> <7> <8> <9> gen * msbm %
<10> <10.95> <12> <14.4> <17.28> <20.74> <24.88> msbm10}{} %
\DeclareSymbolFont{AMSb}{U}{msb}{m}{n} %
\DeclareSymbolFontAlphabet{\Bbb}{AMSb} %
\begin{document}
\newcommand{\dis}{\displaystyle}
\newcommand{\id}{ 1 \hspace{-2.85pt} {\rm I} \hspace{2.5mm}}
\newcommand{\expon}{{\rm e}}
\newcommand{\noi}{\noindent}
\newcommand{\pr}{\prime}
\newcommand{\beq}{\begin{equation}}
\newcommand{\eeq}{\end{equation}}
\newcommand{\bseq}{\begin{subeqnarray}}
\newcommand{\eseq}{\end{subeqnarray}}
\newcommand{\bsq}{\begin{subeqnarray}}
\newcommand{\esq}{\end{subeqnarray}}
\newcommand{\cc}{\mbox{\sc C}}
\newcommand{\la}{\lambda}
\newcommand{\sech}{{\rm sech}}
\newcommand{\cosec}{{\rm cosec}}
\newcommand{\PP}{{\rm P}}
\newcommand{\ga}{\gamma}
\newcommand{\bea}{\begin{eqnarray}}
\newcommand{\eea}{\end{eqnarray}}

\thispagestyle{empty} \vspace*{0.5cm} \centerline{\Large \bf
Superstrings, Gauge Fields and Black Holes.}

\vskip 2cm
\centerline{ B.E. Baaquie {\footnote{Electronic mail:
phybeb@nus.edu.sg }} and L.C. Kwek  } \centerline{{\it Department
of Physics, Faculty of Science, } } \centerline{{\it National
University of Singapore,Lower Kent Ridge,} }
\centerline{{\it Singapore 119260, Republic of Singapore. } }

\vskip 5cm \centerline{\bf Abstract} \vspace{10mm}
\parbox{5in}{\setlength{\baselineskip}{12pt}
{\small \noindent{ There has been spectacular progress in the
development of string and superstring theories since its inception
thirty years ago. Development in this area has never been impeded
by the lack of experimental confirmation. Indeed, numerous bold
and imaginative strides have been taken and the sheer elegance and
logical consistency of the arguments have served as a primary
motivation for string theorists to push their formulations ahead.
In fact the development in this area has been so rapid that new
ideas quickly become obsolete. On the other hand, this rapid
development has proved to be the greatest hindrance for novices
interested in this area. These notes serve as a gentle
introduction to this topic. In these elementary notes, we briefly
review the RNS formulation of superstring theory, GSO projection,
$D$-branes, bosonic strings, dualities, dynamics of $D$-branes and
the microscopic description of Bekenstein entropy of a black hole.
} }}

\newpage
\parbox{11cm}{\setlength{\baselineskip}{12pt}{
\tableofcontents
}}

\newpage
\section{Introduction}

Superstring theories arose as an attempt to unify all the forces
in Nature: the gravitational, the weak, the electromagnetic and
the strong forces. An early attempt in this direction started with
the Veneziano model in which one endeavored to reconcile the
duality between s and t channels with the regularity in the
spin-mass squared plot, namely the Regge behavior of $S$-matrix
theory. The spectrum, which emerged from this early attempt,
exhibited degeneracy at each mass level increasing exponentially
with mass. Furthermore, this spectrum contained positive and
negative norm states. However, the undesirable negative norm
states (ghosts) could be removed if one restricted the dimension
of space-time, $d$, to less than or equal to 26. In particular,
the choice of a critical dimension of $d=26$ made the model
unitary\footnote{However, tachyons still exists in the bosonic
model.}. Veneziano model was formulated with bosons. In fact, the
critical dimension for a fermionic version, called the dual pion
model \cite{schwarz} turned out to be 10.

It was subsequently found that the infinite particle spectrum in
the Veneziano model could be derived in a more consistent manner
in a quantized string theory. Originally proposed by Nambu, Goto,
Nielson and Susskind, this formulation of a
reparametrization-invariant string action showed that excitations
of the one-dimensional (bosonic) string could be identified with
the infinite particle spectrum of Veneziano model. The fermionic
string theory was subsequently formulated by Ramond \cite{ramond},
Neveu and Schwarz \cite{neveu}. An ad-hoc procedure to incorporate
both theories was later proposed by Gervais and Sakita
\cite{gervais}. An interesting feature of these string theories is
that consistent quantization requires the fixation of space-time
dimension to $d=26$ for bosonic string and $d=10$ for fermionic
string, just like the Veneziano model.

Despite these developments, string theory was not readily accepted
in these early attempts primarily due to the emergence of quantum
chromodynamics and the failure of the dual models to describe
experimental results at high energy scattering \cite{schwarz}.
Moreover, the theory predicted a massless sector with a spin 2 particle
and all experimental attempts to identify this particle failed. Fortunately,
Scherk and Schwarz \cite{scherk} proposed a quick fix to the
problem.  They suggested that the spin 2 particle is really the
graviton. Indeed,
there were three subsequent discoveries
\cite{schwarz2}that clearly pointed to the possibility of superstring
theory as a strong candidate for the unifying theory of standard
model with gravity. The first observation was the discovery of the
miraculous anomaly cancellation in superstring gauge theory.
Furthermore, consistency at quantum level requires the gauge group
to be $SO(32)$ or $E_8 \times E_8$. The second discovery was the
development of heterotic string theories. It turned out that there
were two new consistent heterotic superstring theories based on
closed oriented strings with gauge groups $SO(32)$ or $E_8 \times
E_8$. The third discovery was the realization that the heterotic
string with $E_8 \times E_8$ gauge group admits solutions which
results in 4d effective theory at low energies on Calabi-Yau
compactification.

At the end of 1985, there were five totally self-consistent
superstring theories in ten dimensions: Type I, non-chiral Type
IIA, chiral Type IIB, $E_8 \times E_8$ heterotic string and
$SO(32)$ heterotic string theory. One major obstacle remained. To
make the necessary connections to our four dimensional world,
these ten dimensional string theories have to be compactified. It
turns out that there are thousands of ways of performing this
compactification.

To reconcile the five superstring theories, it turned out that it
was necessary to invoke the notion of duality symmetries. Existence
of duality symmetries began as a conjecture and has remained a
conjecture \cite{sen}. One such duality symmetry is T duality. This
duality occurs when closed bosonic string compactified on a circle
of radius $R_1$ has the same mass spectrum of physical states and
scattering amplitudes as another string theory compactified on a
radius of $R_2$ with $R_1 R_2 = \alpha^\prime$, where
$\alpha^\prime$ is the universal Regge slope parameter. Using
duality argument, one then shows that the Type IIA and IIB theories
are T dual. So are the two heterotic string theories. Indeed, T
duality is really a generalized Fourier transformation. Besides T
duality, there were other dualities, namely S duality and U
duality. S duality is essentially non-perturbative and exchanges
weak and strong couplings while U duality describes a larger group
of duality symmetries which encompasses both T and S dualities.

In 1995, Witten \cite{witten1} invoked the power of duality
symmetries to map all the strong coupling regimes of some string
theory to the weak coupling regimes of another.  Thus all the five
superstring theories can be related to one another through duality
symmetries. An important formulation using these duality
symmetries is M theory.  M theory is an 11 dimensional theory
which reduces to 11 dimensional supergravity in its low energy
limit and reproduces Type IIA string theory when compactified on a
circle of vanishingly small radius. Moreover, it is also related
to Type IIB string theory. To see this relation, one observes that
Type IIA and IIB are T dual.  Combining these facts, one sees that
M theory compactified on a torus is dual to Type IIB superstring
theory compactified on a circle.

Type II superstring theories has a number of $(p+1)$-dimensional
membrane solutions that preserve half the supersymmetries and are
called $p$-branes. A large class of these $p$-brane excitations
are the Dirichlet or $D$-branes. Indeed, D-branes are classical
solutions and can also be regarded as topological obstructs in
superstring theories. Moreover, it is found that the longitudinal
components can be gauged away leaving the transverse components as
excitations These in turn can be described by an effective theory
given by supersymmetric Yang-Mill fields. In addition, N D-branes
can be superposed to give rise to an U(N) nonabelain gauge theory
for the lowest energy excitations.

This review arose from a series of lectures on superstring,
$D$-branes and black holes by the first author in 1997. Since
then, there have been many important discoveries. Indeed an entire
arsenal of insightful tools have been invented and recent
investigations into Maldacena conjectures concerning the AdS/CFT
duality have shown that it is possible to study large $N$ gauge
theory using strings and $D$-branes dynamics and vice versa\cite{malde1,
witten3}. M theory was first formulated by Witten to describe the
11-dimensional quantum theory whose effective description at low
energy is the 11-dimensional supergravity. Recently, matrix theory
has been formulated to provide a non-perturbative description of M
theory. Not all the complexities regarding matrix quantum
mechanics have been resolved, especially in the large $N$ limit.

Another recent development is the formulation of U theory
\cite{sen} as the underlying fundamental theory whose limits
result in the various string theories and their compactifications.
Using duality symmetries, attempts have been made to understand
this theory perturbatively through empirical conjectures regarding
their effective actions. Indeed, there are numerous excellent
reviews regarding these recent developments and their results
\cite{dev}.

Indeed duality arguments feature strongly in many superstring
theories.  Besides the tools in geometric engineering, a
generalized Fourier transformation, there has been a tremendous
explosion of activities around Maldacena conjecture which states
that the quantum string theory on backgrounds of the form $AdS_d
\times M_{D-d}$, where $AdS_d$ is an anti de-Sitter space of space
time dimension $d$ and $M$ is a compactification space of
dimension $D-d$, is dual to the conformally invariant quantum
field theory on the boundary of the anti de Sitter space.  In
particular, it has been shown\cite{witten4} that ${\cal N}=4$
super Yang Mills theory on $M^4$ with gauge group $SU(N)$ is dual
to Type IIB superstring theory on $AdS_5 \times M^5$.

Starting with basic quantum field theory, we look at superstring
theory. We consider the RNS formulation of superstring theory and
look at how GSO projections can remove the existence of tachyons
and impose space-time supersymmetry. We then briefly review and
describe $D$-branes and bosonic strings. Duality is a fundamental
concept in string theory. We look into the power of duality
symmetries and review the relations between the various strings
theories,namely the Type I, Type IIA and Type IIB. Dynamics of
$D$-branes has provided much insight into M and U theories and we
take a brief look at these dynamics.  Finally, we review the
microscopic description of Bekenstein entropy and Hawking
radiation of a black hole and briefly describe the AdS/CFT
correspondence.

\bigskip
\section{ RNS Formulation of Superstring Theory}

\bigskip

When a particle traverses spacetime, it sweeps out a worldline. A
similar traversal by a string naturally generate a worldsheet. Let
$X_\mu(\sigma,\tau)$ be the position vector of a string in $d-1$
space-like dimensions.  The simplest bosonic action can then be
written in conformal gauge\cite{hatfield,green1, green2} as
($\alpha^\prime =1$) \beq S_B = \frac{1}{2\pi} \int d \sigma d
\tau
\partial_\alpha X^\mu(\sigma, \tau)
\partial^\alpha X_\mu(\sigma,\tau). \label{bose} \eeq Indeed we
expect the string coordinates $X^\mu(\sigma, \tau)$ to be
invariant under a reparametri\-zation which constrains the state
space. In conformal gauge, this can be realized by a super
Virasoro algebra. The action (\ref{bose}) acts in $d$-Minkowski
spacetime. Consistency of special relativity and quantum mechanics
requires $d=26$.  This consistency requirement arises from the
need to ensure that the Lorentz generators constructed in the
theory can close properly.

Bosonic string has the following problems
\begin{itemize}
\item The particle spectrum contains a tachyon with $m^2 < 0$ but whose
norm is positive definite.
\item It is inconsistent at one loop calculation due to the tachyon.
\item There are no spacetime fermions and so it cannot be used to
describe Nature realistically.
\end{itemize}

The introduction of fermionic string with
supersymmetry\cite{schwarz3, west, muller, sohnius} solves these
problems. There are two ways of introducing superstrings:
\begin{itemize}
\item Green-Schwarz string.  This string has genuine ten-dimensional spinor
fields but the theory is not manifestly covariant.
\item Ramond-Neveu-Schwarz (RNS) string.  This formulation is manifestly
covariant but the ten-dimensional spacetime supersymmetry is not obvious.
\end{itemize}

We follow the RNS-formulation.  To attain supersymmetry, we
introduce a fermionic field $\psi^\mu$ to the bosonic string
action. This fermionic field $\psi^\mu$ is essentially a
worldsheet spinor and can be expressed as a doublet \beq \psi^\mu
= \left(
\begin{array}{c} \psi^\mu_- \\ \psi^\mu_+
\end{array} \right) \eeq but acts as a vector under $SO(1,D-1)$
Lorentz group. The open superstring action is now written as \beq
S = \frac{1}{2\pi} \int_{-\infty}^{\infty} d \tau \int_0^\pi d
\sigma \{
\partial_\alpha X^\mu(\sigma, \tau) \partial^\alpha
X_\mu(\sigma,\tau) - i \bar{\psi^\mu} \rho^\alpha
\partial_\alpha \psi_\mu \}
\eeq
where $ \dis
\rho^0 = \left( \begin{array}{cc} 0 & -i \\
i & 0 \end{array} \right)
$
and
$ \dis
\rho^1 = \left( \begin{array}{cc} 0 & i \\
i & 0 \end{array} \right)
$ and
 $\{ \rho^\alpha, \rho^\beta \} =
- 2 \eta^{\alpha \beta} $ are in the Majorana representation $
\bar{\psi^\mu} = \psi^T \rho^0$.

This action possesses a supersymmetric invariance
\bsq
\delta X^\mu & = & \epsilon \psi^\mu \\
\delta \psi^\mu & = & - i \rho^\alpha \partial_\alpha X^\mu \epsilon.
\esq
We introduce worldsheet light cone coordinates, $\partial_\pm =
\partial_\tau \pm \partial_\sigma$.  In this formalism, the fermionic
part of the action reads \beq S_F = \frac{1}{\pi} \int d \tau \int
d \sigma \{  \psi_-^\mu \partial_+ \psi_{-\mu} + \psi_+^\mu
\partial_- \psi_{+\mu} \} \eeq so that \beq \delta S_F =
\frac{1}{\pi} \int d \tau \int d \sigma \{ \delta \psi_-^\mu
\partial_+ \psi_{-\mu} + \delta \psi_+^\mu \partial_- \psi_{+\mu}
\} + \frac{1}{2\pi} \int d\tau [ \psi_- \delta \psi_- - \psi_+
\delta \psi_+ ]_0^\pi. \eeq 
Field equations of the fermionic action requires $\partial_+\psi_-
= 0 =
\partial_-\psi_+$ and we impose of the boundary conditions at
$\sigma = 0$ and $\sigma = \pi$ so as to kill the surface terms.
We can always choose $\psi_+(\pi, \tau) = \psi_-(\pi, \tau)$.
However, at $\sigma= 0$, we have the following two distinct
boundary conditions:
\begin{enumerate}
\item $\psi_+ (0, \tau) = \psi_-(0,\tau)$, Ramond sector;
\item $\psi_+ (0, \tau) = - \psi_-(0,\tau)$, Neveu-Schwarz sector.
\end{enumerate}

For closed superstrings, the  boundary conditions are given by
\begin{enumerate}
\item Periodic \\
$\psi_{\pm} (0, \tau) = \psi_{\pm} (2 \pi,\tau)$, Ramond sector;
\item Anti-periodic \\
$\psi_{\pm} (0, \tau) = - \psi_{\pm}(2 \pi,\tau)$, Neveu-Schwarz sector.
\end{enumerate}

For the spectrum of states, we recall that superstrings are
invariant under super-diffeomorphisms. Since we have chosen to
work in the conformal gauge, we need to impose constraints on our
state space, and these constraints are provided by the super
Virasoro algebra. The Fourier expansion of the fermion strings in
the Ramond and Neveu-Schwarz sectors are given by \bsq \psi_-^\mu
& = & \sum_{n \in {\rm \bf Z}} d_n^\mu \exp (-2 i n (\tau -
\sigma) )
\\ \psi_+^\mu & = & \sum_{n \in {\rm \bf Z}} \tilde{d}_n^\mu \exp
(-2 i n (\tau + \sigma) ) \esq and \bsq \psi_-^\mu & = & \sum_{r
\in {\rm \bf Z + \frac{1}{2}}} b_r^\mu \exp (-2 i r (\tau -
\sigma)
\\ \psi_+^\mu & = & \sum_{r \in {\rm \bf Z + \frac{1}{2}}}
\tilde{b}_r^\mu \exp (-2 i r(\tau + \sigma) \esq respectively with
the commutation relations \bsq \{ d_n^\mu, d_m^\nu \} & = &
\eta^{\mu \nu} \delta_{n + m}, \mbox{{\rm etc }}\\ \{ b_r^\mu,
b_s^\nu \} & = & \eta^{\mu \nu} \delta_{r + s}, \mbox{{\rm etc }}.
\esq

Let $\cal{F}^B$ and $\tilde{\cal{F}}^B$ be the left and right
movers for the bosons.  Then for open strings, one sees that the
state space is given by \beq {\cal F}_{{\rm open}} = {\cal F}^B
\otimes ({\cal F}^R \oplus {\cal F}^{NS}). \label{eqn10}
\eeq In
the case of closed strings, the space is given by \beq {\cal
F}_{{\rm closed}} = {\cal F}_{{\rm open}} \otimes \tilde{\cal
F}_{{\rm open}}. \eeq

Thus, in the Neveu-Schwarz sector, one demands that \beq b_r |0,
NS> = 0 \eeq for $r = \frac{1}{2}, \frac{3}{2}, \cdots$ and create
higher mass states through $\dis b_{-r_1} b_{-r_2} \cdots b_{-r_N}
|0, NS>$. It will turn out that all excited states in the Neveu
Schwarz sector are spacetime bosons. A similar analysis for the
Ramond sector shows that the vacua transform as a spinor of
Spin$(1,d-1)$.

The energy-momentum tensor in the theory forms a closed algebra.
From the Fourier modes of the energy-momentum tensor, one can
construct the world sheet ${\cal N }= 1$ super Virasoro generators
$(L_m, G_r )$ of the left moving sector for instance, satisfying
\bsq \lbrack L_m, L_n \rbrack & = & (m - n) L_{m + n} +
\frac{c}{8}(m^3 - m)\delta_{m + n, 0} \\ \lbrack L_m, G_r \rbrack
& = & (\frac{m}{2} - r) G_{m + r} \\ \{ G_r, G_s \} & = & 2 G_{r +
s} + \frac{c}{2}(r^2 - \epsilon_A)\delta_{r + s, 0} \esq where
both $L_n, G_r$ belong to either the $R$ or $NS$ sector and \bsq
\epsilon_R & = & 0 \\ \epsilon_{NS} & = & \frac{1}{4}. \esq There
is a similar algebra $(\tilde{L}_m, \tilde{G}_r)$ for the right
moving sector yielding a world sheet ${\cal N}=2$ super Virasoro
algebra. Superconformal invariance is achieved through the
introduction of ghost fields and obtaining a central extension
$c_{total} = 0$. It will be shown that by imposing the
Gliozzi-Scherk-Olive condition, ${\cal N} = 2$ world sheet
supersymmetry yields a ${\cal N} = 2$, $d=10$ spacetime
supersymmetry with $2 \times 16 = 32$ supercharges.
\bigskip

\section{Spectrum of Physical States and GSO Projection}

\bigskip

In the previous lecture, we briefly touched on how to maintain
super-conformal invariance through the application of the
super-Virasoro algebra by combining the NS and R sector.  In this
lecture, we shall first consider the spectrum of the physical
states at low mass and show the existence of a state with mass$^2
< 0$, called a tachyon.  To remove this unphysical state, we shall
consider a technique proposed by Gliozzi, Scherk and Olive in the
seventies, commonly known as the GSO projection.

Consider the Ramond sector for the open string. We shall consider
left movers. The ground state $|0,\alpha,R>$ satisfies the
relation, for bosonic annihilators, $\alpha_n$ ($n > 0$), \beq
\alpha_n^\mu |0,\alpha,R> = 0 = d_n^\mu |0,\alpha,R>. \eeq From
the commutation relations of the operators $G_r$ and $L_n$ in the
previous lecture, one can easily show that $\dis G_0^2 = L_0 -
\frac{c}{24}$.  Since the ground state satisfies $G_0 |0,\alpha,R>
=0$, it follows that (for $c=0$) $L_0 |0,\alpha,R> = 0$. Further,
by defining $d_{0\mu} \equiv \Gamma_\mu$, we have \bsq 0 & = & G_0
|0,k, \alpha,R>, \\ & = & \alpha_0^\mu d_{0 \mu} |0,k,\alpha,R>,
\\ & = & k^\mu \Gamma_\mu |0,k,\alpha,R>, \esq giving us the
massless Dirac equation. \footnote{Historically, Ramond considered
the whole idea in reverse; beginning with the massless Dirac
equation and applying it to the superstring theory.} To obtain
further excited massive states, one needs to consider the
anti-symmetric polarization tensor.

In the Neveu-Schwarz sector, the ground state obeys
\bsq
b_{r+1/2}^\mu |0,k,NS> & = & 0, \mbox{\hspace{1cm}} r > 0, \\
G_r |0,k,NS> & = & 0 \mbox{\hspace{1cm}} r \in {\bf \rm Z} + \frac{1}{2}.
\esq
The physical state is subject to the mass shell condition
\bsq
(L_0 - \frac{1}{2}) |0,k,NS> & = & 0 \\
\Rightarrow (\frac{1}{2} \alpha_0^2 - \frac{1}{2}) |0,k,NS> & = & 0 \\
\Rightarrow (\frac{1}{2} k^2 - \frac{1}{2}) |0,k,NS> & = & 0 \\
\Rightarrow k^2 - 1  =  - m^2 - 1 & = & 0 ,
\esq
so that this state has mass$^2 < 0 $ and is indeed a tachyon.

\subsection{Excited states}
The first excited state $|\zeta, k, NS>$ is obtained
through the relation \beq |\zeta, k, NS> = \zeta_\mu b_{-1/2}^\mu
|0,k,NS> \eeq and satisfies \bsq 0 = (L_0 - \frac{1}{2}) |\zeta,
k, NS> & = & (L_0 - \frac{1}{2}) \zeta_\mu b_{-1/2}^\mu |0,k,NS>
\nonumber
\\ \Rightarrow k^2 & = & 0, \\ 0 & = & L_1 |\zeta, k, NS>
\nonumber  \\ \Rightarrow \zeta \cdot k & = & 0 \label{excite}
\esq Eq(\ref{excite}a) and Eq(\ref{excite}b) give rise to a
massless vector state and a transversality condition. All other
physical states can be obtained through the direct application of
the operators $a_{-r}^i$, $b_{-r}^i$ and $d_{-r}^i$ where $i = 1,
\cdots , 8$.

For closed strings, we need to consider left and right movers in order
to construct the complete spectrum. The fact that the origin for
$\sigma$ is arbitrary for closed strings yields
\beq
p_L = p_R;  \mbox{\hspace{1cm}} \alpha_0^\mu = \tilde{\alpha}_0^\mu
\eeq
The construction of the state space
is easily done.  Here, we shall summarize the result.
\begin{itemize}
\item NS-NS bosons
\begin{itemize}
\item Ground state is a tachyon.
\item The massless states are graviton, $g_{\mu \nu}$,
anti-symmetric tensor, $b_{\mu \nu}$ and the dilaton $\phi$.
\end{itemize}
\item NS-R (or R-NS) fermions
\begin{itemize}
\item No tachyon, since the Ramond sector only allows mass$^2 \geq 0$.
\item The massless ground state is reducible to the gravitino,
and the dilatino.
\end{itemize}
\item R-R bosons
\begin{itemize}
\item No tachyon.
\item The massless states $|0,k,a,R>_L \otimes |0,k,b,R>_R$ decompose as
a sum of the tensor representation of the group $SO(1,9)$.
\end{itemize}
\end{itemize}

By combining the NS and the R sectors, we have constructed a string
theory with bosons (e.g. the massless graviton) and fermions (e.g.
gravitino). Despite the removal of negative norm states through the mass
shell condition of the super Virasoro algebra, we still have states like
tachyons. As it stands, this tachyon has no fermionic partner and the
theory is essentially not space-time supersymmetric.  To produce a
spectrum with space-time supersymmetry, we need  to consider the GSO
projection.

\subsection{GSO Projection}
We define the GSO projection \cite{gliozzi} in terms of the
worldsheet fermion operator $F$ in the form of $(-1)^F$.  For the
NS sector, we may represent $(-1)^F$, for the right movers, by
\beq (-1)^F =- (-1)^{h_{NS}}, \eeq where $\dis h_{NS} = \sum_{r
\in 1/2 + Z^+} b_{-r} \cdot b_r $.  For the Ramond sector, \beq
(-1)^F = \Gamma_{11} (-1)^{h_R} \eeq where $\dis h_R  = \sum_{n >
0} d_{-n} \cdot d_n$. The GSO condition is that, from
eq(\ref{eqn10}), \beq (-)^F {\cal F}_{\mbox{\rm open}} =  {\cal
F}_{\mbox{\rm open}}.\eeq There is a similar representation for
left movers.

To obtain space-time supersymmetry, we note that in the NS sector,
we have 8 physical degrees of freedom for $A_\mu$; since in
general a spinor in 10 dimensions has $2^{d/2} = 32$ complex
components or 64 real components, we need to impose both the
Majorana and Weyl conditions.  This reduces the dimensionality to
16 and the imposition of Dirac equation eliminates half the modes
giving 8 physical modes. These 8 fermionic states are the
superpartners of the 8 bosonic degrees of freedom carried by
$A_\mu$.

For the Ramond sector, we note that in the Clifford algebra
defined by $\{ \Gamma^\mu, \Gamma^\nu \} = 2 \eta^{\mu \nu}$, one
finds, in the Majorana representation, that the real chirality
matrix $\Gamma_{11}$ is defined by \beq \Gamma_{11} \equiv
\Gamma^0 \Gamma^1 \cdots \Gamma^9, ~ ~ \Gamma_{11}^2 = 1 \eeq and
anti-commutes with all the generators in the Dirac spinor
representation.  For $d= 2$ (mod 8), a spinor can be reduced to a
Weyl-Majorana spinor. This Weyl-Majorana condition effectively
reduces the dimensionality of the ground state in the Ramond
sector to an 8-component spinor.

Having analyzed the GSO projection in a single sector of the RNS
string theory, we can next consider closed strings. Although a
naive listing gives four possibilities, a more detailed analysis
using parity operator shows that there are essentially two
inequivalent closed string theories defined by the Ramond ground
states:
\begin{itemize}
\item Type IIA, a non-chiral string theory in which the left-right
Ramond vacua have opposite chirality and the theory is
parity invariant under the exchange of
left-right movers.
\item Type IIB, a chiral string theory in which both the left-right
Ramond vacua have the same chirality.
\end{itemize}

The state space of the two closed superstring theories yields an
irrep of the $d=10$, ${\cal N} = 2$ spacetime supersymmetry.
Conjugation is defined for a Majorana spinor $Q$ by \beq Q^\dagger
= Q^{T \ast} \Gamma_0 \equiv Q^T C. \eeq In the Majorana
representation, $Q_\alpha$ is a real 32-component $d=10$
(Majorana) spinor. A chiral spinor satisfies \beq \Gamma_{11}
Q^{\pm} =\pm Q^{\pm}. \eeq

For Type IIA string theory, since the left and the right movers
have supercharges with opposite chirality, they can be combined
into a single Majorana spinor $Q_{\alpha}$ with $d=10$, ${\cal
N}=2$ supersymmetry algebra ($\alpha = 1, 2, \cdots, 32$) \beq \{
Q_{\alpha}, Q_{\beta} \} = (C \Gamma^\mu)_{\alpha, \beta} P_\mu ~
\mbox{\hspace{2cm}} ~ \mbox{{\rm IIA}} \label{type2a} \eeq where
$P_\mu$ is the $d=10$ translation operator.

For Type IIB string theory, the left and the right supercharges
have the same chirality and yield the chiral $d=10$, ${\cal N}=2$
superalgebra ($I=1,2$, $\alpha = 1, 2, \cdots, 32$) \beq \{
Q_{\alpha}^{+}, Q_{\beta}^{+} \} = \delta^{IJ} (C \Gamma^\mu
\Gamma_+ )_{\alpha, \beta} P_\mu ~ \mbox{\hspace{2cm}} ~
\mbox{{\rm IIB}}\eeq with $\displaystyle \Gamma_+ =
\frac{1}{\sqrt{2}} \left( 1 + \Gamma_{11} \right)$ and
$Q^+_{\alpha}$ are Majorana-Weyl spinors.  One of the many
miracles of string theory is that the GSO projection eliminates
the tachyon state and at the same time yields spacetime
supersymmetry for closed strings.
\bigskip

\section{Coupling of RR Fields to D-Branes}

\bigskip

We recall that the imposition of Majorana-Weyl condition on the
Ramond vacua yields an arbitrary vacuum state of the form \beq
F_{\alpha, \beta} |0,k,\alpha,R>_L |0,k,\beta,R>_R \eeq where
$\alpha, \beta = 1,2,\cdots 16$. The bispinor $F_{\alpha, \beta}$
is a $d=10$ classical background field and specifies the vacua. It
can be thought of as the vacuum condensate of the massless states
of the  RR-sector. Moreover, it can be decomposed in a complete
basis of all antisymmetric gamma-matrix products as \beq
F_{\alpha, \beta} = \delta_{\alpha,\beta} + \sum_{k=1}^{10}
\frac{i^k}{k!} F_{\mu_1 \cdots \mu_k} (\Gamma^{\mu_1 \cdots
\mu_k})_{\alpha,\beta} \eeq where $\dis \Gamma^{\mu_1 \cdots \mu_k
}=  \Gamma^{[\mu_1} \cdots \Gamma^{\mu_k]}$, and $\dis F_{\mu_1
\cdots \mu_k}  $ is an antisymmetric Lorentz tensor.  We also
recall that the RR-vacua has definite chirality and this implies:
\bsq \Gamma_{11} F & =  - F \Gamma_{11} & = F \mbox{\hspace{1cm}
{\rm Type IIA}} \\ \Gamma_{11} F & =  + F \Gamma_{11} & = F
\mbox{\hspace{1cm} {\rm Type IIB}}. \label{chiral} \esq 
Hence the
antisymmetric tensors $ F_{\mu_1 \cdots \mu_k} $ are not
independent.  To write the constraints on $F_{\mu_1 \cdots
\mu_k}$, we note \bsq \Gamma_{11} \Gamma^{\mu_1 \cdots \mu_k} & =
& \frac{(-)^{[k/2]}}{(10-k)!} \epsilon^{\mu_1 \cdots \mu_{10}}
\Gamma_{\mu_{k+1} \cdots \mu_{10}} \\ \Gamma^{\mu_1 \cdots \mu_k}
\Gamma_{11} & = & \frac{(-)^{[(k+1)/2]}}{(10-k)!} \epsilon^{\mu_1
\cdots \mu_{10}} \Gamma_{\mu_{k+1} \cdots \mu_{10}} . \esq 
For Type IIA, to satisfy eq(\ref{chiral}a), we need $k$ to be
even, and similarly for Type IIB, to satisfy eq(\ref{chiral}b),
$k$ has to be odd.  The antisymmetric field tensors satisfy \bsq
F\Gamma_{11} & = & - F \mbox{\hspace{10mm} {\rm Type IIA}} \\
{\mbox{\rm or,} \hspace{10 mm}} F_k & = & - \ast F_{(10- k)}\esq
with $k$ even and which yields $ \dis F^{\mu_1 \cdots \mu_k} =
\frac{(-)^{[(k+1)/2]}}{(10-k)!} \epsilon^{\mu_1 \cdots \mu_{10}}
F_{\mu_{k+1} \cdots \mu_{10}} $.  For Type IIB, the analogous
equations are \bsq F \Gamma_{11} &= &+ F \mbox{\hspace{10mm} {\rm
Type IIB}}
\\ {\mbox{\rm or,} \hspace{10 mm}} F_k & = & + \ast F_{(10-k)} \esq
with $k$ odd.

We next recall that $F_{\alpha, \beta}$ has $16 \times 16$
components. We check that it has the correct components for Type
II A and Type II B strings.

\begin{center}
\begin{tabular}{||l|l|l||}
\hline
\hline
String Type & Independent Tensors & Number of components \\
\hline
Type II A & $F_0$ & 1 \\
& $F_2$ & $\dis \frac{10 \times 9}{2!}$ \\
& $F_4$ & $\dis \frac{10 \times 9 \times 8 \times 7}{4!}$ \\
& & Total: 256 \\
\hline
Type II B & $F_1$ & 10 \\
& $F_3$ & $\dis \frac{10 \times 9 \times 8}{3!}$ \\
& $F_5 = \ast F_5$ (self dual) & $\dis \frac{10 \times 9 \times 8
\times 7 \times 6}{2 \times 5!}$ \\
& & Total: 256 \\
\hline
\hline
\end{tabular}
\end{center}

\subsection{Superconformal Invariance}

The super-Virasoro condition $G_0 |{\rm phys}> =0$ yields \beq
k_\mu \Gamma^\mu F = F k_\mu \Gamma^\mu = 0 \eeq which ultimately
leads to two field equations \beq d F = 0 = d \ast F.
\label{eqn32}\eeq The classical field equations in eq(\ref{eqn32})
arise from the condition of superconformal invariance of the RR
vacuum. We note that for RR background fields, the free massless
equation $dF= 0$ and the Bianchi identity $d \ast F = 0$ appear on
the same footing showing explicit duality.

From Bianchi identity,
$$
F_{\mu_1 \cdots \mu_k} = \frac{1}{(k - 1)!} \partial_{[\mu_1}A_{\mu_2
\cdots \mu_k]}
$$
where $A_{\mu_1 \cdots \mu_k} $ is the completely antisymmetric $U(1)$
gauge field so that $F_{(k)} = dA_{(k-1)}$.  Hence we have the following
RR-fields

\begin{center}
\begin{tabular}{||l|l|l||}
\hline \hline String Type & RR Background Fields & Number of
components
\\ \hline Type II A & $f_{(0)}$ & 0 \\ & $A_\mu^{(1)}$ & 8 \\ &
$A_{\mu \nu \lambda}^{(3)}$ & $\frac{8 \times 7 \times 6}{3!}$ \\
& & Total: 64 \\ \hline Type II B & $A^{(0)}$ & 1 \\ & $A_{\mu
\nu}^{(2)}$ & $\frac{8 \times 7}{2!}$ \\ & $A_{\mu \nu \lambda
\delta}^{(4)}$ self-dual & $\frac{8 \times 7 \times 6 \times 5}{2
\times 4!}$ \\ & & Total: 64 \\ \hline \hline
\end{tabular}
\end{center}

The sources for the $d=10$ RR-fields are the $D$-branes. We list
the $D$-brane couplings of Type-II strings. Note that in general
for tensor field $A^{(p + 1)}$ with electric coupling to $D$-p
brane, we also have the coupling of the dual $\tilde{A}^{q+1}$ to
$D$-q brane via magnetic couplings. Indeed, we have \beq A^{(p+1)}
\rightarrow dA_{(p+2)} \rightarrow d \tilde{A}_{(8 - p)}
\rightarrow \tilde{A}^{(7 - p)} \eeq so that the tensor field
$A^{(p+1)}$ couples to $D$-p brane while its dual $\tilde{A}^{7 -
p}$ couples to a $D-(6-p)$ brane. We have the following $D$-brane
coupling in $D=10$ dimensions:

\begin{center}
\begin{tabular}{||l|l||}
\hline
\hline
String Type & Couplings \\
\hline
Type II A & $F_{(0)} \rightarrow \tilde{F}_{(10)} \rightarrow
\tilde{A}^{(9)}$ couples to 8-brane \\
\hline
& $ A_{\mu}^{(1)} \mbox{\hspace{3mm}}$ couples to 0-brane  \\
& $ \tilde{A}^{(7)} \mbox{\hspace{3mm}}$ couples to 6-brane \\
\hline
& $ A^{(3)} \mbox{\hspace{3mm}}$ couples to 2-brane \\
& $ \tilde{A}^{(5)} \mbox{\hspace{3mm}}$ couples to 4-brane \\
\hline
\hline
Type II B &  $A^{(0)} \mbox{\hspace{3mm}}$ couples to -1-brane \\
& $\tilde{A}^{(8)} \mbox{\hspace{3mm}}$ couples to 7-brane \\
\hline
& $A^{(2)} \mbox{\hspace{3mm}}$ couples to 1-brane \\
& $ \tilde{A}^{(6)} \mbox{\hspace{3mm}}$ couples to 5-brane \\
\hline
& $ A^{(4)} \mbox{\hspace{3mm}}$ couples to 3-brane (self-dual) \\
\hline
\hline
\end{tabular}
\end{center}
\medskip

Finally, we summarize everything in a
`$D$-brane-scan'\cite{townsend}:

\begin{center}
\begin{tabular}{lcccccccccc}
Type II A & & $0_D$ &  & $2_D$ & &$4_D$ & & $6_D$ & & $8_D$\\
Type II B & $-1_D$ & & $1_D$& & $3_D^+$ & &$5_D$ & & $7_D$ & \\
Type I & & &$1_D$  & & & &$5_D$  & & &
\end{tabular}
\end{center}

\subsection{Massive Spectra of Type II A and II B}

Type II A and Type II B strings differ only for the massless
states; the massive states are identical.  While the massless
states can be chiral, the massive  spinor states cannot be chiral.
For example, at the first excited level in the Ramond sector, we
have states of opposite chirality, namely $$\alpha_{-1}^i |
0,k,\alpha,R>, \mbox{\hspace{5mm}} d_{-1}^i|0,k,\alpha,R>  $$
which combine into a massive representation of a Majorana fermion.
This means that the excited massive states are insensitive to the
choice of the massless ground states as they should be since
massive states have both chiralities.

If we compactify one space dimension into $S^1$ with radius $R_A$
for Type II A and with radius $R_B$ for Type II B, then we can
show that the theories possess identical spectrums (including the
massless sector) provided \beq R_A R_B = \alpha^\prime. \eeq This
symmetry is called $T$-duality.

\subsection{Coupling to $D$-branes}

All antisymmetric tensor fields $A^{(k)}$ are $U(1)$ gauge fields
and hence their charges can be defined by Gauss' law.  The field
tensor $A^{(k)}$ couples to a $D-(k-1)$ brane which has
coordinates $X^\mu(\sigma_1, \cdots, \sigma_{k-1}, \tau)$.  Let
$J^{(k)}$ be the $k$-form tangent to the brane.  This acts as the
source for the RR-field\cite{pol4} given by \beq d F^{(k+1)} =
\mu_{k-1} \ast J^{(k)} \eeq where $\mu_{k-1}$ is the charge of the
$(k-1)$-brane. We can define `electric' RR-charge by \beq e_k =
\int_{S^{d-K-1}} \ast F_{(k+1)} \eeq and a dual `magnetic'
RR-charge by its coupling to a $D-k-3$-brane by \beq g_{d-k-2} =
\int_{S^{k+1}} F_{k+1}. \eeq The equivalent Dirac quantization
condition is \beq e_k ~ ~ g_{d-k-2} = 2 \pi n \eeq where $n$ is an
integer.  It turns out that in string theory $n = 1$ for
$D$-branes and the charges $e_k$ and $g_p$ are dimensionless only
for $d= 2(k+1)$. Thus

\begin{center}
\begin{tabular}{lll}
$d= 4$ & $k = 1$ & 0-brane (particle) \\ $d= 6$ & $k = 2$ &
1-brane (string) \\ $d= 10$ & $k = 4$ & 3-brane  \\
\end{tabular}
\end{center}

In the presence of a background gauge field living on the $k$-brane
with field tensor $F= F_{\mu \nu } dx^\mu dx^\nu$, the coupling is
given by
\beq
dF^{(k)} =
\mu_{k-1} J^{(k)} \ast  {\rm tr} \mbox{\hspace{3mm}} \expon^{F/2\pi}
\eeq
which comes from a term in the $D$-brane action of the form
\beq
\mu_{k-1}
\int_{\Sigma_k} c \wedge {\rm tr}
\mbox{\hspace{1mm}} \expon^{F/2\pi} \label{dact}
\eeq
where $c$ is the differential form given by the sum of the
RR-fields. For Type II A theory, for instance,
$$
c = A_\mu^{(1)} dx^\mu + A_{\mu \nu \lambda}^{(3)} dx^\mu dx^\nu
dx^\lambda \equiv A^{(1)} + A^{(3)}.
$$
For example,
using eq(\ref{dact}),
we have $\dis {\rm tr}
\mbox{\hspace{3mm}} \int_{\Sigma^5} A^{(1)} \wedge F \wedge
F$ on a 4-brane so that the $F \wedge F$ term couples to the field
$A^{(1)}$ giving a 0-brane charge equal to $\dis \left( \int_{\Sigma^4}
F \wedge F \right) \int_{\Sigma^1} A^{(1)}$.  In other words, by having
a non-trivial background gauge field in the 4-brane, we have effectively
created a 0-brane embedded within the 4-brane carrying the right charge
to couple to the RR field $A_{\mu}^{(1)}$.

There is a symmetry in Type II strings which is a $U(1)$ gauge symmetry
$\delta F = d\Lambda$.  To preserve this symmetry we need to modify $F$
to $F + B$, where $B_{\mu \nu}$ is the NS-NS field.  The complete action
for the single $(k+1)$-form gauge field $A^{(k+1)}$ is given by
\begin{eqnarray*}
S & =& \frac{1}{2} \int \ast F^{(k+2)} \wedge F^{(k+2)} d^{10}x + \mu_k
\int_{\Sigma } c \wedge
{\rm tr} \mbox{\hspace{3mm}}  \expon^{\frac{F + B}{2 \pi}} \\
& & \mbox{\hspace{10mm}}
+ i \mu_k \int_{\Sigma}
A_{\mu_1 \cdots \mu_{k+1}}^{(k+1)} \partial_1 X^{[\mu_1}
\cdots \partial_{k+1} X^{\mu_k + 1]} d^{k+1} \sigma
\end{eqnarray*}
where $\Sigma$ is the k-brane world volume.  The charges satisfy the
Dirac quantization condition
$$ \mu_{6-k} \mu_k = 2 \pi.$$

\bigskip

\section{Bosonic String}

\bigskip

Let $X^\mu(\sigma,\tau)$ be the coordinates of the string in
$d$-dimension with world volume $M$.  In the conformal gauge, we
have ($\alpha^\prime$) $$ S = \frac{1}{2 \pi} \int_M d^2 \sigma
\partial^\alpha X^\mu
\partial_\alpha X_\mu.
$$
Reparametrization invariance is obtained by imposing the Virasoro
conditions on the state space.  To obtain the field equations we have
$$
\delta S = - \frac{1}{\pi} \int_M d^2 \sigma \delta X^\mu \partial^2
X_\mu + \frac{1}{2 \pi} \int_{\partial M} d \tau \delta X^\mu \partial_n
X_\mu
$$
where $\partial_n$ is the derivative normal to $\partial M$.
We choose the boundary conditions so that the boundary term in
$\delta S $ reduces to zero. Consider Cartesian coordinates:
\begin{center}
\epsfig{file=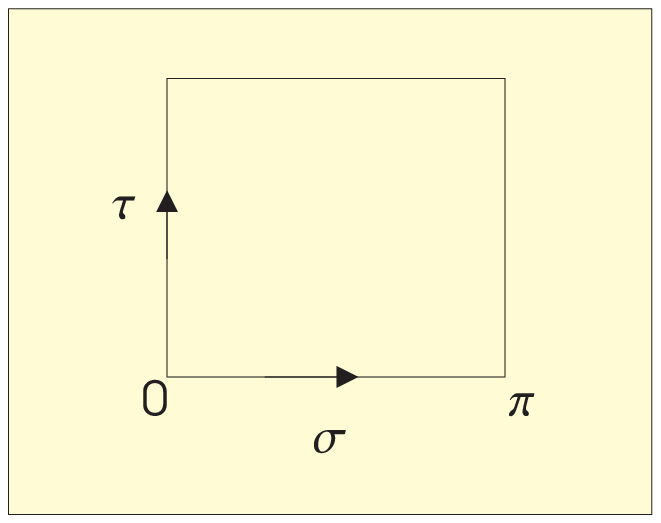, height=.25\textheight}
\end{center}

The boundary term with $\dis \partial_n = \frac{\partial}{\partial
\sigma} = ~^\prime$ is given by $$ \frac{1}{2\pi} \int
\left[\delta X^{\mu} X^\prime_\mu(\pi,\tau) - \delta X^\mu
X^\prime_\mu(0,\tau) \right] d \tau. $$ This gives rise to several
cases\cite{pol2,pol3}:
\begin{itemize}
\item \underline{Closed String} \\
$X_\mu(0,\tau) = X_\mu(\pi,\tau)$, periodic. \\
Usually, in this case, one extends $\sigma$ to $ 0 \leq \sigma \leq 2\pi$.
\item \underline{Open String} \\
There are two possible boundary conditions:
\begin{itemize}
\item Neumann (NN) b.c.: $\dis X^\prime_{|0,\pi} =0$
\item Dirichlet (DD) b.c.: $\dis \delta X_{|0,\pi} = 0$

\end{itemize}
This gives a total of four possible boundary conditions on the two ends
of the open strings,
namely NN, ND, DN, DD.
\end{itemize}

For a $D$-$k$ brane in 10 dimensions,
\begin{center}
\begin{tabular}{lll}
(NN) & $X^\prime_m(0)=0=X^\prime_m(\pi)$,
& $m = 0,1, \cdots k$ \\
(DD) & $X_i(0) =$ constant &  \\
& $X_i(\pi) =$ constant $^\prime$ &  $i = k+1, \cdots 9$
\end{tabular}
\end{center}

\begin{center}
\epsfig{file =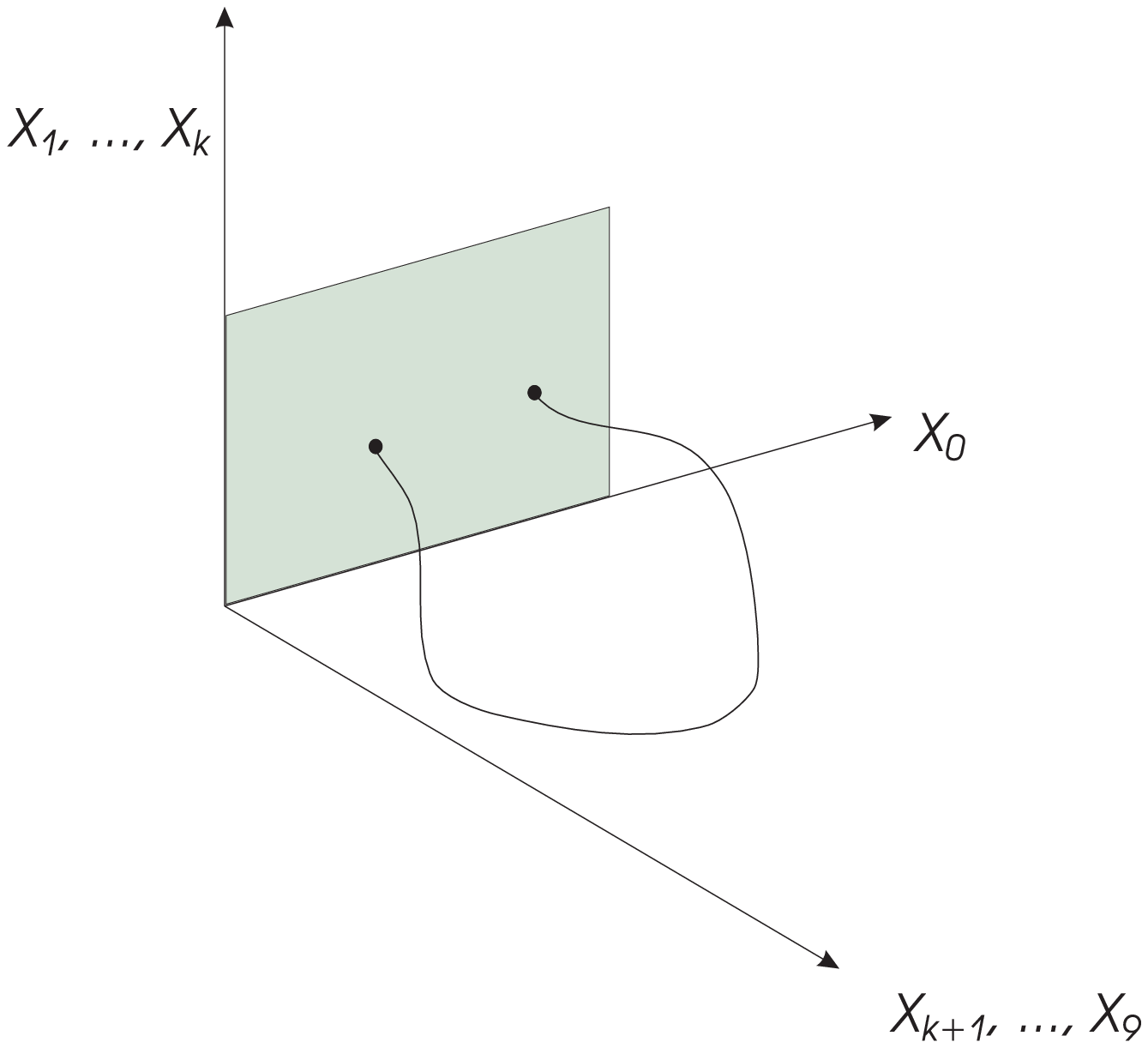, height=.3\textheight}
 \end{center}
The ends of the open strings are free to move in the $(k+1)-$dimensional
world-volume of the $k$-brane.  The $D-$brane is thus a rigid
submanifold of spacetime.

For the usual open string which is part of the type I superstring, we have
$k=9$, i.e. all the components satisfy NN boundary conditions and the
ends are free to move throughout spacetime.

What is the consequence of the NN and DD boundary conditions
on the fermions?  As
before,  for the open strings, the overall sign  is fixed by
$\psi_+^\mu(\pi,\tau) = \psi_-^\mu(\pi,\tau)$.  To preserve worldsheet
supersymmetry, we have
\begin{center}
\begin{tabular}{lll}
NN: & $\psi_+^m(0,\tau) = \pm \psi_-^m (0,\tau)$ & $m = 0,1,\cdots ,
k$ \\
DD: & $\psi_+^i(0,\tau) = \mp \psi_-^i(0,\tau)$ & $i = k+1,\cdots ,
9$
\end{tabular}
\end{center}

$D$-branes are BPS solitonic states and this is a consequence of
the b.c's of the worldsheet fields.  Of the two supercharges of
Type II theory $Q_\alpha^L, Q_\alpha^R$, only the linear
combination $\dis Q_\alpha^L + \expon^{i\phi_k} Q_\alpha^R$ is
conserved in the presence of $D$-branes, with $\phi_k$ being the
phase coming from parity transformations for the transverse
directions $X^i, i = k, k+1, \cdots, 9$.  Since the $D$-branes
have half the supersymmetry of Type II, they form BPS states.

\subsection{Spectrum of states}

For the open string with NN b.c.'s, we solve $\partial^2 X^\mu = 0$ to obtain
$$
X^\mu(0,\tau) = x^\mu + p^\mu \tau + i \sum_{n \neq 0} \frac{\expon^{in\tau}}{n
} \alpha_n^\mu \cos(n \sigma)
$$
with $[X^\mu, p^\nu] = i \eta^{\mu,\nu}$, $[\alpha_m^\mu, \alpha_n^\nu ]
= m \delta_{m+n} \eta^{\mu \nu}$.

The Virasoro operator ($\alpha_0^\mu = p^\mu$), $$ L_0 =
\frac{1}{2} \alpha_0^2 + \sum_{m = 1}^{\infty} \alpha_{-m}
\alpha_m $$ imposes constraint on physical states
$$ (L_0 -1)|\mbox{\rm phy}> = 0 $$
which in turn yields the mass spectrum as $$ M^2 = - p^\mu p_\mu =
\sum_{m=1}^{\infty} \alpha_{-m} \alpha_m - 2 $$ Define $|0,k>_B$
by $\alpha_m |0,k>_B = 0$ for all $ m> 0$. We then have the tachyon
vacua and the massless state given respectively by 
\begin{eqnarray*}
\mbox{{\rm tachyon:}} & |0,k>_B, & M^2 = -1 \\
\mbox{{\rm photon:}} & \alpha_{-1}^\mu |0,k>_B, & M^2 = 0
\end{eqnarray*}

\subsection{Chan-Paton factors}

We can attach non-dynamical degrees of freedom to
the ends of the open string.
 \begin{center}
\epsfig{file =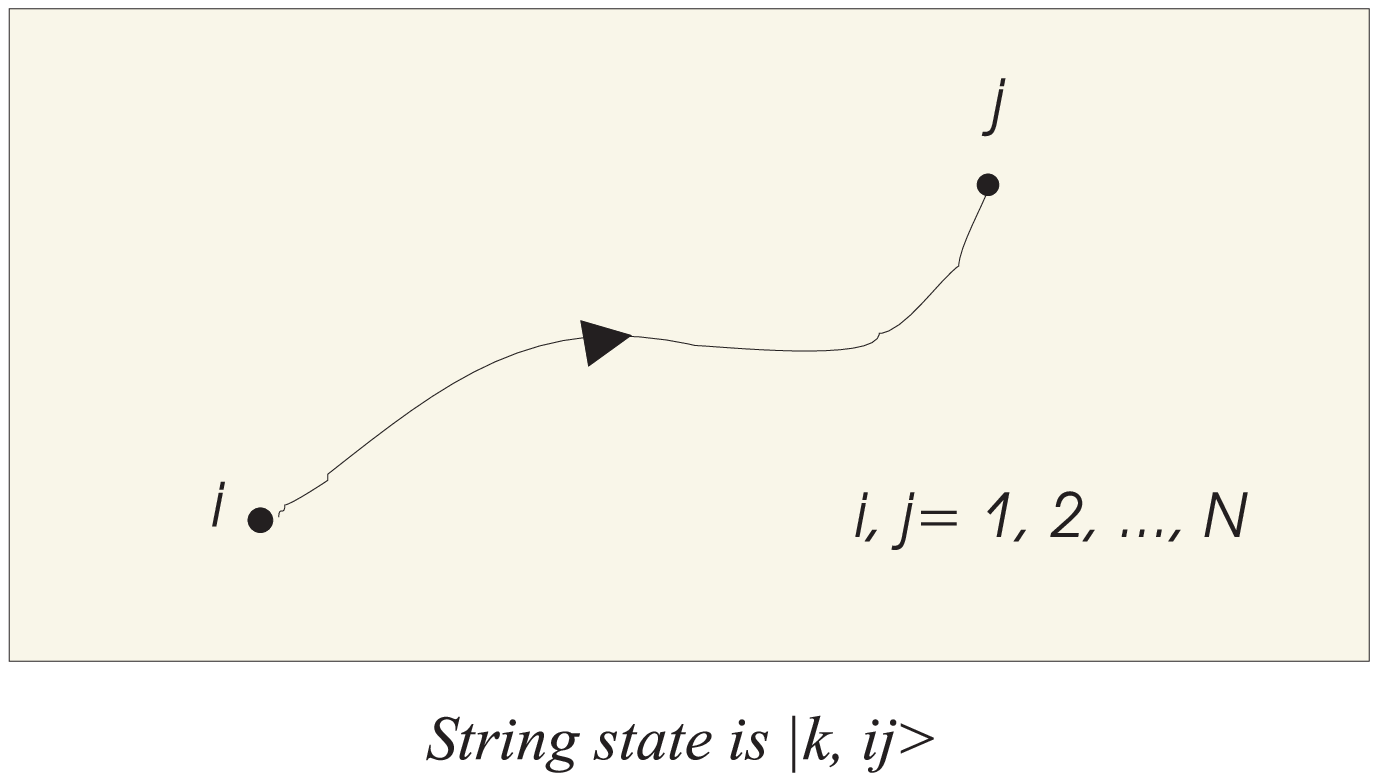, height=.25\textheight}
\end{center}
The $N \times N$ matrices $\lambda_{ij}^a $ form a basis to
decompose the string wave function into an irrep. These matrices
are called the Chan-Paton factors\cite{chan}.
 We have the
irreducible string state given by $$ | k , a> = \sum_{ij}
\lambda_{ij}^a |k, ij> $$ Unitary demands $\lambda^\dagger =
\lambda$, so that $\lambda$ is Hermitian.  The scattering of two
oriented strings appears as:
 \begin{center}
\epsfig{file =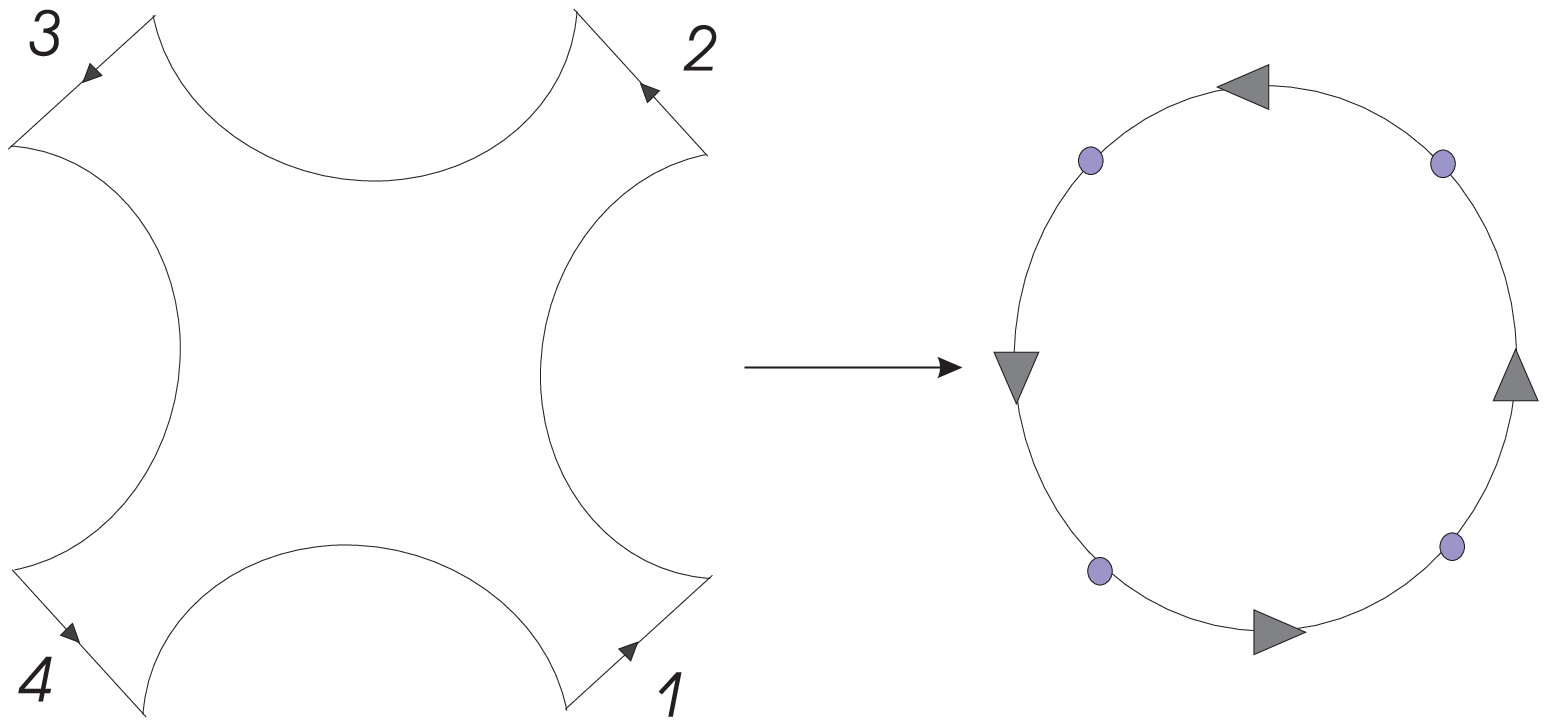, height=.15\textheight}
 \end{center}

This diagram has a factor ${\rm tr} (\lambda^1 \lambda^2 \lambda^3
\lambda^4)$.  For an oriented string, one can think of the two ends
carrying the fundamental $N$ and $\bar{N}$ as:
\begin{center}
\epsfig{file =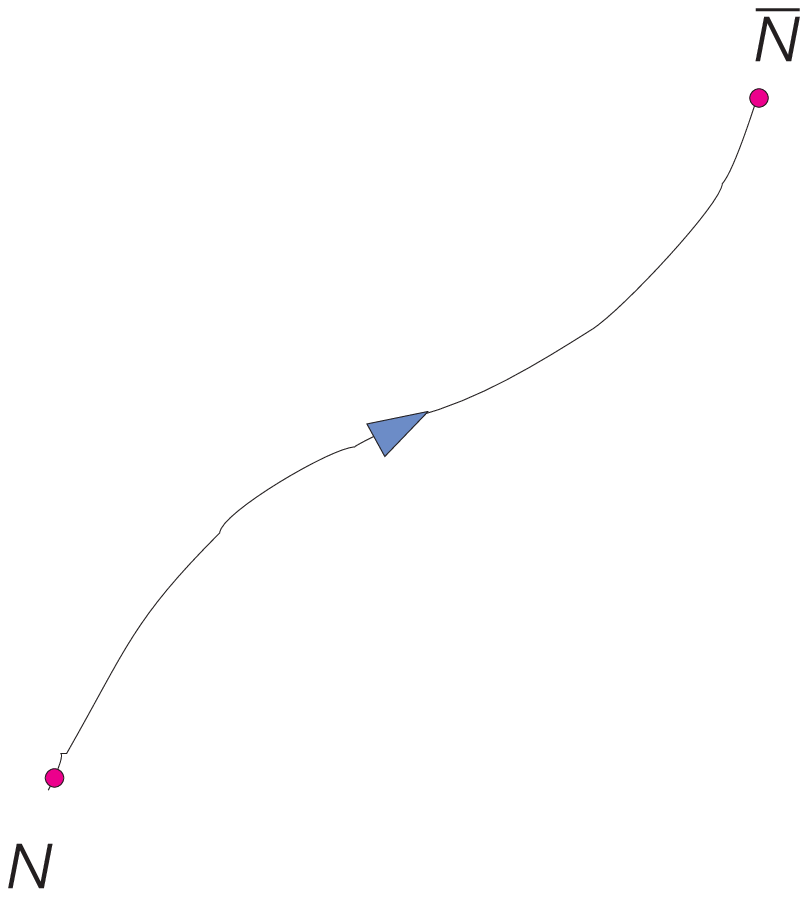, height=.15\textheight}
 \end{center}

\noindent
In order that the massless excitations of the open string can be
precisely those of Yang-Mills gauge fields $N \otimes \bar{N} \sim
\lambda^a$ since the fields have to be expressed in the adjoint representation
of the Lie group, $G$.  There are only three solutions for consistent
open strings given by:

\begin{eqnarray*}
G = U(N) & \bar{N} = N^\ast & {\rm orientable \;\;strings} \\
G = SO(N) & \bar{N} = N & {\rm non-orientable \;\;strings} \\
G = Sp(N) & \bar{N} = N & {\rm non-orientable \;\;strings}
\end{eqnarray*}

We will later show that in $d = 10$ only the $SO(32)$ case is
consistent. Hence we have to analyze the non-orientable strings.

The $D$-brane interpretation is that there are exactly 32 degenerate
9-branes if the open superstring is to avoid tadpole singularities.  The
subscripts $i,j$ in $\lambda_{ij}$ indicates on which of the
degenerate 9-branes
the two endpoints are on.

\subsection{Non-oriented strings}

 \begin{center}
\epsfig{file =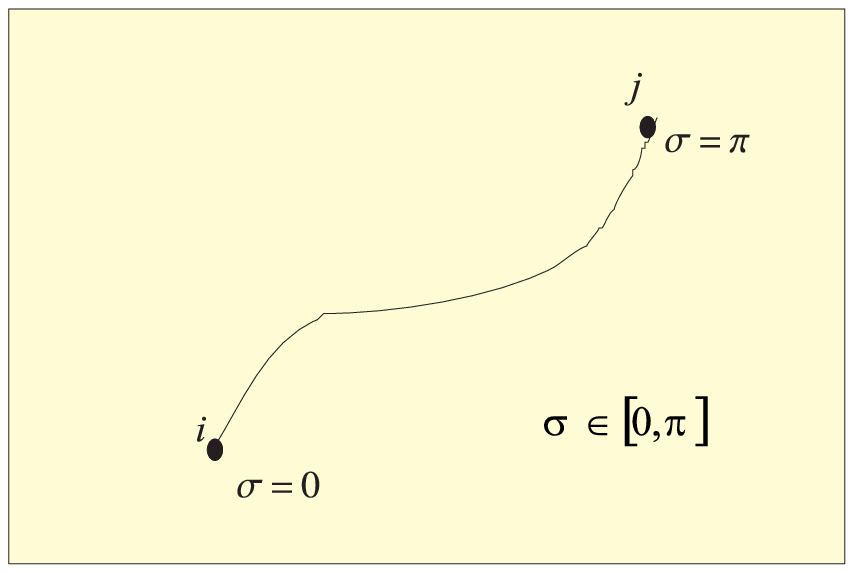, height=.25\textheight}
 \end{center}

For the case above in which the indices $i$ and $j$ are equivalent, we
can associate a vertex operator $V_{\Lambda}(k,\tau)$ for every physical
state $\Lambda$ with momentum $k^\mu$ by
$$
V_{\Lambda}(k,\tau) = \expon^{i \tau L_0} V_{\Lambda}(k,0) \expon^{-i \tau
L_0}.
$$
This vertex operator describes the emission of the state $\Lambda$ from
the $\sigma=0$ end of the open string.  For example, for the string
propagator $\Delta = (L_0 - 1)^{-1}$, the amplitude $<1|V_2(k_2) \Delta
V_3(k_3) \Delta \cdots V_{M-1}(k_M) |M>$ is represented by:
\begin{center}
\epsfig{file =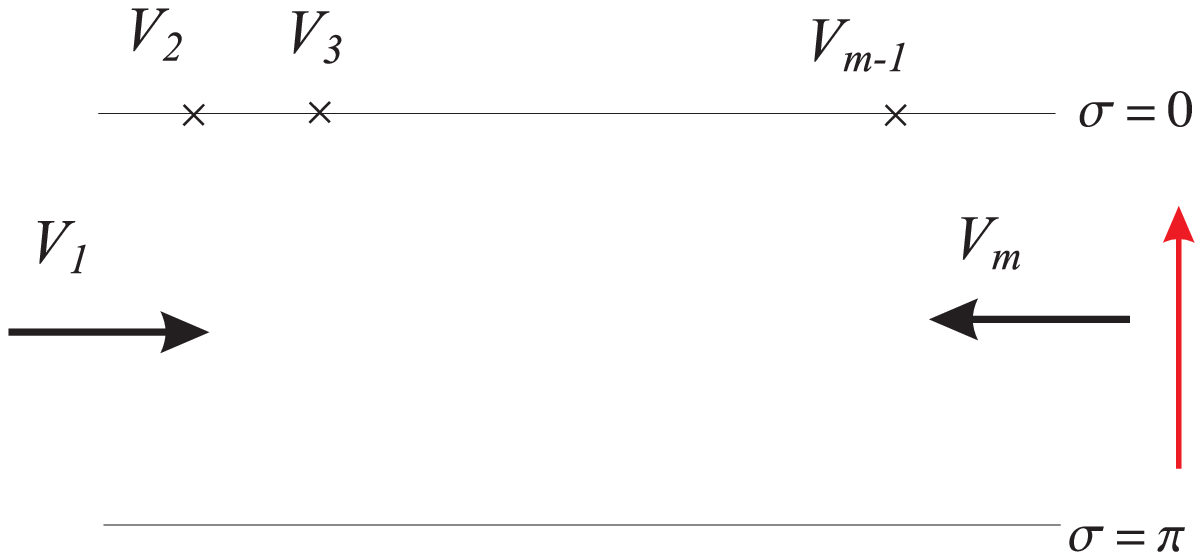,height=.2\textheight, width=.9\textwidth}
 \end{center}

For non-orientable open string, emission from $\sigma =0$ edge
should be symmetric with emissions from $\sigma = \pi$ edge.  This
entails the requirement that the vertex operators be built from
$X(\sigma = \pi , \tau)$ instead of $X(\sigma =0,\tau)$. To
achieve this , we introduce the `twist operator' $\Omega$ which
maps $\sigma$ to $\pi - \sigma$, so that $$ V(\sigma = \pi ) =
\Omega V(\sigma= 0) \Omega^{-1}. $$ Clearly, $\Omega^2 = 1$.
Replacing $\sigma = 0$ by $\sigma = \pi$ for the open string can
be achieved by a change of variables $\sigma \rightarrow \pi -
\sigma = \sigma^\prime$; hence
\begin{eqnarray*}
X(\sigma^\prime, \tau) &= & x^\mu + p^\mu \tau + i
\sum_{n}^\prime\frac{\expon^{-i n\tau} }{2n} \alpha_n^\mu \left[
\expon^{-in\pi} \expon^{i\sigma}\ + \expon^{+in\pi} \expon^{-
i\sigma} \right] \\ & = & ~ ~ \Omega X(\sigma, \tau)\\
\Rightarrow
\Omega \alpha_n & = & (-1)^n \alpha_n, ~ ~  \Omega x^\mu  = x^\mu \\
\Omega p^\mu & = & p^\mu, ~ ~ \Omega(\sigma) = \pi - \sigma
\end{eqnarray*}
Hence $\Omega = (-1)^{N_B}$, where $N_B$ is a bosonic number
operator with $[N_B, \alpha_n] = n \alpha_n$. The vacuum has
$\Omega |0, k>_B = + |0, k>_B$. In terms of these operators, for
the photon, we have $$ \Omega \alpha_{-1}^\mu |0, k> = -
\alpha_{-1}^\mu |0,k> $$ The world sheet parity operator acts
non-trivially on the Chan-Paton factors. $$ \Omega \lambda_{ij}
|k, ij> =  \lambda_{ij}^\prime |k, ij> $$ where $\lambda^\prime
\equiv M \lambda^T M^{-1}$.  Since $\Omega^2 = 1$, we have in
general $$ \lambda^a = M M^{-T} \lambda^{a T} M^T M^{-1}. $$ Since
$\lambda^a$ must form a complete set by CPT, we have by Schur's
lemma $M^T M^{-1} = \pm 1_N$.  There are two possibilities:
\begin{center}
\begin{tabular}{rl}
1. & $M^T = + M = I_N$  \\
tachyon: & $\Omega \lambda_{ij} |0,k,ij> = \lambda_{ij}^T |0,k,ij>$ \\
photon: & $\Omega \lambda_{ij} \alpha_{-1}^\mu |0,k,ij> = -
\lambda_{ij}^T \alpha_{-1}^\mu |0,k,ij>$ \\
& \\
2. & $\dis M^T = - M = i \left(
\begin{array}{ll}
0 & I_{N/2} \\
I_{N/2} & 0
\end{array} \right) \equiv J $ \\
tachyon: & $\Omega \lambda_{ij} |0,k,ij> = (J \lambda^T J)_{ij} |0,k,ij>$ \\
photon: & $\Omega \lambda_{ij} \alpha_{-1}^\mu |0,k,ij> = -
(J \lambda^T J)_{ij} \alpha_{-1}^\mu |0,k,ij>$
\end{tabular}
\end{center}

For non-orientable strings, invariance under $\Omega$ is a symmetry of
the action and all amplitudes.  To obtain a Hilbert space of states
invariant under $\Omega$, we take the oriented string spectrum and
project out states with the projector $\dis P = \frac{1}{2} (1 +
\Omega)$, with $P^2 = P$.

For the open string with Chan-Paton charges, the operator $P$ will
project out the tachyon and retain the photon if we demand for:
\begin{center}
\begin{tabular}{lll}
Case 1 & $\lambda^T = - \lambda$ & i.e. $SO(N)$ gauge group \\
Case 2 & $J\lambda^TJ = - \lambda $ & i.e. $Sp(N)$ gauge group \\
\end{tabular}
\end{center}
Nonorientable worldsheets arise in the following way.  For every
oriented string amplitude with vertex operators, the one loop diagram
gives a sum over a complete set of intermediate states I.
 \begin{center}
\epsfig{file =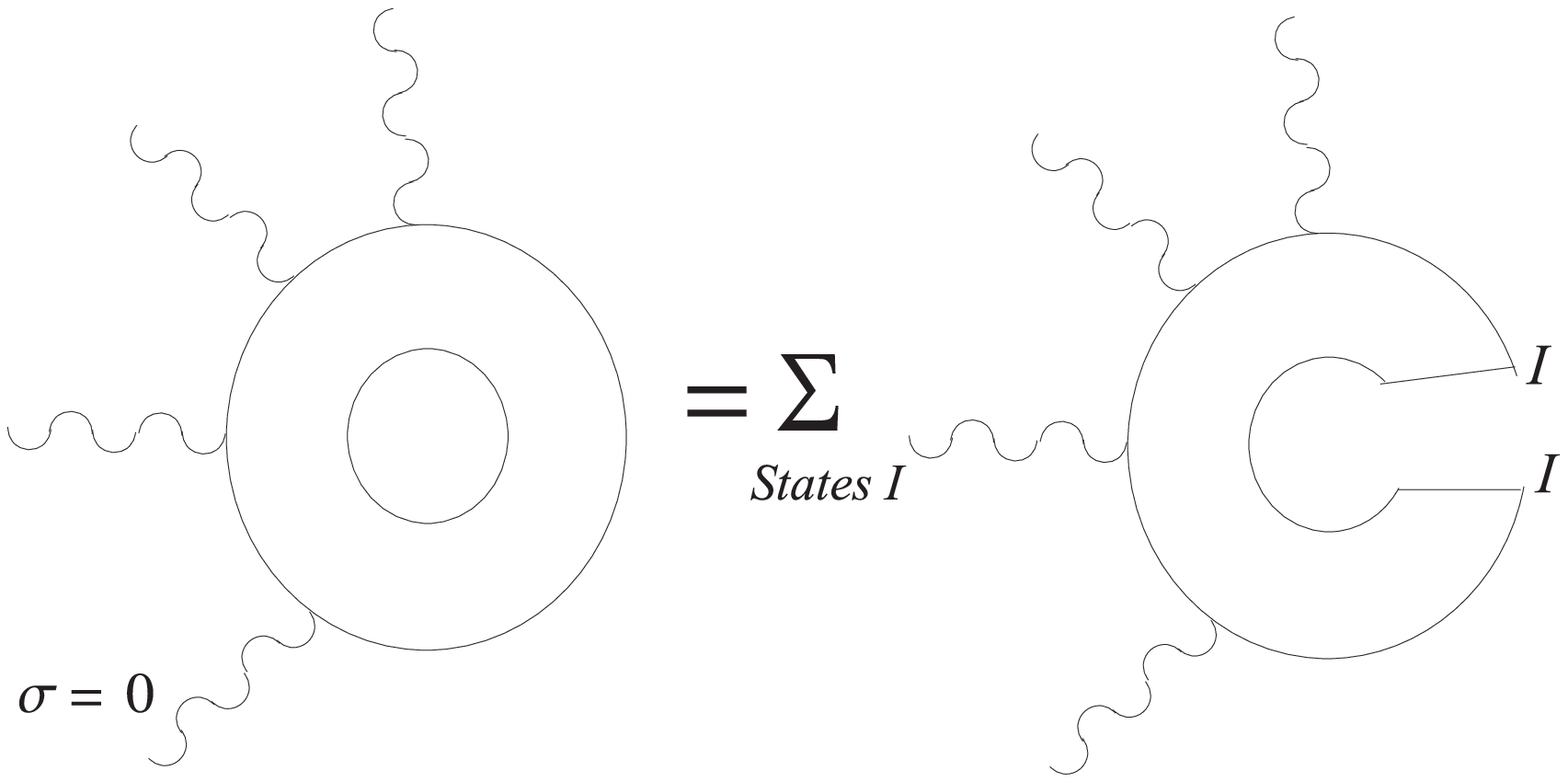, height=.25\textheight}
\end{center}
But when the projector $\dis \frac{1}{2} ( 1 + \Omega)$ is
inserted into the trace it switches the emission operators from
$\sigma = 0$ to $\sigma = \pi$.
 \begin{center}
\epsfig{file =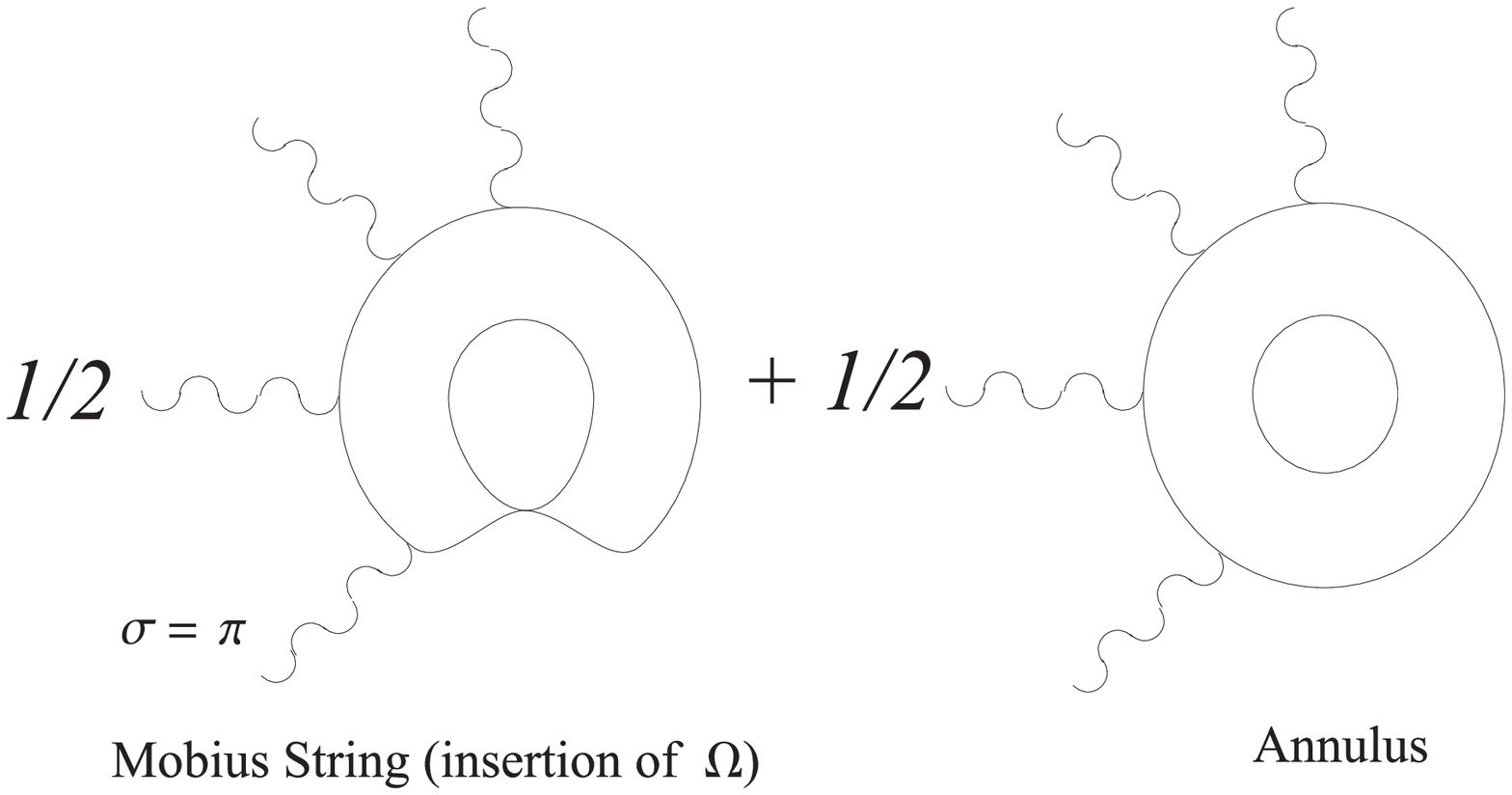, height=.25\textheight}
 \end{center}

\subsection{Type I Superstring}

Type I is constructed from open strings (both oriented and
non-oriented) as well as the closed string sector.  The Hilbert
space ${\cal F}_{open} $ is $\Omega$-symmetric, i.e. $$ \Omega
{\cal F}_{open} = {\cal F}_{open} $$ In GSO projected RNS
superstring, the Type I superstring sector yields an ${\cal N}=1$
supersymmetric Yang-Mills multiplets in $d=10$ with $SO(32)$ being
the only consistent gauge group.  Thus $$ \left.
\begin{array}{rll}
{\rm NS:} & 8 \times 496 & A_\mu \\
{\rm R:} & 8 \times 496 & \psi^a
\end{array} \right\} {\rm space-time \;\;\; susy}
$$
 \begin{center}
\epsfig{file =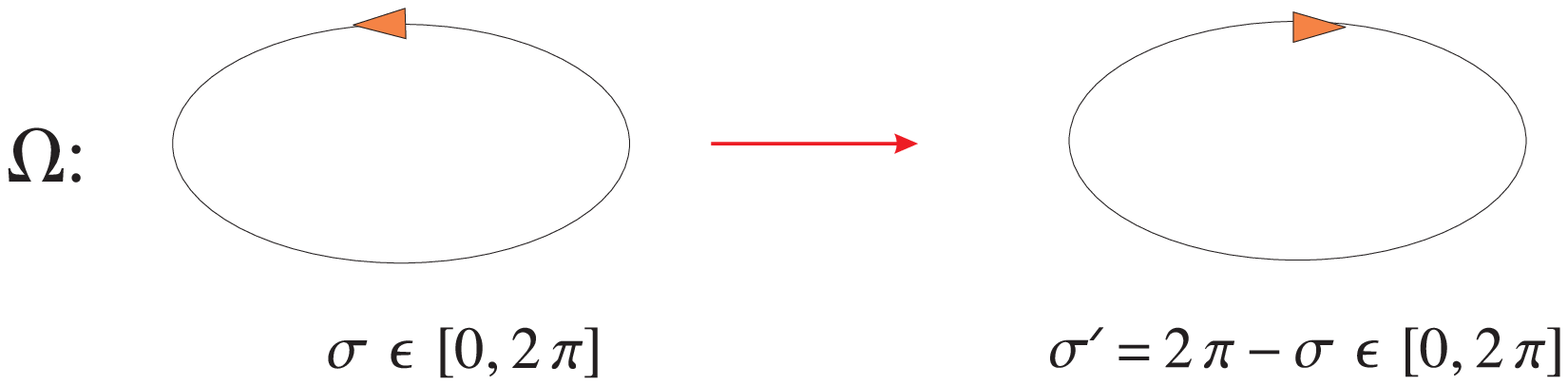, height=.15\textheight,width=.8\textwidth}
 \end{center}
For  the closed string sector, the action of $\Omega$ is to switch the
left and right movers.  To be symmetric under $\Omega$, the left and
right movers must have the same parity.  Hence, we must start with IIB
and retain only states invariant under $\Omega$, i.e. $
{\cal F}_{open,\; closed \; sector} = {\cal F}_{IIB} / \Omega$.

On making this projection, we obtain the Type I closed string sector
with space-time supersymmetric spectrum given by
\begin{center}
\begin{tabular}{llll}
NS-NS & bosons & $G_{\mu \nu}$       & 35  \\
      &        & $\Phi$              & 1  \\
 R-R   & bosons & $A^{(2)}_{\mu \nu}$ & 28 \\
      &        & Total:                    &  64 \\
NS-R  & fermions & $\chi_{\mu}^\alpha$  &  56  \\
R-NS  & fermions & $\lambda_\alpha$              & 8   \\
      &        & Total:                    &     64
\end{tabular}
\end{center}

From the RR-sector, we see that the field $A_{\mu \nu}^{(2)}$ in
Type I couples to a D 1-brane, which is dual to a D5-brane.  It
will turn out that it also couples to a D9-brane.

\bigskip

\section{Background Fields: Supergravity}\label{sectbackgrd}

\bigskip

Recall that the Hilber spaces of Type I and the Heterotic strings
yield an irrep of $d=10$, ${\cal N}=1$ spacetime supersymmetry,
whereas Type II strings yield an irrep of $d=10$, ${\cal N}=2$
supersymmetry. We now study what are the allowed background fields
for the various superstring theories.  We will see that in the
limit $\alpha^\prime \rightarrow \infty$, spacetime supersymmetry
will completely determine the allowed string vacua corresponding
to different background field configurations.  All the background
fields are the result of the condensation of the massless
excitations of the superstrings, since these condensates can be
added to the action without violating superconformal invariance.
Condensates of massive excitations have a mass scale and are
therefore  disallowed as they violate superconformal invariance.

Let us start with the bosonic string.  Instead of expressing the
world sheet action for the string in the conformal gauge as in
eq(\ref{bose}), we write the action, without any particular choice
of gauge, in the Polyakov formulation.  In addition to dynamical
quantum fields $X_\mu (\sigma, \tau)$, we have a dynamical world
sheet metric $\gamma_{\alpha, \beta}$.

Let $g_{\mu \nu} (X)$ be the condensate of spin 2 excitations of
the string, $B_{\mu \nu}(X)$, an anti-symmetric background field
and $\phi(X)$, the dilaton.  Taken together, $g_{\mu \nu} (X)$,
$B_{\mu \nu}(X)$ and $\phi(X)$ are the $d=10$ Neveu-Schwarz (NS)
background fields.

The fundamental open string $X_\mu$ couples to these background
fields and the coupling is given by (restoring $\alpha^\prime$)
\begin{eqnarray} 
S = \frac{1}{4 \pi \alpha^\prime} \int & d^2\sigma & \sqrt{-
\gamma} \{ \gamma^{ij} g_{\mu \nu}(X) \partial_i X^\mu \partial_j
X^\nu \nonumber
\\ & + & \epsilon^{ij} B_{\mu \nu} (X) \partial_i X^\mu \partial_j
X^\nu \nonumber \\ & + & \alpha^\prime {\cal R}^{(2)}(\gamma)
\phi(X) \} \end{eqnarray} where ${\cal R}^{(2)}(\gamma)$ is the
world sheet Ricci scalar curvature.

The quantum field theory is defined by \beq 
Z = \int D \gamma DX \expon^{S\left[ X, \gamma \right]} \eeq On
performing the path integration, one finds that if one demands
that the theory have superconformal invariance, this implies that
all the beta functions must be zero. That is, \beq 
\beta_g = 0 = \beta_B = \beta_{\phi} \eeq To lowest order
(one-loop) in $\alpha^\prime$, one finds that the background
fields $g_{\mu \nu}$, $B_{\mu \nu}$ and $\phi$, together with some
additional fields, must satisfy the classical field equations of
$d=10$, ${\cal N }=1$ supergravity!  In other words, as seen
earlier for the case of RR-fields, superconformal invariance
demands that the background fields must satisfy certain classical
field equations which in turn can be obtained from an effective
lagrangian field theory.

For concreteness, we analyze Type IIA in some detail.  The irrep
of the supersymmetry algebra for IIA given in eq(\ref{type2a})
yields the following bosonic fields

\begin{tabular}{lll}
& $g_{\mu \nu}, \phi, B_{\mu \nu}$ \hspace{2cm} & NS-NS \\ IIA & & \\ 
& $A_{\mu}^{(1)}, A_{\mu \nu \lambda}^{(3)}$ & RR 
\end{tabular}
\medskip

The fermionic superpartners come from the $NS-R$ sector.  The
bosonic component of the non-chiral ${\cal N}=2$ supergravity
lagrangian, in the string frame, is given by \begin{eqnarray} 
S_{IIA} & = & \int d^{10} X \sqrt{-g} \left\{ \expon^{-2 \phi}
\left[ R + 4 |d\phi|^2 - \frac{1}{3} |H|^2 \right] - |K|^2 -
\frac{1}{12}|G|^2 \right\} \nonumber \\ & & \mbox{\hspace{2cm}} +
\frac{1}{144} \int d^{10} X G \wedge G \wedge B \label{bkg1}
 \end{eqnarray}
where $H= dB$, $K= dA^{(1)}$ and $G= dA^{(3)} + 12 B \wedge K$.
Note the significant fact that the $RR$-fields had no direct
coupling to the fundamental string $X_\mu$, and they were `dragged
into' the action purely by demanding supersymmetry.  These
$RR$-fields, of course, exist in closed superstring theory, and
are excited by the presence of $D$-branes in the ground state of
the theory.

Actions similar to $S_{IIA}$ exist for the other consistent string
theories. For Type IIB, we have the background fields given by

\begin{tabular}{lll}
& $g_{\mu \nu}, \phi, B_{\mu \nu}$ \hspace{2cm} & NS-NS (as in Type IIA) \\ IIB & & \\ 
& $A^{(0)}, A_{\mu \nu}^{(2)}, A_{\mu \nu  \lambda \delta}^{(4) +}$ (self-dual)& RR 
\end{tabular}

\noindent and the ${\cal N}=2, d = 10$ effective chiral
supergravity action can be expressed as
\begin{eqnarray} 
S_{IIB} & = & \int d^{10} X \sqrt{-g} \left\{ \expon^{-2 \phi}
\left[ R + 4 |d\phi|^2 - \frac{1}{3} |H|^2 \right] - 2 |d
A^{(0)}|^2 \right. \nonumber \\ & & \left. - \frac{1}{3} |H^\prime
- A^{(0)} H|^2 - \frac{1}{60}|M^+|^2 \right\} - \frac{1}{48} \int
d^{10}X C^+ \wedge H \wedge H^\prime \label{bkg2}
 \end{eqnarray}
where $H = dB$, $H^\prime = dA^{(2)}$ and the full non-linear
Bianchi identity satisfied by the self-dual 5-form $M^+$ is now
$dM^+ = H \wedge H^\prime$ with $M^+ = dA^{(4) +}$.  For Type I
action, the background fields are

\begin{tabular}{lll}
& $g_{\mu \nu}, \phi$ \hspace{2cm} & NS-NS \\Type I &  $A_{\mu}^{(1)}$ & YM 1-form \\
& $A_{\mu \nu}^{(2)}$ & RR 
\end{tabular}

\noindent so that the bosonic sector of the Type I effective
action is
\begin{eqnarray} 
S_{I} & = & \int d^{10} X \sqrt{-g} \left\{ \expon^{-2 \phi}
\left[ R + 4 |d\phi|^2 \right] - \expon^{-\phi} {\mbox{\rm
tr}}|{\cal F}|^2 - \frac{1}{3}|H^\prime|^2 \right\} \label{bkg3}
 \end{eqnarray}
where ${\cal F}$ is the YM 2-form and $H^\prime = dA^{(2)}$ as
before.

There are two important applications of the effective field theory
for the background fields. One direction is compactification based
on the classical solutions of eq(\ref{bkg1})- eq(\ref{bkg3}) for
which all the background fields are zero except for a constant
dilaton and the metric which has the form $M_4 \times K_6$. Note
that $M_4$ is 4-dimensional Minkowski space and $K_6$ is some
6-dimensional (compact) space with Euclidean signature.  If one
demands $d = 4, {\cal N}=1$ supersymmetry for the state space
\footnote{More details regarding this aspect can be found in the
discussion on F-theory} on $M_4$, it can be shown that the
manifold $K_6$ must be Calabi-Yau space.

The other direction is to look for solitonic solutions of
eq(\ref{bkg1})- eq(\ref{bkg3}).  In fact, it is known that the
$Dp$-branes are solitonic solutions of the supergravity
lagrangian.  The solution for an arbitrary $Dp$-brane with the
brane volume in the $(1, 2,
\cdots, p)$ directions in $M$ is given by 
\beq 
ds^2 = f^{-1/2} (-dt^2 + dx_1^2 + \cdots + d x_p^2) + f^{1/2} (
dx_{p+1}^2 + \cdots + dx_{q}^2 ) 
\label{branemetric} \eeq with 
\bseq
\expon^{-2 \phi} & = & f^{\frac{p-3}{2}} \\ A_{0 \cdots p}^{(p)}
&=& - \frac{1}{2} (f^{-1} -1) \\ f & = & 1 + \frac{N c}{r^{7 - p}}
\\
r^2 & = & x_{p+1}^2 + \cdots + x_{q}^2 \\ c & = & \frac{(2 \pi
\sqrt{\alpha^\prime})^{7 -p}}{(7 - p) \Omega_{8-p}} g_s \\ 
\Omega_q
& = & \frac{2 \pi^{(q + 1)/2}}{\Gamma\left[ (q + 1)/2\right]}
\label{breq1}\eseq 
Note that for $p=3$, the dilaton decouples from the D3-brane. The
RR-field $A^{(p)}$ is excited since the D $p$-brane carries
RR-charge equal to $N$.  This solitonic solution is a $BPS-$state
with mass/volume $= N$. A black $p$-brane is a non-extremal black
hole with two horizons. The D $p$-brane solution given above in
eq(\ref{branemetric}) is a BPS-state since it is a charged object
having the structure of an extremal black hole in that both the
horizons have coalesced.  On can also solve for a D3-brane on the
space $AdS_5 \times S^5$.

\bigskip

\section{T-Duality }

\bigskip

For the closed string, we have
$$
\partial^2 X^\mu = 0
$$
with the boundary condition $X^\mu(\sigma + 2 \pi) = X^\mu(\sigma,
\tau)$ and yields normal mode expansion.

\bea X^\mu (\sigma, \tau) &=& x^\mu + p^\mu \tau + \frac{i}{2}
{\sum_n}^\prime \frac{1}{n} (\alpha_{n_i} \expon^{- i n (\tau -
\sigma)} + \tilde{\alpha}_{n_i} \expon^{- i n (\tau + \sigma)}) \\
&=& x_L^\mu + x_R^\mu + \sqrt{\frac{\alpha^\prime}{2}}
\alpha_0^\mu (\tau - \sigma) + \sqrt{\frac{\alpha^\prime}{2}}
\tilde{\alpha}_0^\mu (\tau + \sigma) + {\rm  oscillators} \\ &=&
X_R^\mu (\tau - \sigma) + X_L^\mu (\tau + \sigma) \eea %
where $$
p^\mu = \sqrt{\frac{1}{2\alpha^\prime}} (\alpha_0^\mu +
\tilde{\alpha_0^\mu}) $$

Under $\sigma \rightarrow \sigma + 2\pi, X^\mu (\sigma, \tau)$ changes
by $\sqrt{\frac{\alpha^\prime}{2}} (\tilde{\alpha}_0^\mu - \alpha_0^\mu) 2\pi$.
For non-compact spatial dimension $X^\mu$ is single-valued and hence
$$
\tilde{\alpha}_0^\mu = \alpha_0^\mu = \sqrt{\frac{\alpha^\prime}{2}} p^\mu
$$

Suppose $X^{25}$ in a circle with radius $R$. Then, under a shift of
$\sigma \rightarrow \sigma + 2\pi$, we can have
$$
X^{25} (\sigma + 2\pi, \tau) = X^{25} (\sigma, \tau) + 2\pi R w
$$
where $w$ is the winding number $\in Z$.

Hence
\bea
\alpha_0^{25} + \tilde{\alpha}_0^{25} &=& \frac{2n}{R}
\sqrt{\frac{\alpha^\prime}{2}} \\
\alpha_0^{25} - \tilde{\alpha}_0^{25} &=& wR \sqrt{\frac{2}{\alpha^\prime}}
\eea
where the Kaluza-Klein (KK) momentum is
$\displaystyle{p^{25} = \frac{n}{R}}$.
Furthermore, we have
\bea
\alpha_0^{25} &=& (\frac{n}{R} + \frac{wR}{\alpha^\prime})
\sqrt{\frac{\alpha^\prime}{2}} \\
\tilde{\alpha}_0^{25} &=& (\frac{n}{R} - \frac{wR}{\alpha^\prime})
\sqrt{\frac{\alpha^\prime}{2}}
\eea

The mass spectrum is, putting back $\alpha^\prime$,
\bea M^2 = -
p^\mu p_\mu &=& \frac{2}{\alpha^\prime} (\alpha_0^{25})^2 +
\frac{4}{\alpha^\prime} (L_0 - 1) \\ &=& \frac{2}{\alpha^\prime}
(\tilde{\alpha}_0^{25})^2 + \frac{4}{\alpha^\prime} (\tilde{L}_0 -
1) \eea

Note $$ M^2 (n, w, R) = M^2 (w, n, R^{\prime} =
\frac{\alpha^\prime}{R}) $$ %
i.e. identical mass spectrum for ${\displaystyle n \leftrightarrow
w, R \leftrightarrow \frac{\alpha^\prime}{R}}$.  Under this
transformation \bea \alpha_0^{25} &\rightarrow & \alpha_0^{25} \\
\tilde{\alpha}_0^{25} &\rightarrow & -\tilde{\alpha}_0^{25} \eea

In other words, the $T$-dual theory is described by a parity
transformation on left-movers, i.e. $$ T_{25} [X^{25}] = X^{\prime
25} (\sigma, \tau) = X_R^{25} (\tau - \sigma) - X_L^{25} (\tau +
\sigma) $$ $T$-duality is an exact symmetry of string theory. What
this means is that the Hilbert spaces of the two $T$-dual theories
are identical. In general, for $k$-compact directions we have
$T_{\mu_1, \cdots, \mu_k}$ for the $T$-duality
transformation\cite{giveon, alvarez}.

The effective coupling in the $25^{th}$-dimension is
$\expon^{\phi} R^{-\frac{1}{2}}$ ($\phi$: dilaton).  Hence in the
$T$-dual theory, \bea \expon^{\phi^\prime} R^{\prime - 1/2} & = &
\expon^\phi R^{-1/2} \\ \mbox{{\rm or,} \hspace{1cm}}
\expon^{\phi^\prime} & = & \sqrt{\alpha^\prime} \expon^\phi R^{-1}
\eea

\subsection{Open Strings and $D$-branes}

Recall with $NN$-conditions, we have for the open string,%
\bea X^\mu(\sigma, \tau) & = & x^\mu + p^\mu \tau(2 \alpha^\prime)
+ \frac{i}{2} {\sum_n}^\prime \frac{\alpha_n^\mu}{n} \left(
\expon^{- i n
(\tau - \sigma)} + \expon^{- i n (\tau + \sigma)}\right) \\ %
& = & X_R^\mu
(\tau - \sigma) + X_L^\mu (\tau + \sigma). \eea %
For the open string, left and right movers are reflected at the
ends and form standing waves.  We have \bea X_R^\mu(\tau - \sigma)
& = & \frac{x^\mu}{2} + c + \alpha^\prime (\tau + \sigma) p^\mu +
\frac{i}{2} {\sum_n} \frac{\expon^{-i n (\tau - \sigma)}}{n}
\alpha_n^\mu \\ X_L^\mu(\tau - \sigma) & = & \frac{x^\mu}{2} - c +
\alpha^\prime (\tau - \sigma) p^\mu + \frac{i}{2} {\sum_n}
\frac{\expon^{-i n (\tau + \sigma)}}{n} \alpha_n^\mu \eea Suppose
$X^{25}$ is compact with radius $R$; this implies $\dis p^{25} =
\frac{n}{R}$. For closed strings, $T$-duality was a symmetry
because we could interchange windings of the closed string in
$X^{25}$ namely $w$ with the $KK$-modes due to the compactness of
$X^{25}$.  For the open string there is no winding number.  So how
do we obtain $T$-duality?

As in the case of closed strings, we define the $T$-duality for
open string by $$ T_{25} [X^{25}] = X^{\prime 25} (\sigma, \tau)
\equiv X_R^{25} (\tau - \sigma) - X_L^{25} (\tau - \sigma) $$ or,
$\dis  X^{\prime 25}(\sigma, \tau) = 2 c + 2 \alpha^\prime p^{25}
\sigma - {\sum_n}^\prime \frac{\expon^{- i n \tau}}{n} \sin(n
\sigma) \alpha_n^{25} $. Note at the boundaries $\sigma = 0, \pi$,
the oscillators terms disappear and since there is no other
dependence on $\tau$, the boundaries do not move!  By $T$-duality
we have switched the $NN$-condition $\dis \frac{\partial
X^{25}}{\partial \sigma}|_{0, \pi} = 0$ to $DD$-condition $\dis
\frac{\partial X^{25}}{\partial \tau}|_{0, \pi} = 0$!  Thus \bea
X^{\prime 25}(0, \tau) & = & 2 c \\ X^{\prime 25}(\pi , \tau) & =
& 2 c + 2 \pi \alpha^\prime p^{25} \\ & = & 2 c  + 2 \pi n
R^\prime \label{eqn60} \eea From eq(\ref{eqn60}), we see that in
the dual theory with $R^\prime = \alpha^\prime / R$, the
$X^{\prime 25}$ ends are fixed, i.e. $X^{\prime 25}(0, \tau) =
X^{\prime 25}(\pi, \tau)$ mod $2 \pi R^\prime$.

Since the $X^{\prime 25}$ boundaries cannot move, it is meaningful if
the $X^{\prime 25}$ coordinates \underline{winds} about the $25^{\rm th}$
direction since
the $DD$-conditions prevent it from un-winding.

 \begin{center}
\epsfig{file =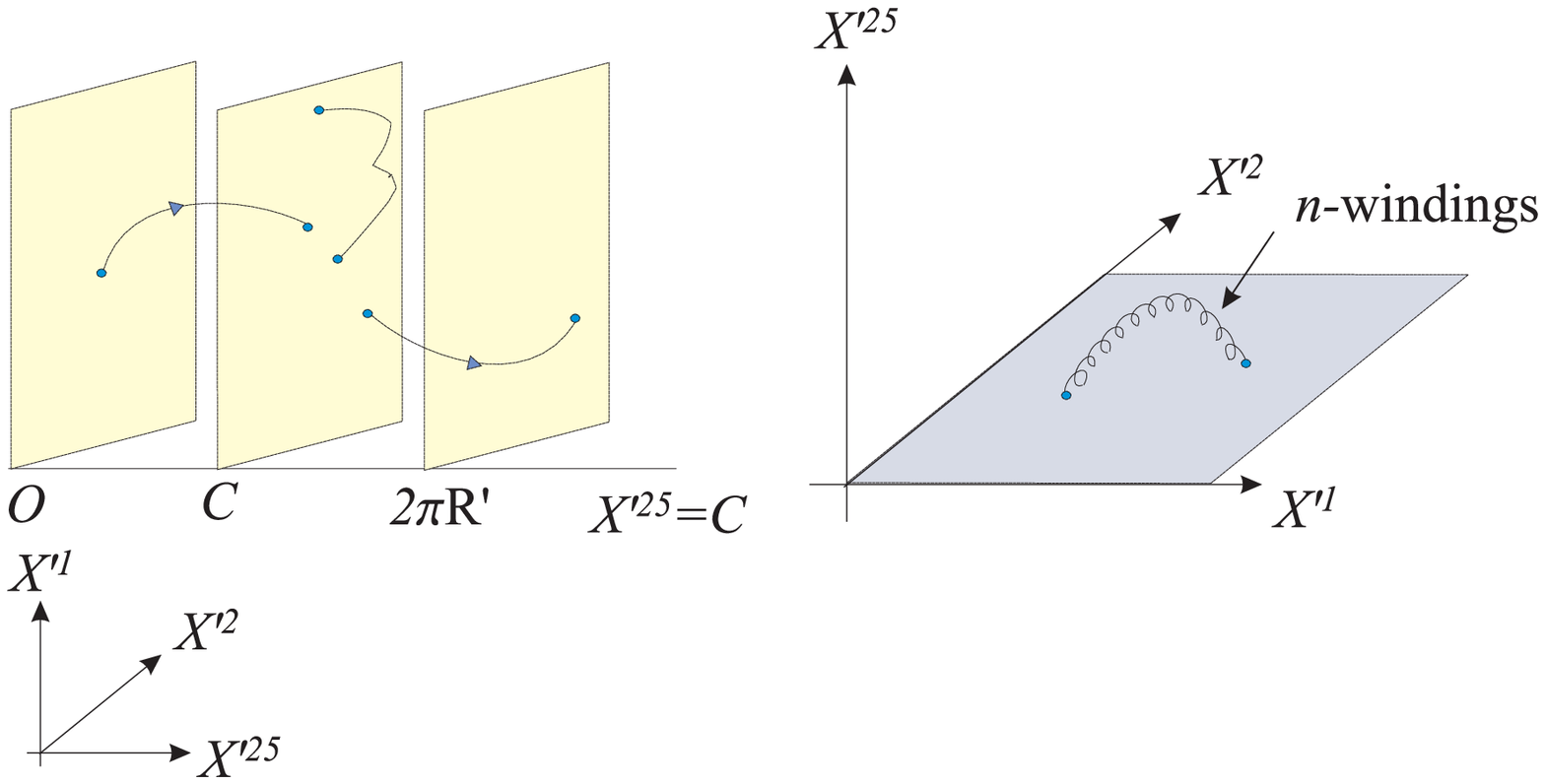, height=.3\textheight}
 \end{center}

We see that open strings have $T$-duality symmetry in a manner
very different from closed strings. Instead of interchanging
winding number with $KK$-momenta as is the case for closed
strings, for open strings the $DD$-boundary conditions effectively
switch the $KK$-momenta into a winding number in the dual theory!

\subsection{$R \rightarrow 0$ Limit}

For \underline{closed strings}, as $R \rightarrow 0$ only the $n =
0 \; KK$ mode survives and we have \bea \lim_{R \rightarrow 0} M^2
&=& \frac{2}{\alpha^\prime} (\frac{n}{R} +
\frac{wR}{\alpha^\prime})^2 + \frac{4}{\alpha^\prime} (L_0 - 1) \\
& \rightarrow & \frac{2}{\alpha^{\prime 3}} (wR)^2 +
\frac{4}{\alpha^\prime} (L_0 - 1) \\ &=& \frac{2}{\alpha^{\prime
3}} W^2 + \frac{4}{\alpha^\prime} (L_0 - 1) \eea where $W = \omega
R \in [-\infty, +\infty]$ is the effect on the mass spectrum due
to the compact direction. This effect is a purely string effect
since as $R \rightarrow 0$ the closed string finds it
energetically more and more favorable to wind around a small $R$
and winding number $w$ becomes arbitrarily large yielding real
continuous variable $W = wR$ : a new continuum of quantum states
labeled by $W$.

For open strings, as $R \rightarrow 0$, the $KK$ momenta is $n =
0$ and hence in the dual picture winding number is also zero.
Unlike the closed string, there is no new continuum of states.
Recall $$ X^{\prime 25} (\pi) - X^{\prime 25} (0) = 2 \pi n
R^\prime \rightarrow 0 $$ That is, the open string is only free to
move in $(d - 1)$-dimensions as $R \rightarrow 0$ unlike closed
strings which continue to vibrate in $d$-dimensions even as $R
\rightarrow 0$.

\subsection{Wilson Lines, Open Strings and $T$-Duality}

Recall for open strings we have non Abelian gauge fields $A_\mu^a (x)$
in the expansion for the open string.
$$
A_\mu^a \lambda_{ij}^a \alpha_{-1}^{\mu} | k, ij >
$$

\vspace{1cm}
 \begin{center}
\epsfig{file =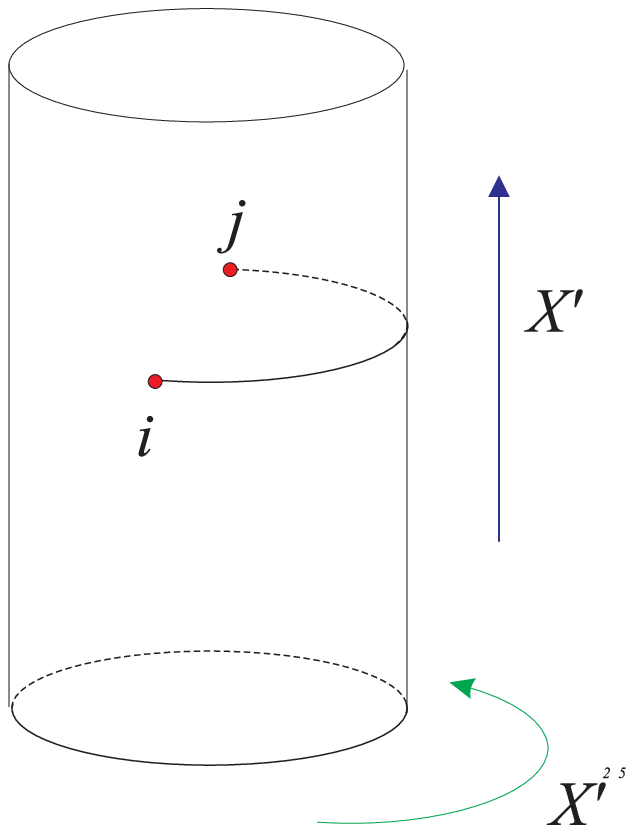, height=.25\textheight}
 \end{center}

For compact direction $X^{25}$ we can consider the gauge field
component $A_{ij}^{25} = A^{a25} \lambda_{ij}^{a}$ to be a
background gauge field. Consider a $U(N)$ oriented string with end
points carrying $N$ and $\bar{N}$ irreps. One can consider Wilson
line by $$ P \expon^{i \int_{X^{25}(0)}^{X^{25}(\pi)} A^{25}
dx^{25}} $$ with the background field in a fundamental irrep $$
A^{25} = \frac{1}{2 \pi R} \left( \begin{array}{ccccc} \theta_1 &
& & & \\ & . & & & \\ & & . & & \\ & & & . & \\ & & & & \theta_N
\end{array} \right)
= -i \Lambda^{-1} \frac{\partial \Lambda}{\partial x^{25}}
$$
which breaks the gauge symmetry from $U(N)$ to $U(1)^N$.

We have in the fundamental representation, $$ \Lambda = {\rm diag}
(\expon^{\frac{i X^{25} \theta_1}{2 \pi R }}, \cdots ,
\expon^{\frac{i X^{25} \theta_N}{2 \pi R }}) $$ We can set
$A^{25}$ to zero by a gauge transformation $\Lambda$.  We also
have to simultaneously gauge transform  the state by \bea |ij> &
\rightarrow & \Lambda_{\alpha i}^{-1} (X^{25}(0)) \Lambda_{\alpha
j} (X^{25}(\pi)) |i j > \\ & = & \expon^{\frac{i}{2 \pi R}
(\theta_j X^{25} (\pi) - \theta_i X^{25}(0)) } |ij > \eea For the
string state $$ |{\rm string}> = \expon^{i p^{25} X^{25}} |ij> $$
and under $X^{25} \rightarrow X^{25} + 2 \pi R$, $|ij> \rightarrow
\expon^{i(\theta_j - \theta_i)} |ij>$.  Hence $\dis p^{25} =
\frac{1}{2 \pi R} (\theta_j - \theta_i + 2 n \pi R)$.  In other
words, a Wilson line formed by breaking $A^{25} \rightarrow
<A^{25}> = -i \Lambda^{-1}
\partial_{25} \Lambda$ is effectively imparting fractional KK-momenta to the open
string given by $p^{25}$.

Recall under $T$ duality
\bea
X^{\prime 25} (\pi) - X^{\prime 25} (0) &=& 2 \pi \alpha^{\prime} p^{25} \\
&=& \frac{\alpha^\prime}{R} (2 \pi n + \theta_j - \theta_i)
\eea
The state of the string at $\sigma = 0$ should depend only on $i$ and at
$\sigma = \pi$ on only $j$ (otherwise we would violate locality). Therefore,
\bea
X^{\prime 25} (\pi) &=& 2 \pi n R^\prime + \theta_j R^\prime \\
X^{\prime 25} (0)   &=& \theta_i R^\prime
\eea

In other words, for $\theta_i \neq \theta_j$, in the dual spacetime, the
open string endpoint for state $i$ is located on a $D$-brane placed at
$\theta_i R^\prime$ along $X^{\prime 25}$ and the other endpoint is located
on a $D$-brane at $\theta_j R^\prime$
(both ends located modulo $2 \pi R^\prime$).

 \begin{center}
\epsfig{file =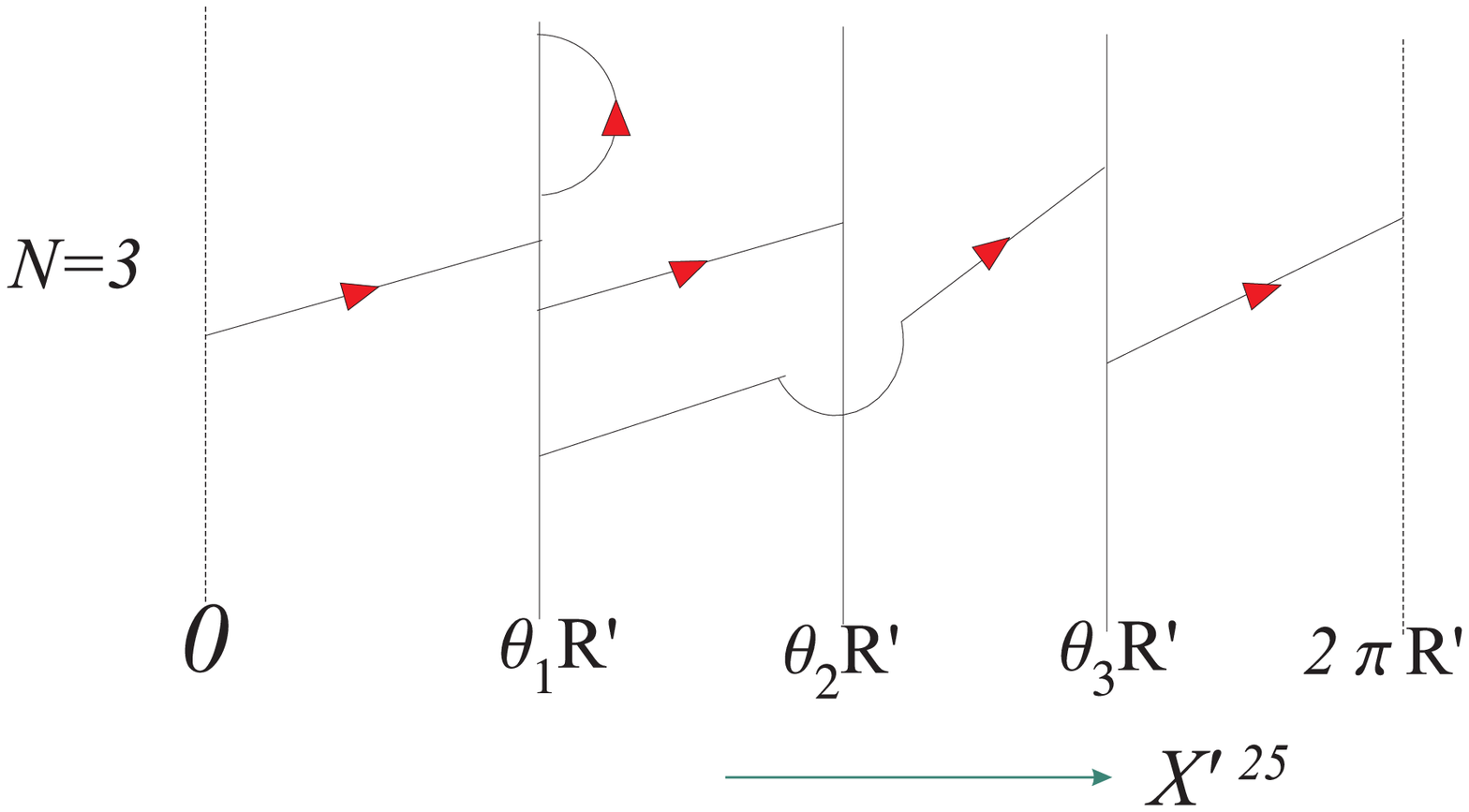, height=.25\textheight}
 \end{center}

A Wilson line introduces, in the dual theory, $D$-branes located
at points inside the compact dimension $X^{\prime 25}$.

\bigskip

\section{D-Brane Dynamics and Gauge Fields}

\bigskip

In this section, we look into the fluctuations of D-p
branes\cite{douglas1,douglas2,taylor}. We recall that for the case
of open strings carrying fractional momenta $p^{25}$, the $T$-dual
of the compact coordinate yields \bea M^2 &=& (p^{25})^2 +
\frac{1}{\alpha^\prime} (L_0 - 1)
\\
    &=& \{ \frac{R^\prime}{2\pi \alpha^\prime} [ 2\pi n + (\theta_i -
\theta_j)]\}^2 + \frac{1}{\alpha^\prime} (L_0 - 1)
\eea

Massless states can only arise for $n = 0$, i.e. for non-winding
strings whose ends are on the same hyperplane $(\theta_i =
\theta_j)$ shown below. Massive string states are given by open
strings stretching between $D$-branes with mass of the state given
by the product of the string tension and the length of the string.

 \begin{center}
\epsfig{file =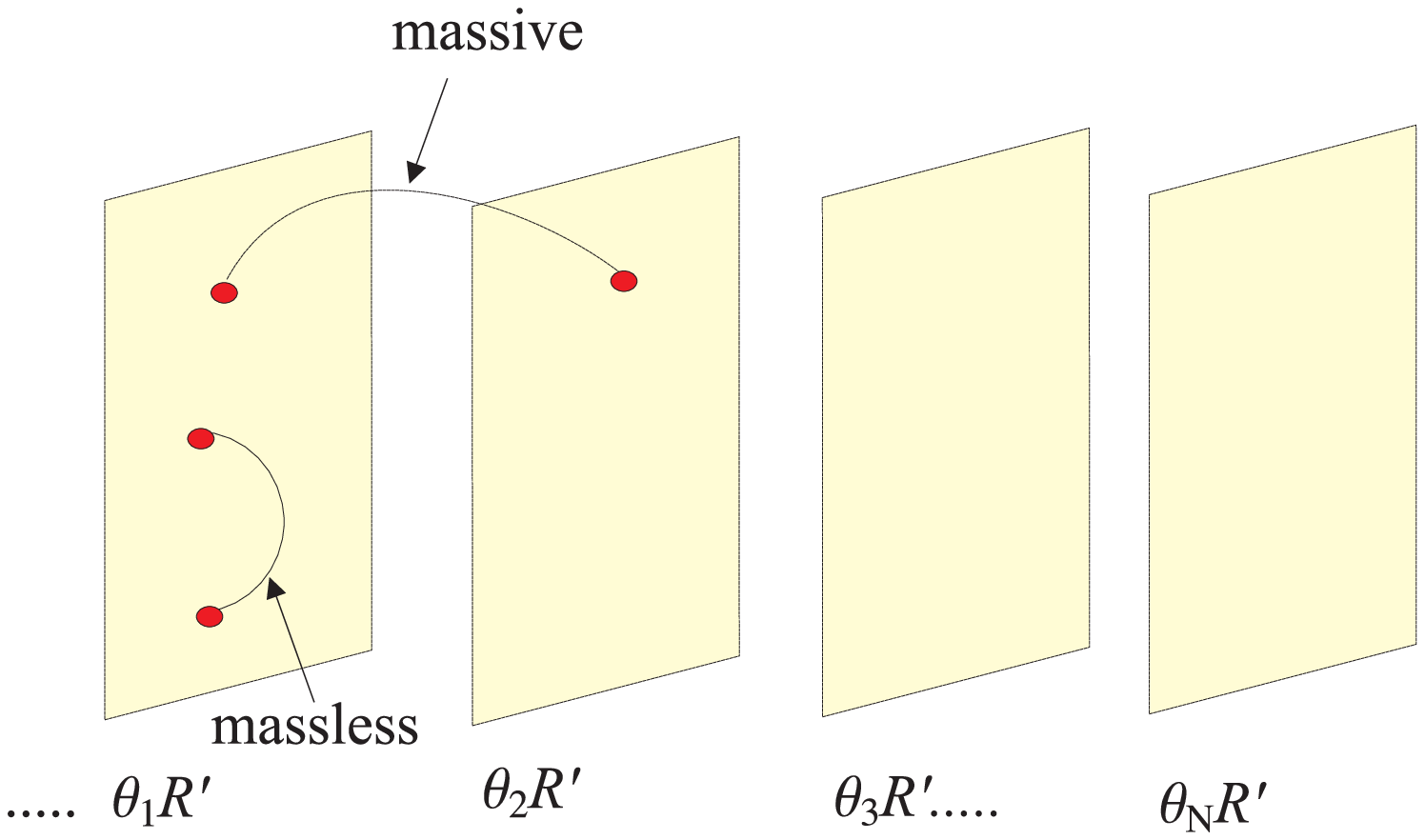, height=.3\textheight}
 \end{center}

Continuing in the $T$-dual picture, when none of the $N$ D-branes
coincide, there is just one massless $U(1)$ vector field in each D-brane
with $U(1)^N$ being the unbroken group. If $m < n$ of the D-branes
coincide, i.e. $\theta_1 = \theta_2 = \cdots = \theta_m$, there are
$m^2$ massless vectors with $m^2 - m$ new massless states since open
strings stretched between these branes have zero length. These $m^2$
massless vectors form the adjoint representation of a $U(m)$ gauge theory.
This is a reflection of $U(m)$ subgroup being unbroken by the Wilson
line.

It has been shown by Witten that $N$ parallel coinciding $D-p$
branes have low energy excitations described by a $U(N)$ Susy
YM-theory dimensionally reduced from $d = 10$ to $p+1$ dimensions.

\subsection{Super Yang-Mills from $D$-Branes}

Let $\psi$ be Majorana-Weyl spinor of $SO(1,9)$ acted on by $32
\times 32$ gamma matrices; let $A_\mu$ be the $U(N)$ gauge field
given by an $N \times N$ Hermitian matrix. We then have the YM
Lagrangian $$ S = \int d^{10} \xi (-\frac{1}{4 g^2} {\rm tr}
F_{\mu \nu} F^{\mu \nu} + \frac{i}{2} \bar{\psi} \Gamma^\mu D_\mu
\psi) $$

This action is invariant under supersymmetric transformation
\bea
\delta A_\mu &=& \frac{i}{2} \bar{\epsilon} \Gamma_\mu \psi \\
\delta \psi &=& -\frac{1}{4} F_{\mu \nu} \Gamma^{[\mu}
\Gamma^{\nu]} \epsilon \eea

Moreover, 10-$D$ SYM has fields:
\begin{center}
$A_\mu$ : 8 bosonic degrees of freedom. \\ $\psi$  : 8 fermionic
degrees of freedom. \\$Q_{\alpha}$ : 16 supercharges.
\end{center}
\noindent Consider $N$ parallel $D$-$p$ branes. To describe the
$D-p$ branes low energy dynamics, we make all the fields
independent of the $9-p$ transverse (to the $D-p$ brane)
directions. The transverse oscillations of the $N$ parallel $D-p$
branes is described by the $A_\alpha$ field.

\begin{center}
\epsfig{file=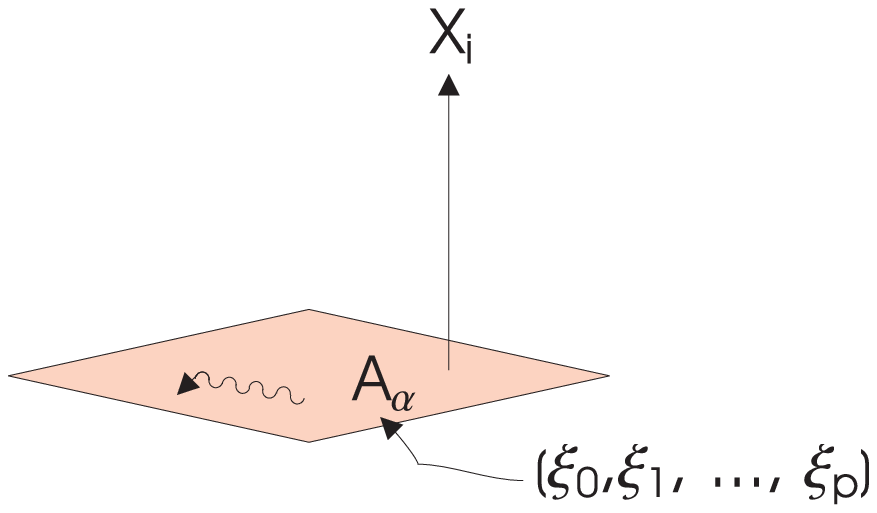, height=.2\textheight}
\end{center}

Next, we notice that

\begin{tabular}{lll}
$A_\alpha (\xi)$, & $\alpha = 1, 2, \cdots p+1$: & propagate only
inside the $D-p$ brane  and \\ $A_i \equiv X_i (\xi)$,  & $i =
p+2, \cdots 10$: & transverse position of the $N$ parallel $D-p$
\\ & & brane move in the $9-p$ transverse dimensions.
\end{tabular}
\medskip

\noindent Ignoring fermions, we have $$ S = \frac{1}{4 g^2} \int
d^{p+1} \xi \{ - F_{\alpha \beta} F^{\alpha \beta} - ( D_\alpha
(A) X^i )^2 + [ X^i, X^j ]^2 \} $$

\subsection{Classical Vacua}

\begin{itemize}
\item Fermions vanish.
\item $X^i$'s are constant and
\item $[X^i, X^j] = 0$
\end{itemize}

Thus we can simultaneously diagonalize all the $X^i$'s,
$$ X^i = \mbox{\rm diag}(x_{1}^i, x_2^i, \cdots, x_N^i ),
\mbox{\hspace{1cm}} i = 1,2, \cdots, k.$$ or
\begin{center}
\epsfig{file=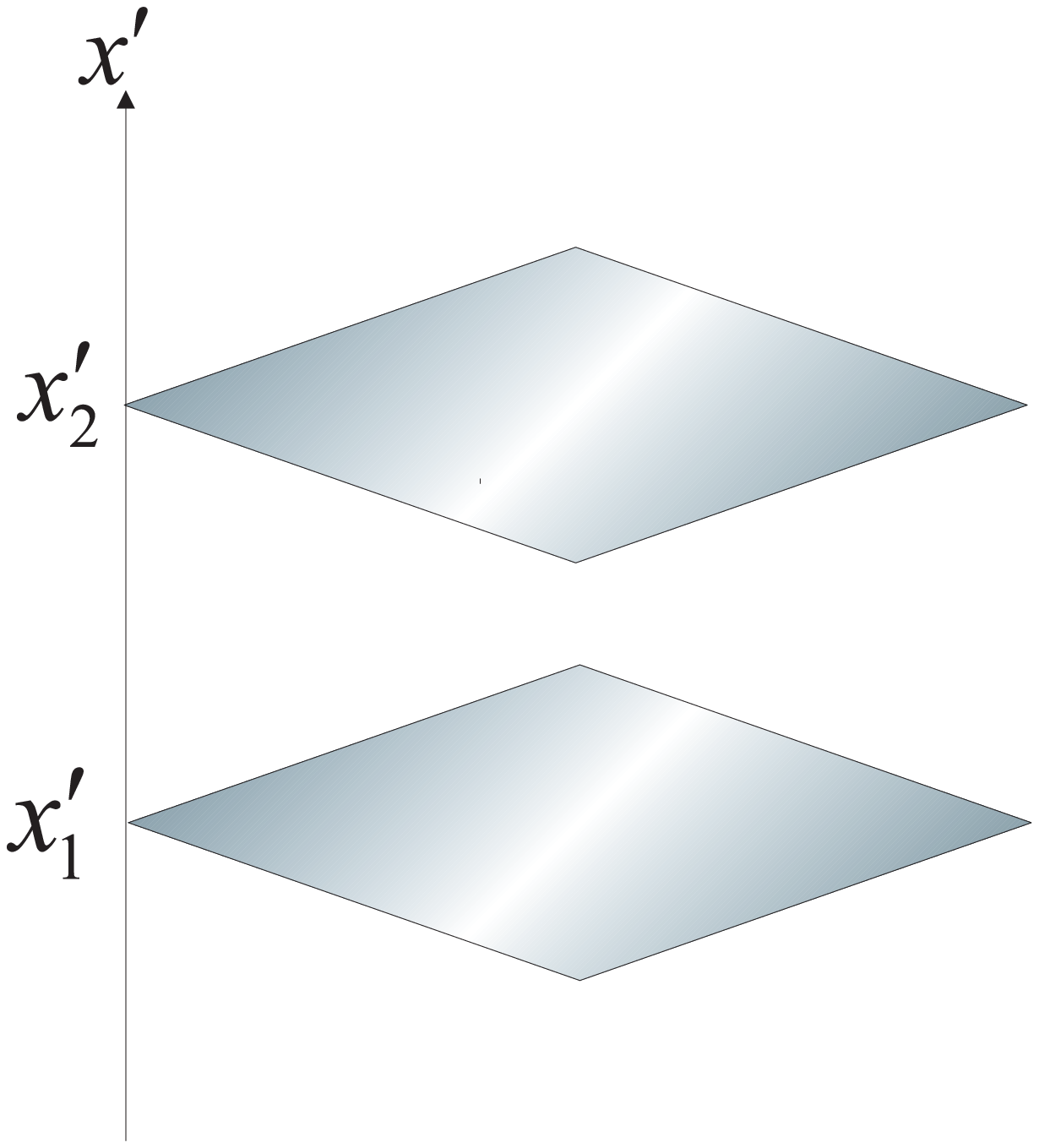, height=.3\textheight}
\end{center}
or
$$ $$
where $\vec{x_k}$ is the (transverse) position of the $k$-th
parallel $D-p$ brane. Configuration space is $({\Bbb R}^{9 - p})^N
/ S_N$ due to permutation symmetry (Branes are identical-bosons).

If $m$ of the branes are coincident $\vec{x} = \vec{x_1} = \vec{x_2} =
\cdots = \vec{x_m}$ and
$$ \vec{X} = \mbox{\rm diag} \left( \vec{x}, \cdots, \vec{x},
\vec{x_{m+1}}, \cdots, \vec{x_N} \right)$$ It is instructive to
look into an example. Consider $N$ 0-branes. In gauge $A_0 = 0$,
we have $A^a \rightarrow X^a, \;\;\; a = 1, 2, \cdots, 9$. Hence,
$$ L = \frac{1}{2 g^2} \{ \dot X^a \dot X_a + \sum_{a < b} {\rm
tr} [X^a, X^b]^2 + 2 \theta^T ( \dot \theta + \Gamma_a [X^a,
\theta]) \} $$ where $\theta$ is 16-component real spinor.

\noindent The classical vacua is given by $[X^a, X^b] = 0$ where %
$$ \vec{X} = \mbox{\rm diag} \left( \vec{X_1}, \vec{X_2}, \cdots,
\vec{X_N} \right) \mbox{\hspace{1cm} : positions of the ~} N 0
\mbox{-branes} $$ \vspace{0.5cm}
\begin{center}
\epsfig{file=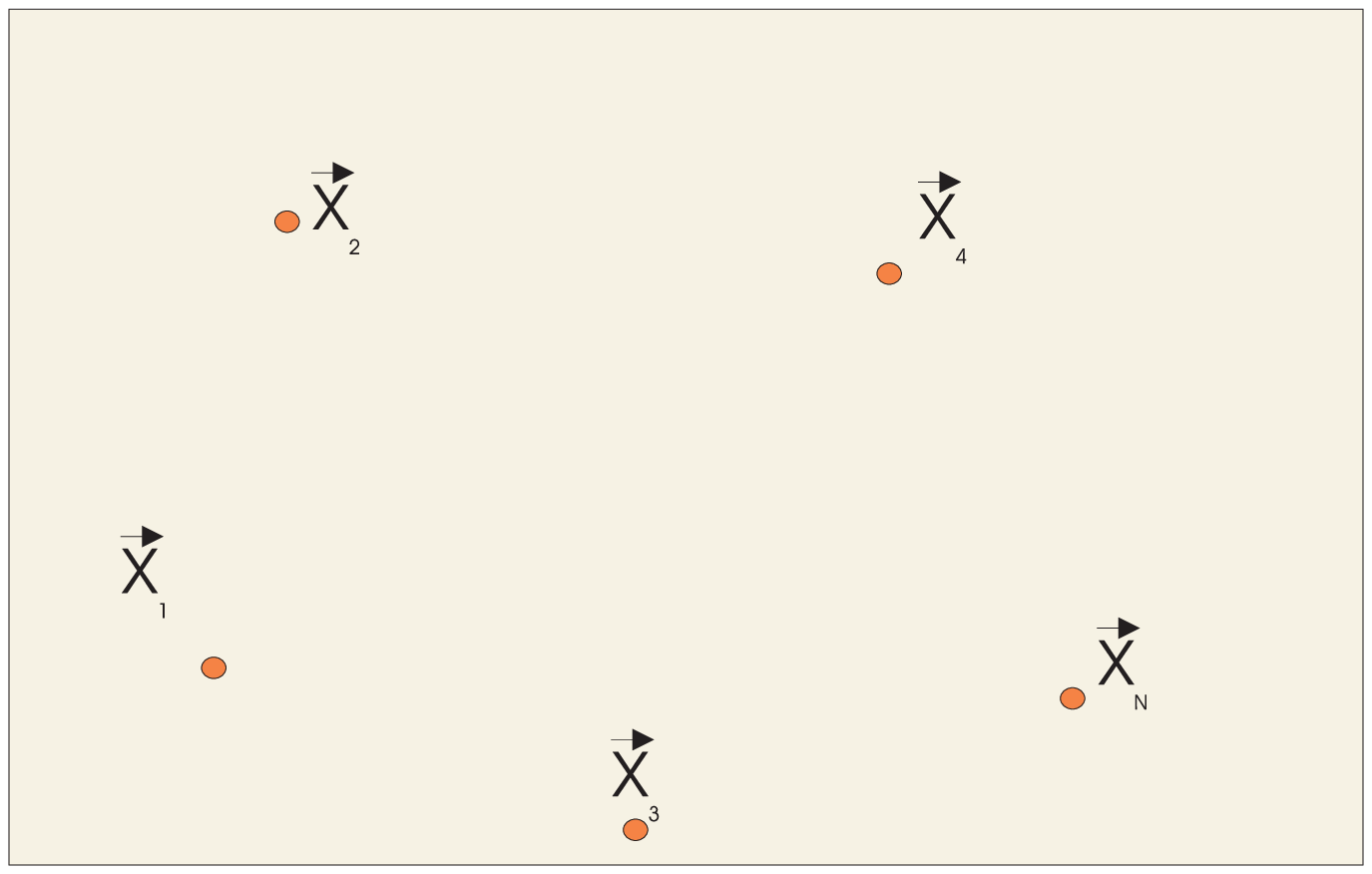, height=.2\textheight}
\end{center}

Configuration space = $({\rm \bf R}^9)^N / S_N$.

Off-diagonal $X^a$ give a realization of  non-commutative
geometry.

\subsection{$T$-Duality and Super Yang-Mills Theory}

Consider strings on a circle of radius $R$ giving an $S^1 \times
{\rm \bf R}^9$ spacetime.

\begin{center}
\epsfig{file=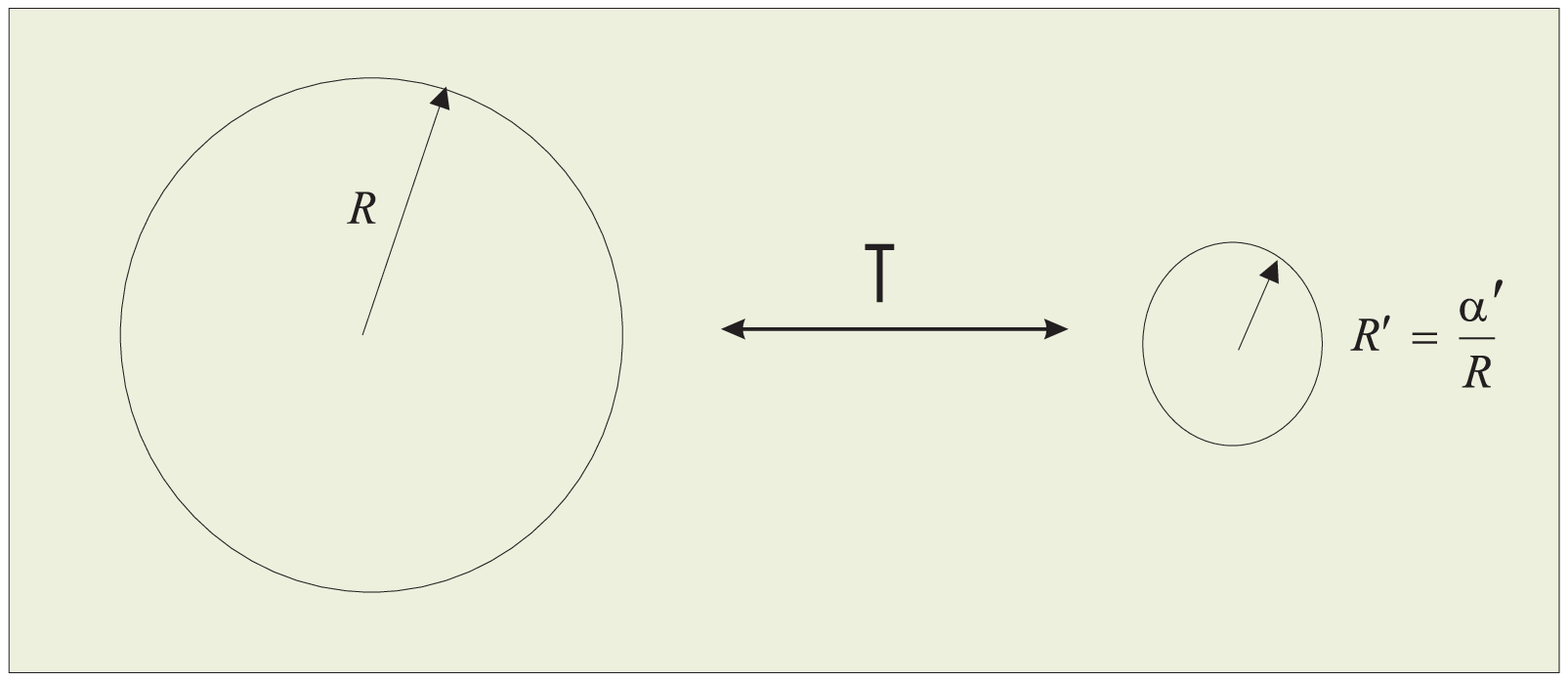, height=.2\textheight, width=.8\textwidth}
\end{center}

$T$-duality maps Neumann $\leftrightarrow$ Dirichlet.

Consider a $D-p$ brane. Under $T$-duality we have

\begin{center}
\epsfig{file=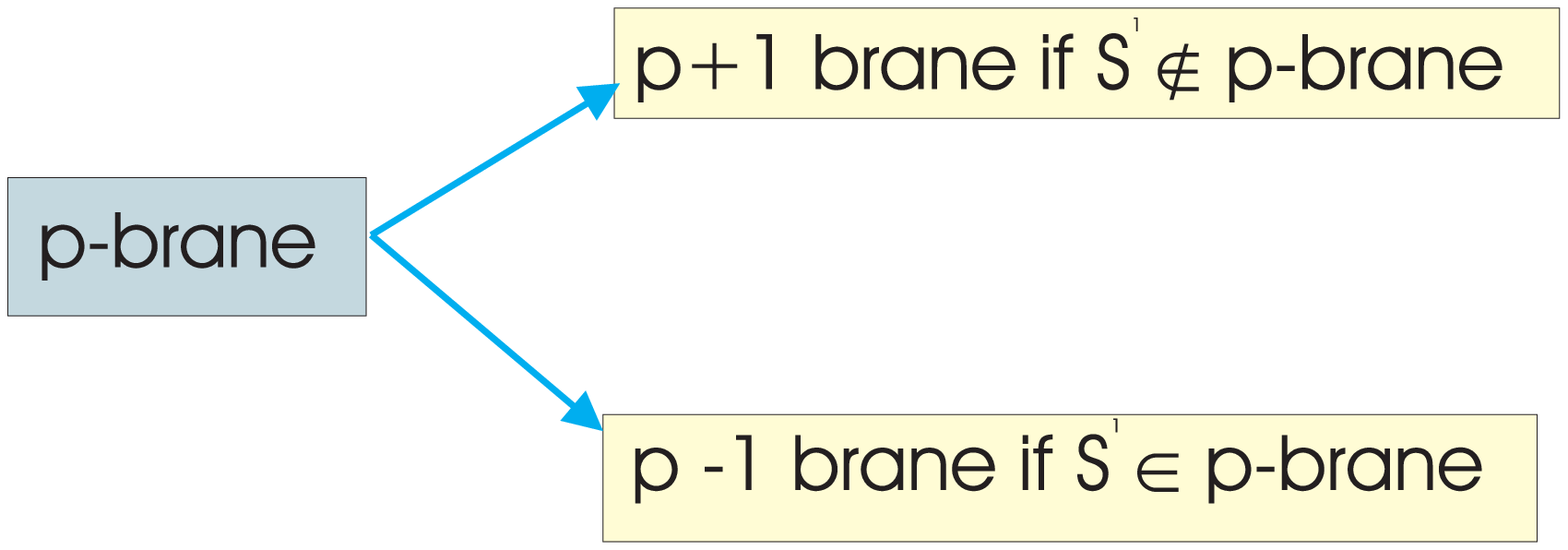, height=.15\textheight,width=.5\textwidth}
\end{center}

How do we realize $T$-duality for Super-Yang-Mills? Consider for
simplicity $N$ 0-branes. What is the realization for $X^a, a = 1, 2,
\cdots, 9$ for space $S^1 \times {\rm \bf R}^9$? ($A_0 = 0$ gauge) We construct
$S^1 = {\rm \bf R} / 2\pi R {\rm \bf Z} = {\rm \bf R} / \Gamma$.

\begin{center}
\epsfig{file=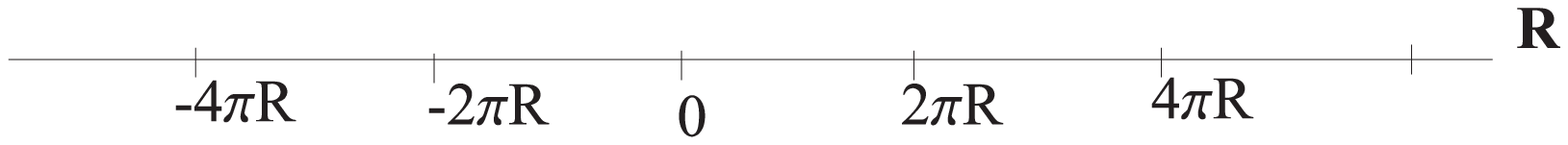,width=.8\textwidth}
\end{center}

The $N \times N$ matrices $X^a$ are now realized as infinite dimensional
matrices which are infinitely many copies of $X^a$ in ${\rm \bf R}^{10}$; we will
quotient these matrices by $\Gamma$ to obtain $S^1 \times {\rm \bf R}^9 =
{\rm \bf R}^{10} / \Gamma$.

Organize the $\infty \times \infty$ matrices $X^a$ into an infinite
collection of $N \times N$ matrices each labeled by $m, n \in {\rm \bf
Z}$, i.e.
$$
X^a \rightarrow \{X_{mn}^a \}
$$
namely,
$$X^a = \left( \begin{array}{cccc}
 & \vdots & \vdots & \\
\cdots & X_{01}^a & X_{02}^a & \cdots \\
\cdots & X_{11}^a & X_{12}^a & \cdots \\
 & \vdots & \vdots &
\end{array} \right)
$$

Suppose $X^1$ corresponds to $S^1$. Then quotient by $\Gamma$ leads to
the following symmetry
\bea
X_{mn}^i &=& X_{(m-1)(n-1)}^i, \;\;\; i > 1     \\
X_{mn}^1 &=& X_{(m-1)(n-1)}^1, \;\;\; m \neq n  \\
X_{nn}^1 &=& 2\pi R I + X_{(n-1)(n-1)}^1
\eea
From above it follows that all information is contained in the $N \times
N$ matrices $X_{0 n}^a \equiv X_n^a$ with $(X_n^a)^\dagger = X_n^a$.
Using notation $X_k = X_{0k}^1$ we have
$$
X^1 \sim M = \left( \begin{array}{cccccc}
\cdots &X_1 &X_2 &X_3 &\cdots &\\
X_{-1} &X_{0} - 2 \pi R &X_1 &X_2 &X_3 &\cdots\\
X_{-2} &X_{-1} &X_0 &X_1 &X_2 &\cdots\\
\cdots &X_{-2} &X_{-1} &X_0 + 2 \pi R &X_1 &\cdots
\end{array} \right)
$$

\noindent $T$-duality is now defined by the transformation \bea
X^1 & \rightarrow & M = \sum_{n \in {\rm \bf Z}} \expon^{inx^1 /
R^\prime} X_n^1 \\ X^i & \rightarrow & X^i = \sum_{n \in {\rm \bf
Z}} \expon^{inx^1 / R^\prime} X_n^i , \;\;\; {\rm for} \;\; i > 1.
\eea and $M = i \partial^1 + A^1$ since by applying Fourier
transformation to $i \partial^1$ we have $$ i \partial^1 = {\rm
diag} (\cdots, -4 \pi R, -2 \pi R, 0, 2 \pi R, 4 \pi R, \cdots).
$$ This is equivalent to the following for $T$-duality. \bea
X^\alpha & \leftrightarrow & (2 \pi \alpha^\prime) (i
\partial^\alpha + A^\alpha), \;\;\; \alpha: {\rm compact \;
directions} \\ X^\beta  & \leftrightarrow & X^\beta,
\;\;\;\;\;\;\;\;\; \beta: {\rm noncompact ~ directions} \eea

Consider the following example.
$$ A^1 = \frac{1}{2 \pi R} \mbox{\rm diag} \left(
 \theta_1, \cdots,  \theta_N \right)
$$

 Under $T$-duality %
$$ 2 \pi \alpha^{\prime} \left( i
\partial^1 + A^1 \right) \rightarrow \frac{2 \pi \alpha^\prime}{2
\pi R} \mbox{\rm diag} \left(
 \theta_1, \cdots,  \theta_N \right)  = X^1 $$ That
is, $\partial^1 \equiv 0$ as this has been dualized and lies
inside the 1-brane and 
$$ X^1 = N ~ {\rm positions ~ of ~ the ~ } D 1-{\rm ~ branes} =
\mbox{\rm diag } \left( \theta_1 R^\prime, \cdots, \theta_N
R^\prime \right) $$ On taking the $T$-dual of the 0-brane we have
obtained a 1-brane since the open string has $DD$-b.c's on dual
$S^1$. Note \bea [X^1, X^a] &\rightarrow & \int [ i
\partial^1 + A^1, X^a] dx^1 \\&=& i \oint ( \partial^1 X^a - i [
A^1, X^a])
\\&=& i \oint (D^1 X^a)^2 dx^1\eea
\bea(D^0 X^1)^2 &=& \{ \partial^0 X^1 - i [A^0, X^1] \}^2
\\ &=& \oint \{
\partial^0 (i \partial^1 + A^1) - i [A^0, i \partial^1 + A^1]\}^2
\\&=& \oint \{ \partial^0 A^1 - \partial^1 A^0 - i [A^0, A^1] \}^2
\\&=& \oint (F_{0 1})^2 dx^1 \eea Hence by putting the prefactors
in $(R^\prime = \alpha / R)$ \bea L &=& \frac{{\rm tr}}{2 g^2}
(\dot X^1 \dot X_1 + ( [X^1, X^a] )^2 +\cdots ) \\ &\rightarrow &
\frac{1}{R^\prime} \frac{1}{2 g^2} \oint dx_1 \{ {\rm tr} F_{01}^2
- (D^1 X^a)^2 + \cdots \} \eea Hence the $T$-dual to 0-branes
yields a $2D$ SYM with 16 supercharges. This result can be
generalized to show that a $D-p$ brane on $T^{|p - q|}$ is
$T$-dual to a D-q brane.We can also generalize the result to many
compact dimensions. For directions $X^a$ and $X^b$ with radii
$R_a$ and  $R_b$ we have $$ ([X^a, X^b])^2 \rightarrow
-\frac{1}{R_a^\prime R_b^\prime} \oint dx^adx^b (F^{ab})^2 $$

\bigskip
\section{M- and F-Theory}

\subsection{M-Theory}
\bigskip

The highest dimension with a unique and consistent classical
theory of ${\cal N}=1$ supergravity is  $d=11$ dimensional
spacetime. The bosonic fields are

\begin{tabular}{ll}
$G_{MN}:$ \hspace{2cm}& metric \\ 
$C_{MNP}:$ & antisymmetric tensor potential
\end{tabular}

\noindent with $M, N, P = 0, 1, 2, \cdots, 10$.  Note $C_{MNP}$
couples to an $M$2-brane and its dual couples to an $M$5-brane.

On dimensionally reducing this theory from 11 to 10 dimensions, we
obtain the ${\cal N}=2$, $d=10$ non-chiral supergravity Lagrangian
of Type IIA strings given in eq(\ref{bkg1}).  In particular,  the
field reduction yields, for $\mu, \nu, \rho = 0, 1, 2, \cdots, 9$,
$$ G_{MN} \rightarrow \left\{
\begin{array}{l}
G_{10,10} = \expon^{\phi} \\ G_{10,\mu} = A^{(1)}_{\mu} \\ G_{\mu
\nu} = g_{\mu \nu}
\end{array}
\right. $$ and $$ C_{MNP} \rightarrow \left\{ \begin{array}{l}
C_{\mu \nu \rho } = A_{\mu \nu \rho}^{(3)} \\ C_{\mu \nu (10) } =
B_{\mu \nu}
\end{array} \right.$$
We see that the field content of Type IIA string theory emerges
naturally from $d=11$ supergravity fields.

Consider a $d=11$ manifold $M_{10} \times S^1$, with radius $R$
for $S^1$.  The conjecture of M-Theory is that $R =
\sqrt{\alpha^\prime} \expon^{\phi}$; and that for $R \rightarrow
0$, M-Theory reduces to Type IIA string theory.  For $R
\rightarrow \infty$, M-Theory is a Lorentz invariant quantum
theory in $d=11$ such that its low energy infra-red limit is given
by ${\cal N}=1$, $d=11$ supergravity.

Since string coupling constant $g_{IIA} = \expon^{\phi}$, we see
that Type IIA is the weak coupling (small radius) limit of
M-Theory. Indeed, M-Theory has been subjected to several
successful tests so far. For example, all the D-branes in Type IIA
are seen to emerge from dimensional reduction and wrappings of the
M2- and M5-branes of M-Theory.  The compact direction $S^1$ has
KK-momentum modes for a
given point particle given by \beq 
p_{10} = n \frac{2 \pi}{R} = \frac{2 \pi n}{\expon^{\phi}
\sqrt{\alpha^\prime}}, ~ ~ ~ n \in {\bf Z} \eeq What are the
states in Type IIA which correspond to this?  Recall the mass of a
BPS D0-brane in Type IIA is given by \beq 
m_0 = \frac{2 \pi }{g_{IIA} \sqrt{\alpha^\prime}}. \eeq Consider a
collection of $n$ D0-branes.  Since this is a BPS state, their
bound state has energy which is additive and yields
total energy \beq 
n m_0 = \frac{2 \pi n}{g_{IIA} \sqrt{\alpha^\prime}} = p_{10} \eeq
Hence, although Type IIA does not explicitly have an eleventh
dimension, imprints of this dimension are seen in the spectrum of
states.

We have seen that Type IIA results from a $d=11$ higher
dimensional theory.  What about Type IIB? Is this the result of
compactifying some higher dimensional theory?  Consider Type IIB
compactified on $M_9 \times S^1$ with radius $\displaystyle
R^\prime = \frac{\alpha^\prime }{R}$ is dual to Type IIA on $M_9
\times S^1$ with radius $R$. It can be shown that M-Theory
compactified on $M_9 \times T^2$ is equivalent to Type IIB with
$\displaystyle g_{IIB} = \frac{R_{10}}{R_9}$. A simultaneous limit
$R_9, R_{10} \rightarrow 0$ yields an arbitrary $g_{IIB}$.  We see
that the exact $SL(2,{\bf Z})$ symmetry of Type IIB results from
the symmetry of exchanging $R_9, R_{10}$ within the
compactification. M-Theory thus ``geometrizes" the $SL(2,{\bf Z})$
symmetry of Type IIB.

\subsection{F-Theory}

So far, all compactifications of string theory from $M_{10}
\rightarrow M_{10-n} \times K_n$ have been based on making all the
background fields except the metric independent of the compact
manifold $K_n$.

F-Theory stands for a more general compactification of Type IIB,
where the background fields have non-trivial dependence on $K_n$.
For usual compactifications, one solves for Einstein equation for
$K_n$, which gives the Ricci flat condition, \beq 
R_{ij}^{(K)} = 0. \eeq For even $n$, unbroken ${\cal N}=1$
supersymmetry for $M_{10-n}$ requires that $K_n$ be a Calabi-Yau
manifold with $SU(n/2)$ holonomy.

Suppose in addition to the 10-dimensional metric $g_{\mu \nu}$,
other background fields also depend on $K_{n}$.  The simplest case
is to allow this field be the complex scalar field $\displaystyle
\lambda = A^{(0)} + i \expon^{- \phi},$ where $A^{(0)}$ is the
RR-scalar of Type IIB theory. The Lagrangian for these classical
background fields is given by the Type IIB classical supergravity
action given in eq(\ref{bkg2}). We have from eq(\ref{bkg2}) \beq
{\cal L} = \int d^{10}X \sqrt{g} \left( R - \frac{1}{2}
\frac{g^{\mu \nu}
\partial_{\mu} \lambda
\partial_{\nu} \bar\lambda}{\mbox{{\rm Im ~}}(\lambda)^2} \right)
\label{clfld} \eeq Suppose $g_{\mu \nu}$ is the metric for $M_8
\times S^2$. Clearly, $S^2$ is not a Calabi-Yau manifold since it
is not Ricci flat . Hence, Einstein equations for this
compactification must give a non-trivial dependence for the other
background fields on $S^2$. We assume that $\lambda$ is a
function only of $S^2$ 
The field equation for the $\lambda-$field, from eq(\ref{clfld}),
is given by the Einstein field equation \beq 
R_{i j}^{(2)} - \frac{1}{2} g_{i j }{(2)} R^{(2)} =
T_{ij}^{\lambda}, \label{eins} \eeq $(i,j = 1,2)$ where
$T^{\lambda}_{i j}$ is the energy-momentum stress tensor of
$\lambda$.

Let $z = X_8 + i X_9$, then eq(\ref{eins}) yields \beq 
\frac{\partial \lambda(z, \bar z)}{\partial \bar z} = 0 \eeq and
hence $\lambda$ is holomorphic.  However, since Type IIB has
$SL(2, {\bf Z})$ symmetry, $\lambda(z)$ is not an arbitrary
holomorphic function but rather must lie in the fundamental domain
of the torus.
Typically, we obtain, near $z=0$, \beq 
\lambda = \frac{1}{2 \pi i} \ln z. \eeq The monodromy of $\lambda$
about $z$ is given by $z \rightarrow \expon^{2 \pi i} z$, $\lambda
\rightarrow \lambda + 1$.  In other words, at $z = 0$, there is a
single RR-charge.  Since $\lambda$ couples magnetically to a
$D$7-brane, this implies that a single $D$7-brane is located at
the ``point" $z=0$, and fills up the entire space transverse to
$(X_8, X_9)$, i.e. fills up $(X_1, X_2, \cdots, X_7)$
\begin{center}
\epsfig{file=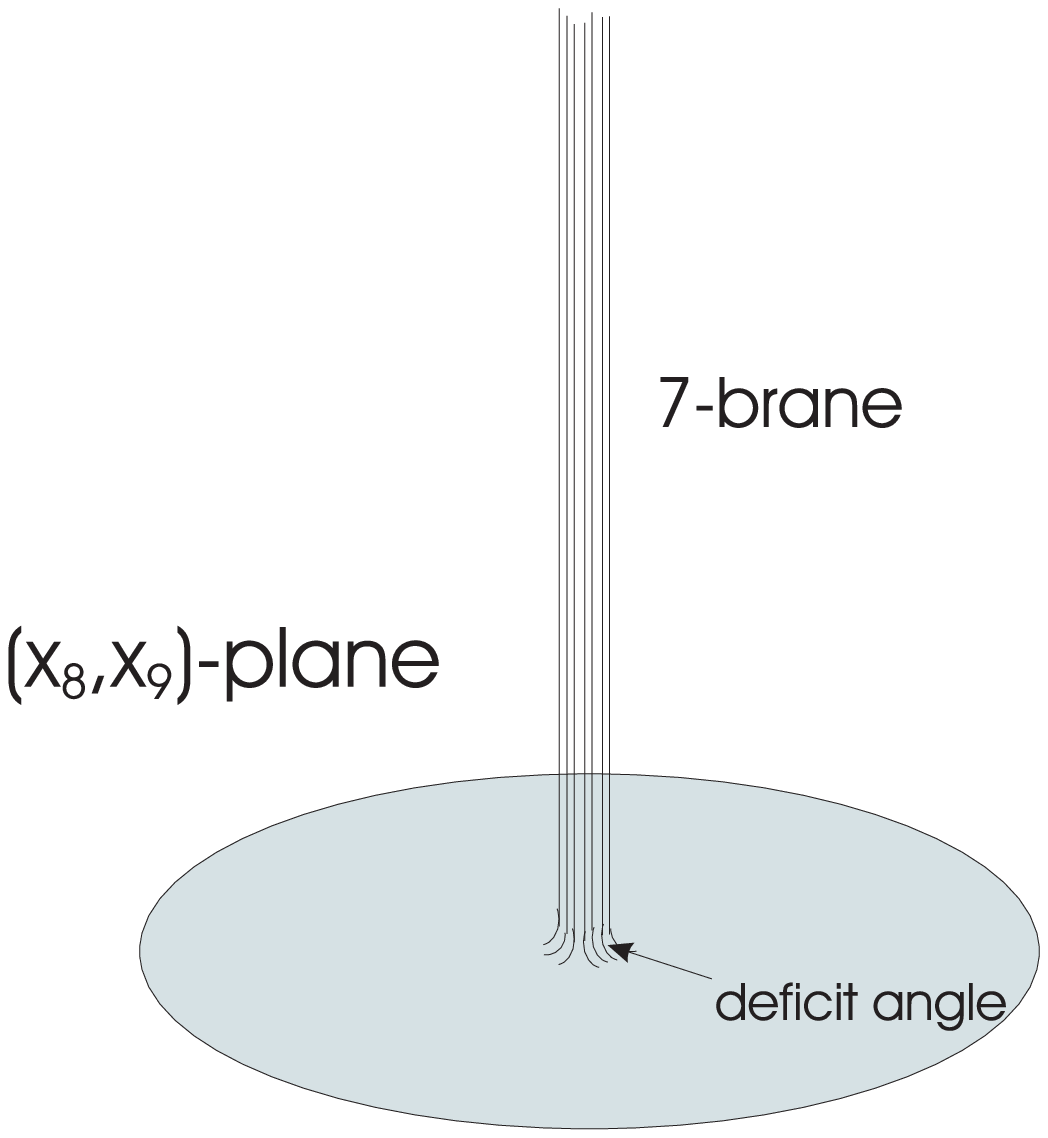, height=.25\textheight}
\end{center}
It is known that the $D$7-brane couples to the metric on $(X_8,
X_9)$ and induces a deficit angle of $\pi/6$ at $z=0$.  Since the
total solid angle of a sphere is $4 \pi$, we need $\displaystyle 4
\pi \div \frac{\pi}{6} = 24$ $D$7-branes placed on the plane
$(X_8, X_9)$ to make it `curl up' into $S^2$.
\begin{center}
\epsfig{file=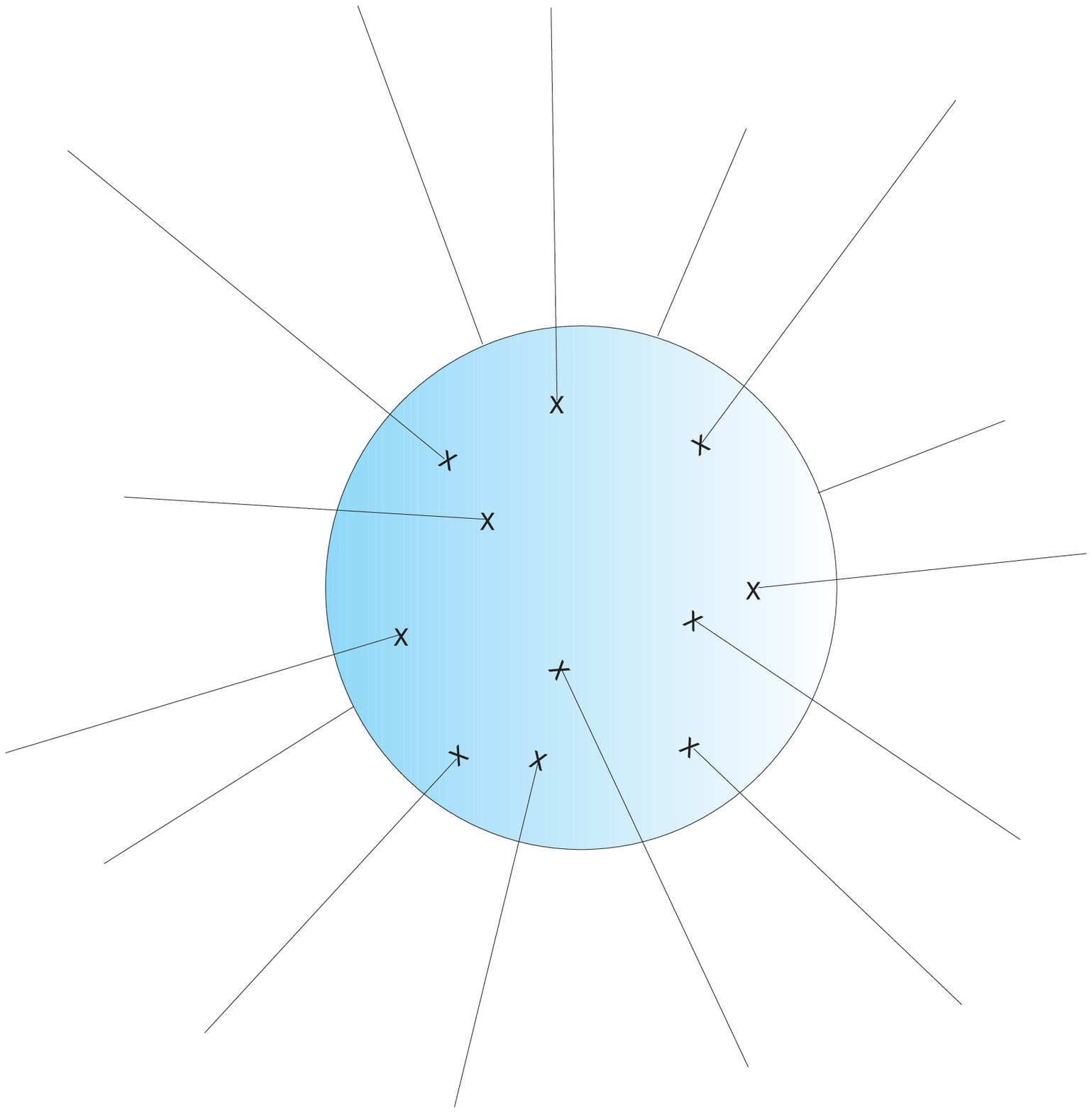, height=.25\textheight}
\end{center}

In other words, near $z=z_i$, \beq  
\lambda \approx \frac{1}{2 \pi i} \ln (z - z_i), ~ ~ ~ i = 1,2,
\cdots 24 \eeq In summary, only by placing 24 $D$7-branes in the
manifold $M_{10}$ transverse to $(X_8, X_9)$ can Type IIB theory
be compactified on $M_8 \times S^2$.  These $D$7-branes appear as
points on $S^2$.

The variation of $\lambda(z)$ over $S^2 = \mbox{{\rm CP}}^1$ is
known exactly, and indeed it is an elliptic fibration of $S^2$
with the total bundle space being equal to $K3$, the Calabi-Yau
manifold with 2 complex dimensions.

In analogy with M-Theory where the dilaton $\expon^{\phi}$ of Type
IIA was seen as the moduli of M-Theory defined on $M_{10} \times
S^1$, one can view the complex dilaton $\lambda$ of Type IIB
theory as the moduli of F-Theory defined on $M_8 \times K3$, a
12-dimensional manifold.

F-Theory refers to the as yet unknown theory in 12 dimensions
which when compactified on $M_8 \times K$, where $K$ is an
elliptic fibration by a torus of some manifold $B$, yields Type
IIB string theory on $M_8 \times B$ with $\lambda$ identified as
the complex structure of the torus.  F-Theory is not as well
understood as M-Theory since it is not clear whether the low
energy supergravity Lagrangian for it even exists.

\bigskip

\section{$D$-Branes from Gauge Fields}

In this section, we briefly review the topological classification
of gauge field configurations. Consider an n-dimensional base
space $M$. Let the fiber $F$ be $k$-dimensional ${\bf R}^k$ and
the bundle space $E$ locally is $M \times {\bf R}^k$. $E$ has dim
= $n + k$.

\begin{center}
\epsfig{file=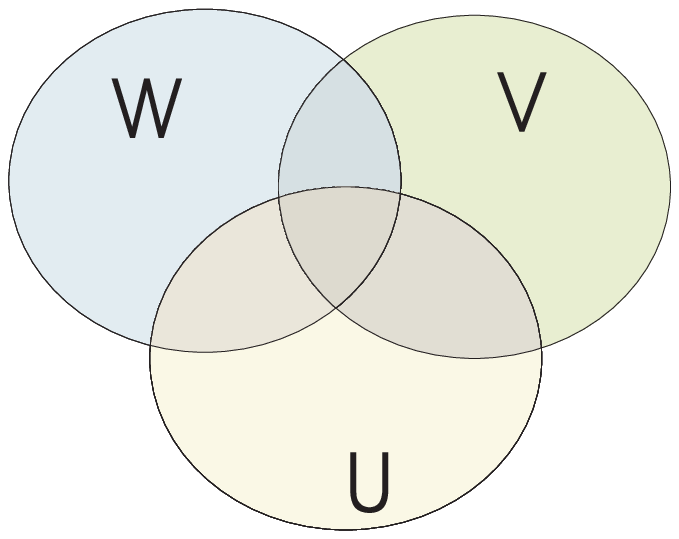, height=.2\textheight}
\end{center}
Consider three patches of $E$. The transition function between
patches $U$ and $V$ is given by $g_{u v}$ such that for $(x, f)
\in U$ and $(x^\prime, f^\prime) \in V$ $$ f^\prime = g_{u v} f $$
Bundle $E$ is called a vector bundle if $\{g\} \in G L (k, {\bf
R})$. $E$ is a principal bundle if the fiber $F = G$ and
transition functions $\{g\} = G$, where $G$ is a compact Lie
group.

Transition function on a triple overlap satisfy the cocycle consistency
condition
$$
g_{u v} g_{v w} g_{w u} = I
$$
This yields the Dirac quantization condition for monopoles.

A YM-connection $A = A_\mu^a T^a d x^\mu$ defines parallel
transport for the principal bundle. On two overlapping patches $$
A^\prime = g A g^{-1} - i d g g^{-1} $$ A cross section or simply
a section of a fiber bundle $E$ assigns a specific point $f(x) \in
F $ for each point $x \in M$.
\begin{center}
\epsfig{file=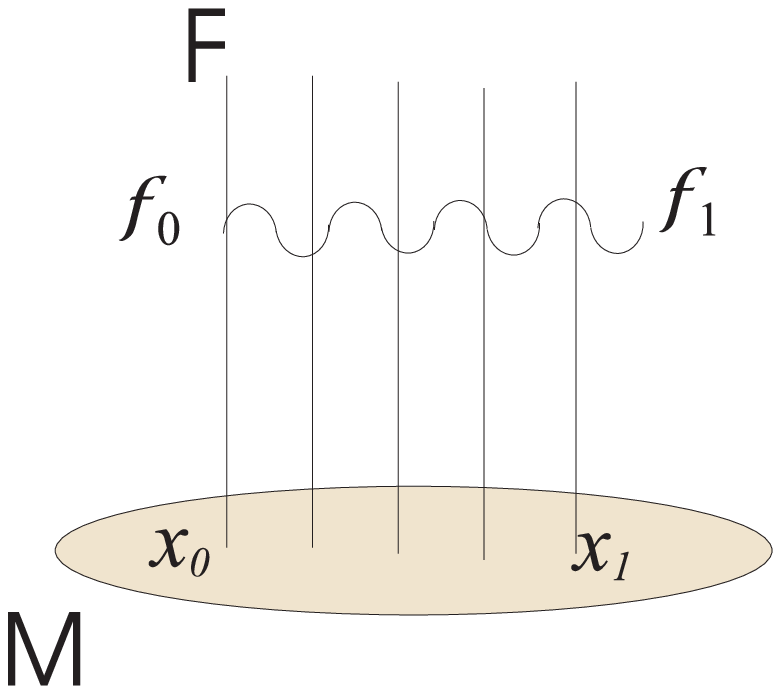, height=.2\textheight}
\end{center}
Matter fields are sections of associated bundles.

Suppose the base manifold $M$ is a compact space, say $T^{2n}$. How do
we classify all the non-trivial principal bundles that can be defined on
$M$? This is done by determining all the topological invariants of the
bundle space $E$. In particular the Chern classes classify all principal
bundles. Let $\Omega = F_{\mu \nu}^a \; T^a \; d x^\mu \wedge d x^\nu$;
then the total Chern form is given by
\bea
c(\Omega) &=& {\rm det} \; (1 + \frac{i}{2 \pi} \Omega) \\
&=& 1 + c_1(\Omega) + c_2(\Omega) + \cdots + c_N(\Omega).
\eea

\noindent The individual Chern classes are given by \bea c_1 &=&
\frac{i}{2 \pi} {\rm tr} F \\ c_2 &=& \frac{1}{8 \pi^2} {\rm tr} F
\wedge F - ({\rm tr} F) \wedge ({\rm tr} F)) \\ c_3 &=& \cdots
\eea Note the $c_i$'s are elements of integer-valued cohomology
classes.

Recall that for a D-p brane, we have an interaction term
$$
S_{cs} \; \sim \; \int_{\sum_{p + 1}} c \wedge {\rm tr} \; \expon^\Omega
$$
where $c = \sum_{k = 1}^N A_{\mu_1, \cdots, \mu_k}^{(k)} d x^{\mu_1}
\cdots d x^{\mu_k}$ with $A_{\mu_1, \cdots, \mu_k}^{(k)}$ being the $R -
R$ fields.

Consider the field tensors $F$ to be a background field on a $D-p$
brane. Then $$ S_{cs} \; \sim \; \int_{\Sigma_{p + 1}} c \; {\rm
tr}
\; F \; \sim \; \int_{\Sigma_{p -1}} \; \int_{T ^2} {\rm tr} \; F %
\; \sim \; c_1 \int_{\Sigma_{p -1}} \; A^{(p-1)} $$
In other words
$c_1$ acts as a charge for a D-$(p - 2)$ brane which couples to
the appropriate $RR$ field $A^{(p - 2)}$. We consequently have
that
\begin{itemize}
\item $\int F$ carries charge of a $(p - 2)$ brane
\item $\int F \wedge F$ carries charge of a $(p - 4)$ brane
\item $\int F \wedge F \wedge F$ carries charge of $(p - 6)$ brane,  etc.
\end{itemize}

Hence, by having background gauge fields with non-trivial topology
we in effect are creating D-branes embedded in the world volume of
a $p$-brane. For example, $\displaystyle \int F \neq 0$ is
equivalent to have a $(p -2)$-brane embedded in a $p$-brane etc.

\subsection{Gauge Fields on $T^d$ (${}^\prime$t Hooft)}

In addition to the instanton number (= $c_2$ = Second Chern
class), for gauge fields in the absence of fermions in the
fundamental representation, the $U(N)$ theory is invariant under
${\Bbb Z }_N$, where $$ Z_N = \{ \expon^{2 \pi i n / N}, \; n = 0,
1, \cdots, N-1 \}. $$ Since $Z_N^{-1} \; A_\mu \; {\Bbb Z}_N \; =
\; A_\mu$, $Z_N$ is a special gauge transformation and the theory
is invariant under $U(N) / {\Bbb Z}_N$.

A consequence of ${\Bbb Z}_N$ symmetry is the existence of another
class of topological quantum numbers in addition to $c_2$. Let
$\Omega \; A_\mu \; \equiv \; \Omega \; A_\mu \; \Omega^{-1} \; -
\; i \; \Omega \;
\partial_\mu \; \Omega^{-1}$. On a
torus, periodicity yields the following.
\begin{center}
\epsfig{file=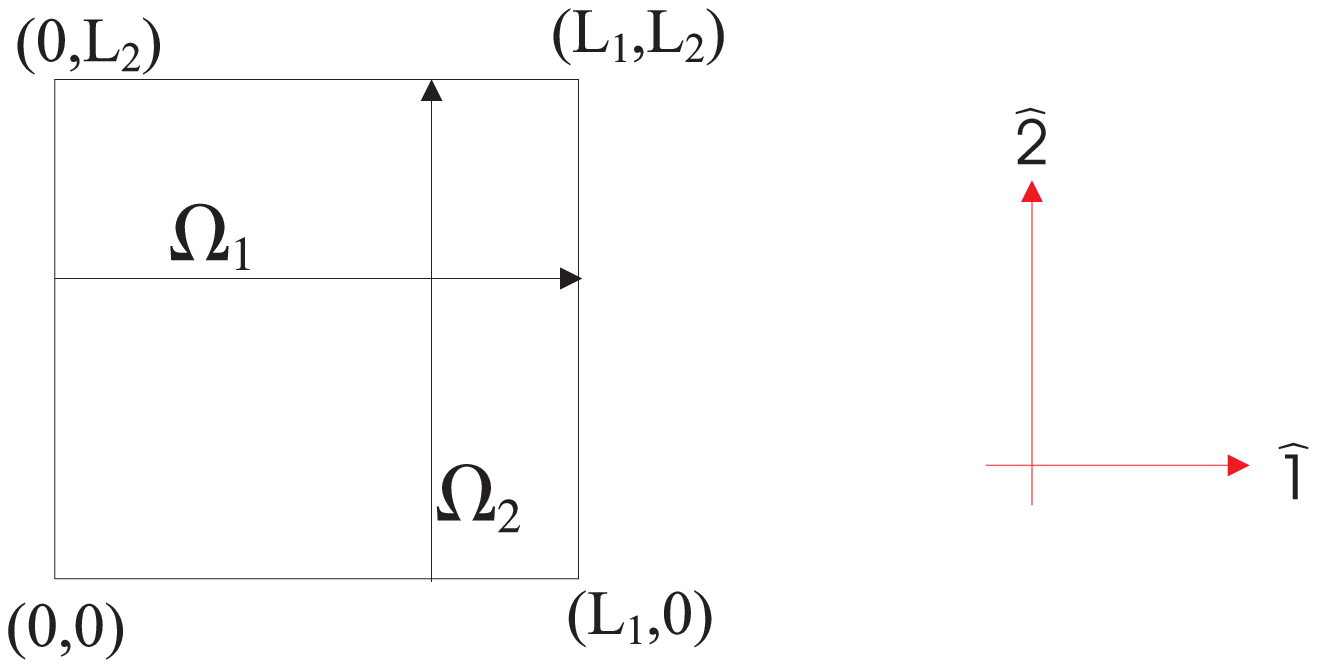, height=.2\textheight}
\end{center}
Then \bea A_\mu (L_1, X_2)  &=& \Omega_1 (X_2) A_\mu (0, X_2) \\
A_\mu (X_1, L_2)  &=& \Omega_2 (X_1) A_\mu (X_1, 0) \eea Consider
two paths $I$ and $II$ from $(0,0)$ to $(L_1, L_2)\equiv 0$.
\begin{center}
\epsfig{file=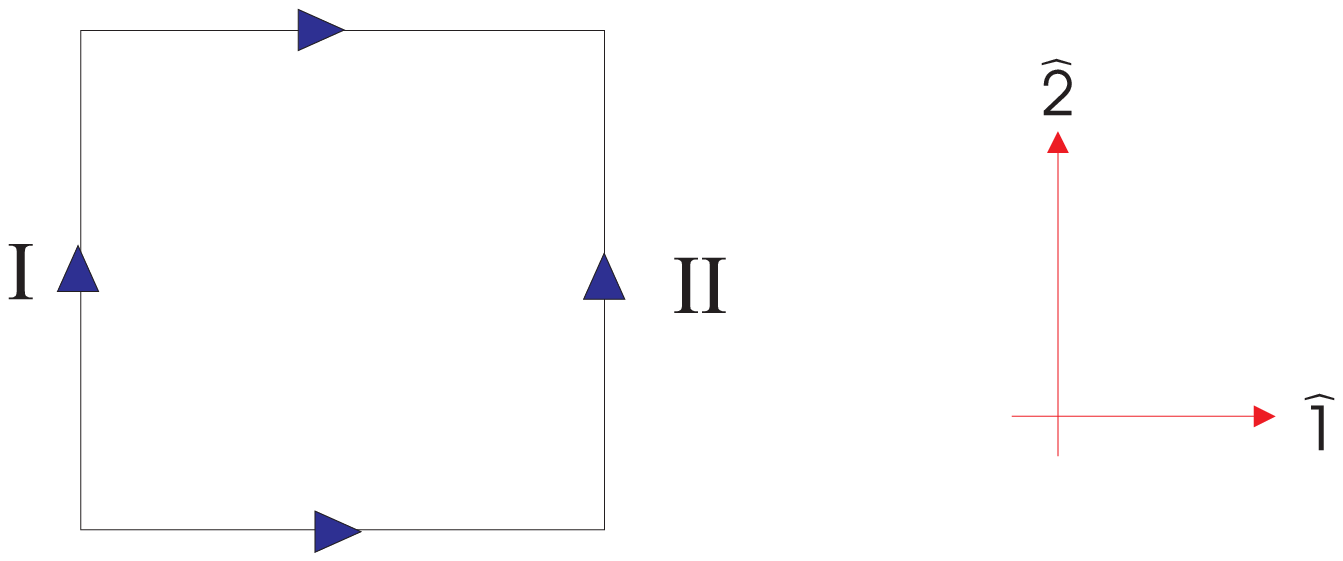, height=.2\textheight}
\end{center}
\bea
{\rm I} :  A_\mu (L_1, L_2) &=& \Omega_2 (L_1) A_\mu (L_1, 0) \\
                            &=& \Omega_2 (L_1) \Omega_1 (0) A_\mu (0, 0)
\eea

\bea
{\rm II} : A_\mu (L_1, L_2) &=& \Omega_1 (L_2) A_\mu (0, L_2) \\
                           &=& \Omega_1 (L_2) \Omega_2 (0) A_\mu (0, 0)
\eea Since $(0, 0) \equiv (L_1, L_2)$, we have $A_\mu (L_1, L_2) =
\Omega A_\mu (L_1, L_2) = z A_{\mu} (L_1, L_2)$ and hence, for
non-trivial topology, consistency requires
\begin{equation}
 \Omega_1 (L_2)
\Omega_2 (0) = \Omega_2 (L_1) \Omega_1 (0) z, \;\; {\rm for } \; z
\in {\Bbb Z}_N  \label{eqno1}
\end{equation} where $\Omega_1, \Omega_2$ reflects
the (non-trivial) topology of $A_\mu$.

For an arbitrary gauge transformation $$ A_\mu (X_1, X_2)
\rightarrow \Omega (X_1, X_2) A_\mu (X_1, X_2) $$ where we can
have $\Omega_2 (X_1)$ and $\Omega_1 (X_2)$ arbitrary but
constrained by eq(\ref{eqno1}). That is, we cannot take both
$\Omega_1 (X_2) \rightarrow I$ and $\Omega_2 (X_1) \rightarrow I$
(trivial topology) since this violates eq(\ref{eqno1}).

Recall in eq(\ref{eqno1}) we choose plane (1, 2) and hence $z =
z(1, 2)$ by continuity. Since there are $\frac{1}{2} d (d - 1)$
independent planes in $T^d$, and $[z] = N$, the number of
topological classes of gauge field configuration is $\displaystyle
N^{\frac{d(d-1)}{2}}$. Introducing fermionic fields in the
fundamental representation destroys the ${\Bbb Z}_N$ symmetry and
consequently wipes out the $\displaystyle N^{\frac{d(d-1)}{2}}$
topological charges.

Let us consider an example. Consider a 2-brane wrapped $N$-times
on $T^2$ giving a $U(N)$ bundle on $T^2$. These bundles are
classified by $\displaystyle c_1 = \frac{\rm tr}{2 \pi} \int F$.
Since $\displaystyle U(N) = \frac{U(1) \times SU(N)}{{\Bbb Z}_N}$,
consider the $U(1)$ component. Suppose $F_{12}$ = constant; then
we can take $A_1 = 0, A_2 = X F_{12}$. On $T^2$, we have \bea X &
\rightarrow & X + 2 \pi R_1 \\ Y & \rightarrow & Y + 2 \pi R_2 \\
A_2 & \rightarrow & A_2 + 2 \pi R_1 F_{12} \eea
\begin{center}
\epsfig{file=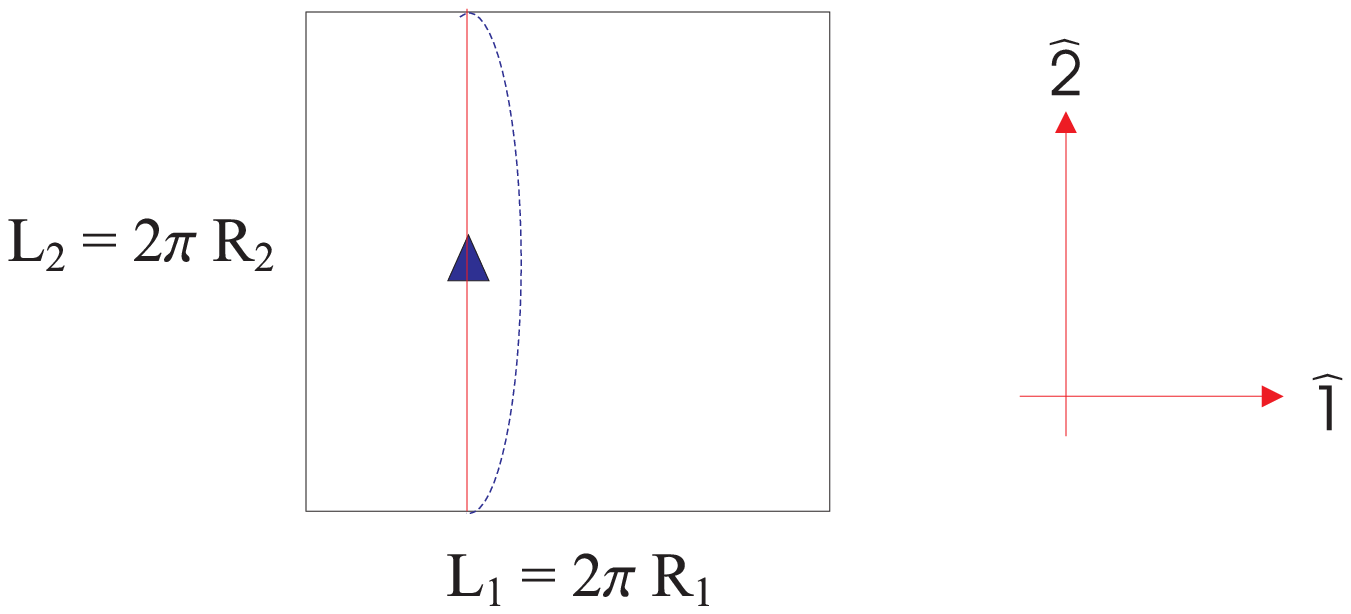, height=.2\textheight}
\end{center}
Hence
\bea {\rm Wilson \; loop\;} &=& \expon^{i \oint d y A_2}
\\
                       &=& \expon^{2 \pi i R_2 X F_{12}} \\
   & \rightarrow & \expon^{2 \pi i R_2 F_{12} (X + 2
\pi R_1)}. \eea Hence, since Wilson loop must be single-valued, we
have \bea R_1 R_2 F_{12} 2 \pi &=& k \in {\Bbb Z}
\\ {\rm or} \;\;\;\;\;\;\;\;\;\;\;\
    F_{12} &=& \frac{k}{2 \pi R_1 R_2} \;\;\;\;\;\;\; : \;\; {\rm
Quantization} \\ \mbox{\rm which yields ~ ~ } c_1 &=&
\frac{F_{12}}{2 \pi} (2 \pi)^2 R_1 R_2 = k \eea For $c_1 = k$, the
$U(1)$ flux is $\dis F= \frac{k}{N} I$ and the $U(N)$ bundle has
twist $z_k = \expon^{2 \pi i k /N}$; in $d=2$, there are $\dis
N^{\frac{2(2-1)}{2}} = N$ topological charges for $U(N)$ given by
$z_k$. . Let us choose $k=1$; then the $U(N)$ principal bundle on
$T^2$ is specified by $$ \frac{{\rm tr }}{2 \pi} \; \; \int F = 1
$$ Choose the boundary conditions \bea \Omega_1 (x_2) & = & V
\expon^{2 \pi i (\frac{x_2}{L})T} \\ \Omega_2 (x_1) & = & I \eea
where \bea V & = & \left( \begin{array}{cccccc} 0 & 1 & 0 & 0 &
\cdots & \cdots
\\ 0 & 0 & 1 & 0 & \cdots & \cdots \\ \vdots & \vdots & \vdots &
\vdots & \vdots & \vdots \\ 1 & 0 & 0 & \cdots & \cdots & 0
\end{array} \right) \\
T & = & {\rm diag} \; (0, 0, \cdots , 1/N)
\eea

For this choice of $\Omega_1$ and $\Omega_2$, $F =$ constant and
\bea A_1 & = & 0 \\ A_2 & = & x_1 F + \frac{2 \pi}{L_2} {\rm diag}
\; (0, \frac{1}{N}, \cdots \frac{N-1}{N}) \eea with $$ F = \frac{2
\pi}{N L_1 L_2} I_{N \times N} $$ $$ c_1 = \frac{1}{2 \pi} \int
{\rm tr} \; F = \frac{2 \pi N}{2 \pi N L_1 L_2} \cdot L_1 L_2 = 1
$$ Hence $c_1 = 1$ is the charge for a $(2-2) = 0$ brane; that is,
due to the flux in the D2-brane, there is a $0$-brane embedded in
$T^2$ together with $N$ 2-branes. $c_1$.
\begin{center}
\epsfig{file=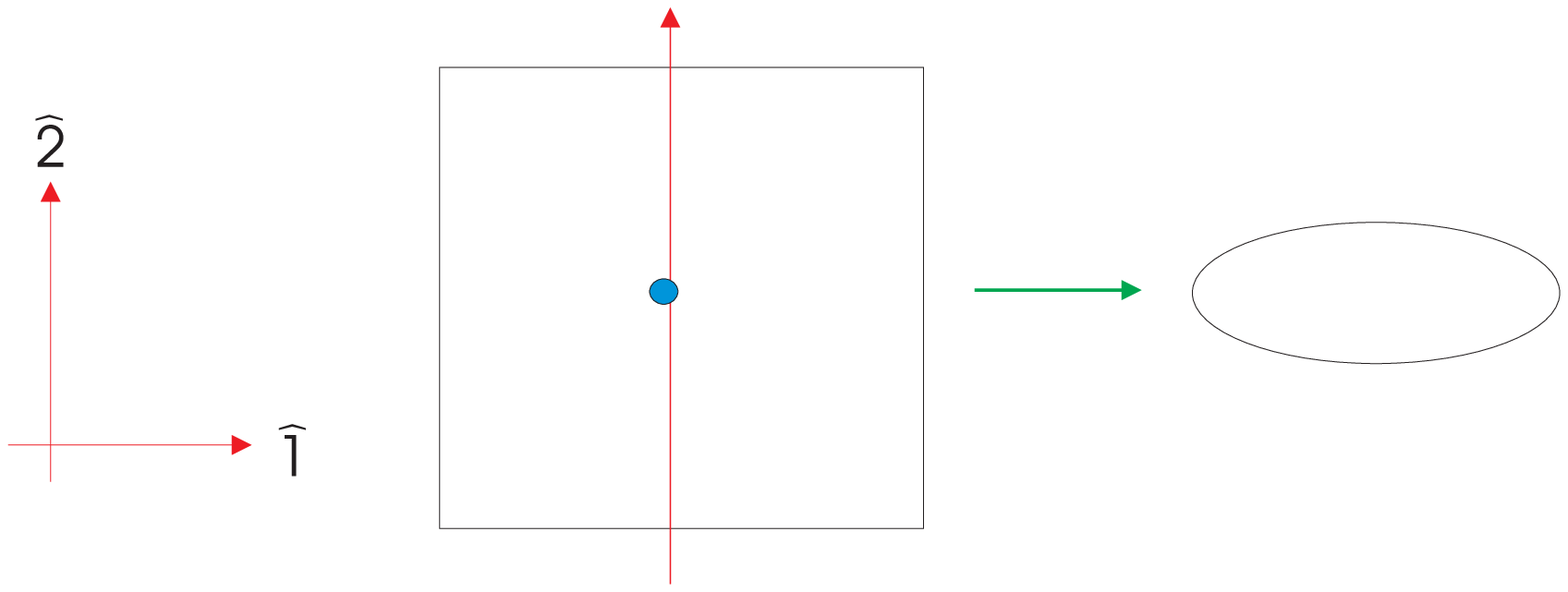, height=.2\textheight}
\end{center}

Let us $T$-dualize the 2-direction inside the 2-brane on $T^2$. We
expect the following in the $X^2$-direction:
\begin{tabular}{ll|ll} $N$ &  2-branes &  $N$ & 1-brane \\ 1 &
0-brane & 1 & 1-brane \end{tabular}
$T$-duality yields \bea X^2 &=& (2 \pi \alpha^\prime) (i
\partial^2 + A^2) \\ &=& (2 \pi \alpha^\prime) A^2 \eea where
$\partial^2 = 0$, being orthogonal to brane. Hence,
\bea X^2 &=& \frac{4 \pi^2 \alpha^\prime}{L_2} \frac{1}{N}
\mbox{\rm diag} \left(
 \frac{x_1}{L_1}, \frac{x_1}{L_1} + 1,
\cdots, \frac{x_1}{L_1}+ N - 1 \right) \label{144}\\ &=& {\rm
coordinates \; of \; the \; brane \; in \;} \hat{2} - {\rm \;
direction.} \eea Since we are on $T^2$ we have from eq(\ref{144})
\begin{center}
\epsfig{file=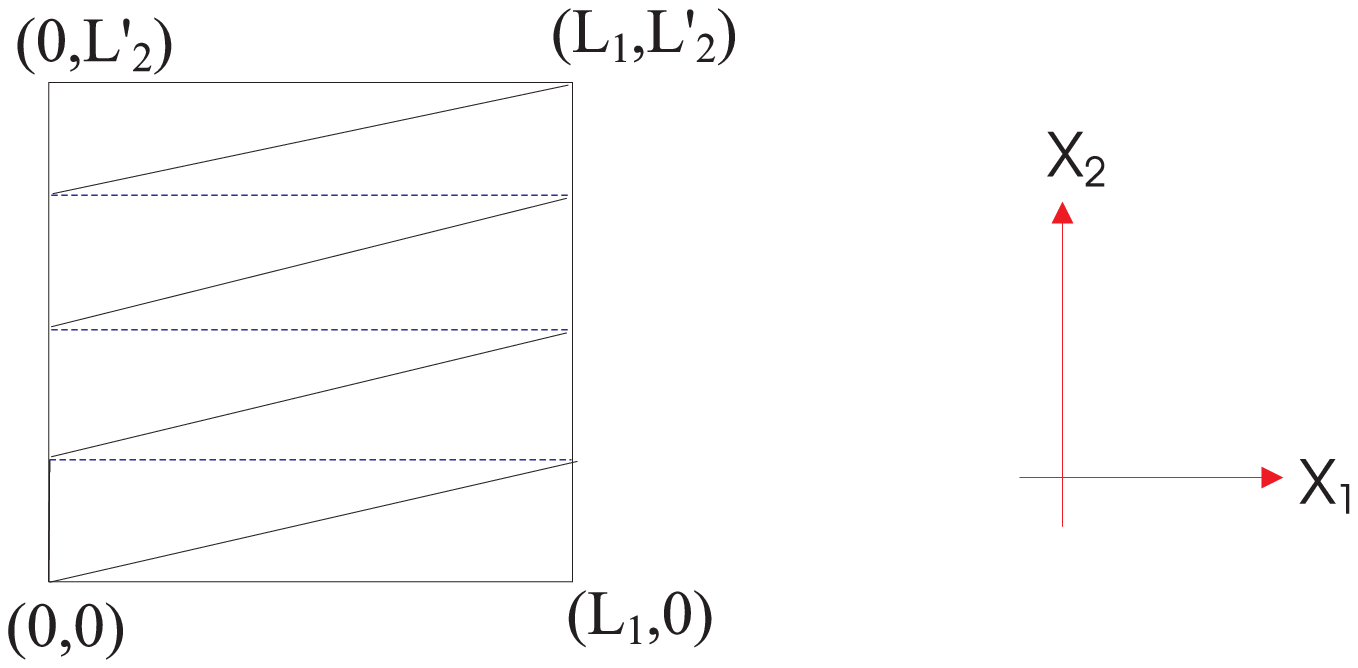, height=.2\textheight}
\end{center}
The 1-brane coming from the $T$-dual of the 0-brane is in the
$X_2$-direction and winds only once since $c_1 = 1$ around the $X_2$-direction.

\begin{center}
\epsfig{file=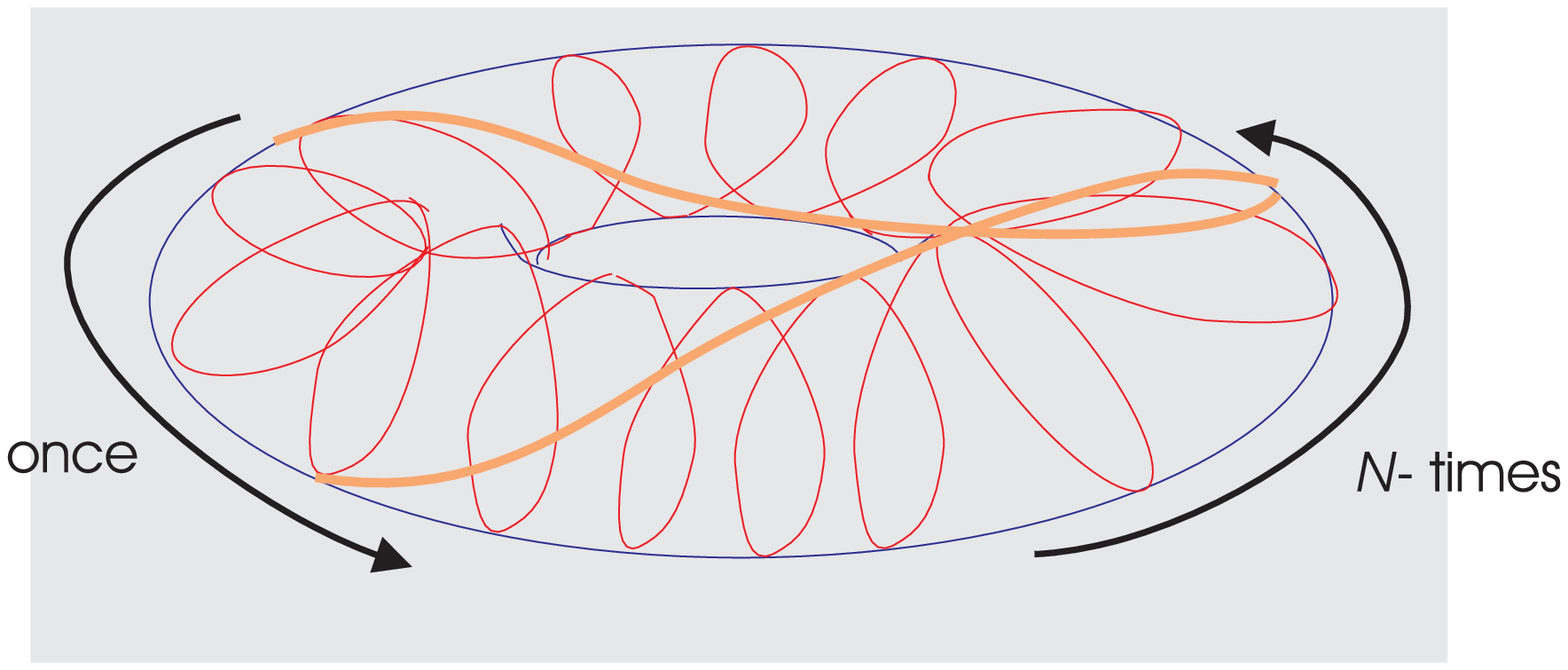, height=.2\textheight}
\end{center}
The $N$-winding is due to the original brane configuration. The 1-brane
$T$-dual to the 2-brane winds $N$-times since the 2-brane was wrapped
$N$-times on $T^2$ to start with.

What happens when $\displaystyle c_1 = \frac{1}{2 \pi} \int \;\mbox{\rm tr}
\; F \; = \; k$? We first of all expect $k$ 0-branes to be embedded in
$T^2$. On $T$-duality thus we expect each 0-brane to go to its $T$-dual
1-brane. But since all these $k$ 0-branes come from the same non-trivial
background gauge field, we expect these $k$ 1-branes to wind in the
$X^2$-direction $k$-times.

\begin{center}
\epsfig{file=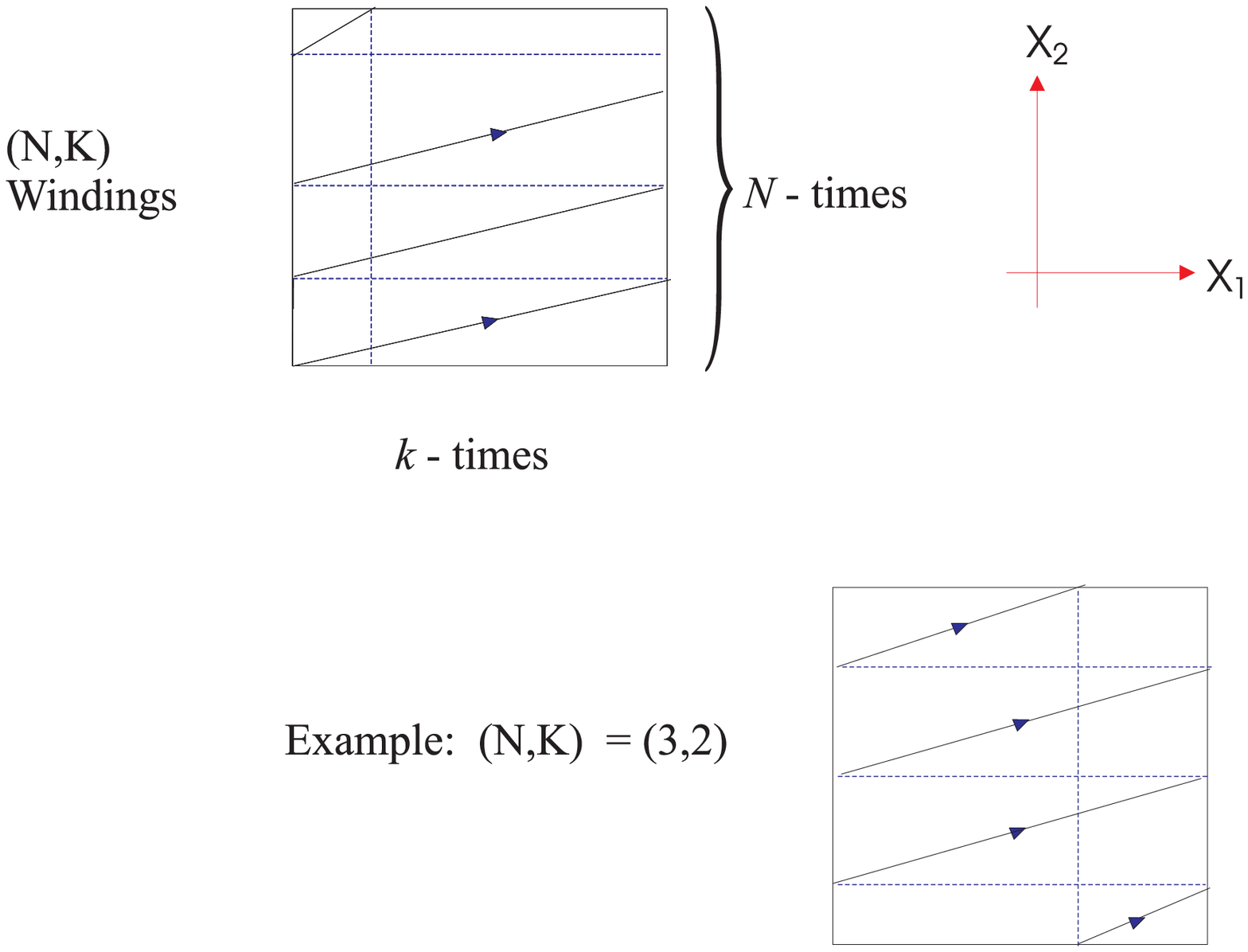, height=.4\textheight, width=.9\textwidth}
\end{center}

\bigskip

\section{D-Brane Dynamics}

The complete description of D-brane physics is given by the
Born-Infeld lagrangian $$ S = - \tau_p \int d^{p +1} \xi \; {\rm
det}^{\frac{1}{2}} (G + B + 2\pi \alpha^\prime F) \;+\; {\rm
fermions} \;+\; CS $$ where $G_{\mu \nu}, B_{\mu \nu}$ are the
NS-fields and $F_{\mu \nu}$ is constructed from the
ten-dimensional RR-fields. $\tau_p$ is the `tension' of a p-brane.

In the static gauge, flat background and $U(1)$ gauge fields we have to
dimensionally reduce
$$
S = - \int d^{10} x \sqrt{- \;{\rm det}\; (\eta_{\mu \nu} + F_{\mu \nu} +\;
{\rm fermions})}
$$
After dimensional reduction
$$
S = - \tau_p \int d^{p+1} \xi \sqrt{ \;{\rm det}\; (\eta_{\mu \nu} +
\partial_{\alpha} X^a \partial_{\beta} X^a + 2 \pi \alpha^\prime
F_{\alpha \beta}) +\; {\rm fermions}}
$$
This simplified action can describe simple aspects of D-brane dynamics
such as energy, fluctuations and scattering amplitudes.

Consider the following special D $p$-brane configurations
\begin{itemize}
\item $X^a = 0$, for $a = p+1, \cdots, 9$
\item $F_{\alpha \beta} =$ constant
\item $[F_{\alpha \beta}, F_{r \delta}] = 0$
\end{itemize}
This yields
$$
S = - \tau_p \int d^{p + 1} \xi \; {\rm tr} \; \sqrt{- \; {\rm det} \;
(\eta_{\alpha \beta} + 2 \pi \alpha^\prime F_{\alpha \beta}) }
$$
The energy is given by
$$
E = \tau_p \int d^{p} \xi \; {\rm tr} \; \sqrt{ \; {\rm det} \;
(\delta_{\alpha \beta} + 2 \pi \alpha^\prime F_{\alpha \beta})}
$$
where the trace refers to non-abelian index and determinant
is on Lorentz indices.
Note for a p-brane
$$
\tau_p = \frac{ {\rm Energy}} { {\rm Volume \;of \;p-brane}}
$$
From duality, for $S^1$ with radius $R$, we have on taking $T$-dual, $p
\rightarrow p \pm 1$ and
\bea
\tau_p &=& \tau_{p + 1}^\prime (2 \pi R^\prime), \;\;\;\;\;\tau_p (2 \pi R) =
\tau_{p -1}^\prime \\
R &=& \alpha^\prime / R^\prime
\eea

\noindent Examples

(a) $N$ 2-branes with $c_1 = q$ units of flux.
\beq {\rm det} \; (\delta_{\alpha \beta} + 2 \pi \alpha^\prime
F_{\alpha \beta}) =  \left( 1 + (2 \pi \alpha^\prime F_{12})^2
\right) \eeq
Hence, \bea E &=& \sqrt{ 1 + (2 \pi \alpha^\prime F_{12})^2} \;
\tau_2 \int d^2 \xi \;{\rm Tr} \; ({\rm I}) \\ &=& \tau_2 N L_1
L_2 \sqrt{ 1 + 4 \pi^2 \alpha^{\prime 2} F_{12}^2 }. \eea 
Since $\displaystyle F_{12} = \frac{q}{N L_1 L_2}$, we have $$ E =
\sqrt{ (N \tau_2 L_1 L_2)^2 + (q \tau_0)^2} $$ where $\tau_0 = 2
\pi \alpha^\prime \tau_2 =$ mass of 0-brane. Note for a system of
$N$ 2-branes and $q$ 0-branes that are well separated we expect $$
E_{\infty} = N \tau_2 L_1 L_2 + q \tau_0 $$ Since $E <
E_{\infty}$, the system forms a bound state.

On $T$-dualizating the $\hat{2}$-direction \bea \tau_2 L_2 &=&
\tau_1^\prime \\ \tau_0 &=& \tau_1^\prime L_2^\prime \eea Since
$E$ is unchanged we have \bea E &=& \sqrt{ (\tau_1^\prime N L_1)^2
+ (\tau_1^\prime L_2^\prime q)^2} \\ &=& \tau_1^\prime \sqrt{ (N
L_1)^2 + (q L_2^\prime)^2} \eea where from the diagram, we see
that $E$ is the energy for $(N,q)$ windings on $T^2$.
\begin{center}
\epsfig{file=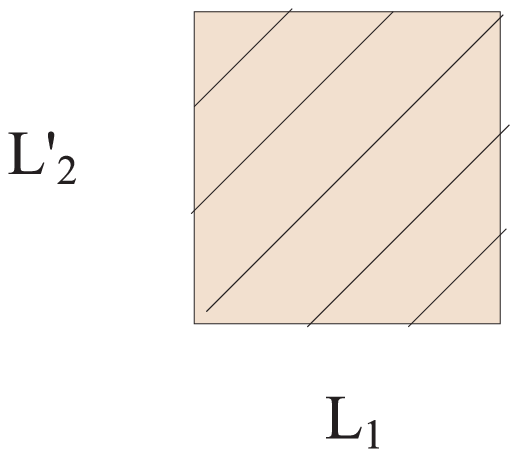, height=.2\textheight}
\end{center}

(b) Consider 2 4-branes wrapped on $T^4$ with instanton number $k = 2$
and volume $V = L_1 L_2 L_3 L_4$.
Choose linear connection
\bea
A_1 &=& A_3 = 0 \\
A_2 &=& \frac{2 \pi X_1}{L_1 L_2} \tau_3, \;\;\;\;\; A_4 = \frac{2 \pi
X_3}{L_3 L_4} \tau_3
\eea
Then
\bea
F_{12} &=& \frac{2 \pi}{L_1 L_2} \tau_3, \;\;\;\;\; F_{34} = \frac{2
\pi}{L_3 L_4} \tau_3 \\
E &=& \tau_4 V \;{\rm Tr}\; \sqrt{ {\rm det}\; (\delta_{\alpha \beta} + 2
\pi \alpha^\prime F_{\alpha \beta})}
\eea
\bea
& & {\rm det} \left( \begin{array}{cccc}
I & 2 \pi \alpha^\prime F_{12} \tau_3 & 0 & 0 \\
-F_{12} \tau_3 & I & 0 & 0 \\
0 & 0 & I & F_{34} \tau_3 \\
0 & 0 & - F_{34} \tau_3 & I
\end{array} \right) \nonumber \\
& = & I (1 + (2 \pi \alpha^\prime F_{12})^2)
(1 + (2 \pi \alpha^\prime F_{34})^2) \\
\Rightarrow E & = & 2 \tau_4 V \sqrt{(1 + (2 \pi \alpha^\prime
F_{12})^2) (1 + (2 \pi \alpha^\prime F_{34})^2)}
\eea

If $L_1 L_2 = L_3 L_4, F_{12}= F_{34}$ : self-dual, and hence \bea
E &=& 2 \tau_4 V + 2 \tau_0 \\ &=& {\rm energy \; of \; 2 \;\;
4-branes \; + \; energy \; of \; 2 \;\; 0-branes} \eea since $F$
is self-dual, the system satisfies the $BPS$ condition and energy
being additive is a reflection of the fact that $\frac{1}{2}$ susy
is preserved.

\bigskip

\section{Black Hole Entropy and D-Branes}

General Relativity predicts gravitational collapse with the
formation of a space-time singularity covered by an event horizon
with Schwarschild radius $R_s$. The area of the event horizon
$A_H$ always increases, i.e. $\Delta A_H \geq 0$.

Quantum fields in a black hole classical background exhibit
Hawking thermal radiation emanating from the black hole of mass
$M^5$ with temperature given by \footnote{Restoring the $\hbar$}
\bea T_H & = & \mbox{\rm Hawking \ Temperature}
\\ & \sim & \frac{\hbar}{R_S} \sim \frac{\hbar}{GM} \mbox{\rm
in} \ d = 4
\\ S_{BH} & = & \mbox{\rm entropy \ of \ black \ hole} \\ & = &
\frac{A_H}{4 G \bar{h}} \eea

We also have $dM = T_H dS$ (First Law of Thermodynamics) where
\bea S & = &  \mbox{\rm total \ entropy} \\ & = & S_{BH} + S_{{\rm
Radiation}} \\ \Delta S & \geq & 0 \eea

\subsection{Charged Black Holes}

For a black hole with mass $M$ and charge $Q$, the cosmic
censorship hypothesis states that naked singularities do not form
from generic smooth initial configurations.  This hypothesis
implies that $$ M \geq Q $$ For the extremal case $M= Q$, we have
the BPS condition that electric and gravitational forces exactly
cancel giving no interaction energy.

Starting from $M > Q$, the black hole radiates off energy until it
reaches equilibrium with $M = Q$. This violates the third Law of
Thermodynamics which says that $S \rightarrow 0 $ as $
T\rightarrow 0$.  For $M > 0$, the black hole emits mostly
electrons since in appropriate units $m_e << e$.
\begin{center}
\epsfig{file=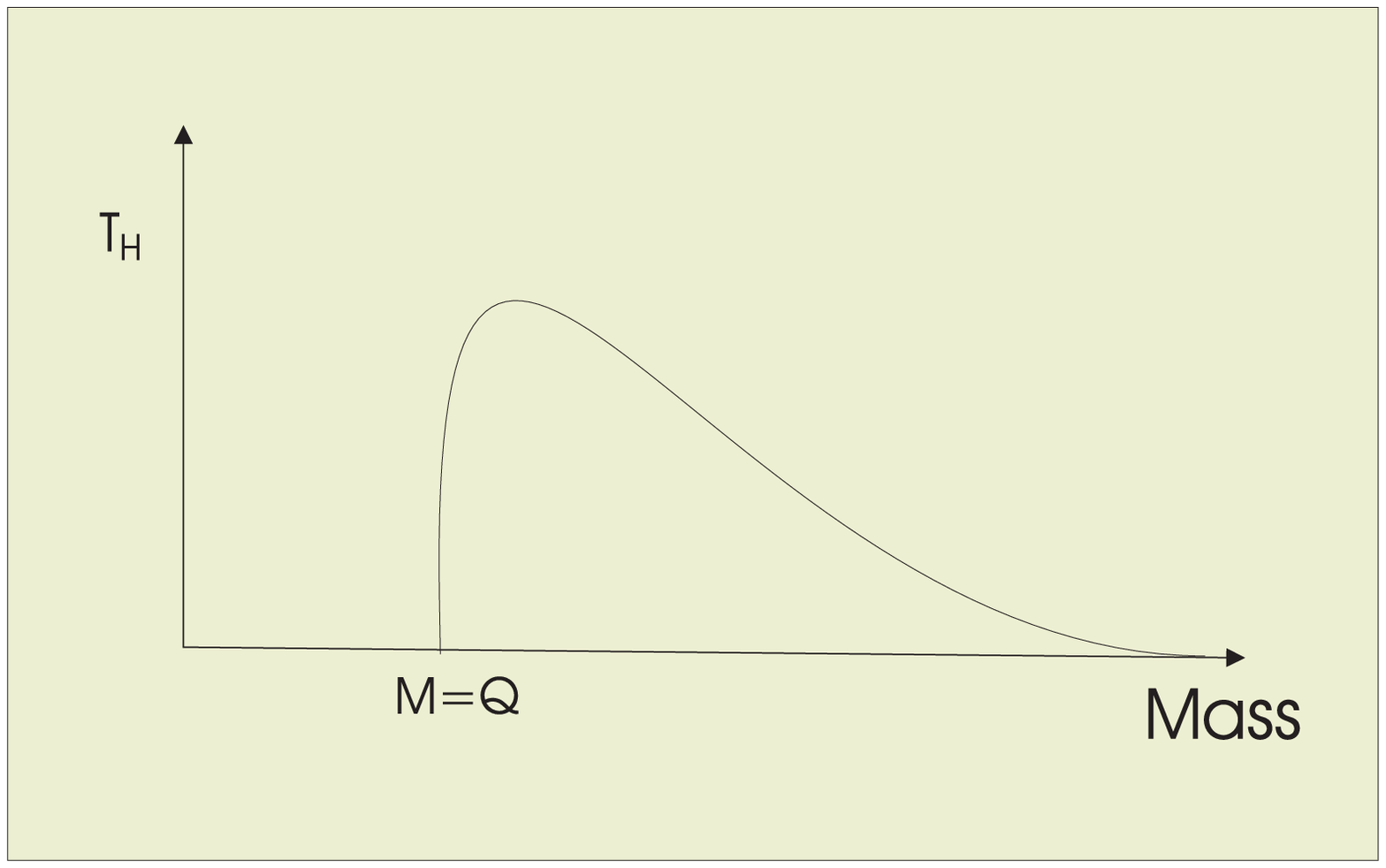, height=.3\textheight}
\end{center}
The extremal black hole, with $M = Q$, has a finite horizon as
well as non-zero entropy.  This is a particularly useful case
since there is no time dependence in the problem and its entropy
can directly be calculated from its density of states.

\subsection{Black Holes and Strings}

One might at first sight think that black holes being macroscopic
objects have nothing to do with strings.  More precisely, since
the Schwarschild radius is $R_S \sim G M_{BH}$ (which for $M_{BH}
= M_\odot$ gives $R_S \sim 1.5$ km ) and strings are typically of
Planck length $\ell_p \sim \sqrt{G} \sim 10^{-35}$ m, the scales
do not match. Also, for an excited string state at level $N$, its
mass is $M_{str} = \sqrt{N} /\ell_{str}$, where $\ell_{str}$ is
the average length of a string, the density of states
$\rho_{str}(M_{str}) \sim \expon^{M_{str}}$ whereas since $\dis
S_{BH} = \frac{A_H}{4G} \sim \frac{R_S^2}{G} \sim G M_{BH}^2$, we
have $\rho_{BH}(M_{BH}) \sim \expon^{M_{BH}^2}$. Hence, naively,
$\rho_{BH}$ for a black hole of mass $M$ does not match the
$\rho_{str}$ for a string of the same mass.

So how can we resolve this contradiction?  To start with, note
that $\dis G= g^2 \ell_{str}^2$, where $g$ is the dimensionless
effective string coupling constant.  We can study the string in
two limits, namely
\begin{itemize}
\item the string limit where $\ell_{str}$ is kept fixed at $\ell_p$
as $g$ varies and
\item the Planck limit where $G$ is held fixed as $g$ varies.
\end{itemize}
Both approaches give the same answer, but we consider the Planck
limit for clarity and treat $G=$ as a constant.

Consider $$ \frac{M_{BH}}{M_{str}} = \frac{R_S}{G}
\frac{\ell_{str}}{\sqrt{N}} = \frac{R_S}{g^2 \ell_{str} \sqrt{N}}
$$ Clearly, for an excited state $N$, the mass of the black hole
and string in general are not equal.  Where should we match them?
Clearly, we should set them equal where $R_S \sim \ell_{str}$,
i.e. $$1 = \frac{M_{BH}}{M_{str}} = \frac{1}{g^2 \sqrt{N}}  $$
That is, for a given $N$, consider coupling $g$ such that $g^2
\sqrt{N} = 1$. For this coupling $M_{BH} = M_{str}$.  Note that we
have not assumed that $\ell_{str} \sim \ell_p$.  On the contrary,
since, $$ \ell_{str} = \frac{\sqrt{G}}{g} =  N^{1/4}\sqrt{G} $$ we
can consider highly excited states such that $\ell_{str} \sim R_S
\sim 1$ km. For such large $N$, the description of the black hole
as an excited string state becomes appropriate. The reason being
that general relativity is valid only if the curvature of space is
much less than $\dis \frac{1}{\ell_{str}}$.  When $R_S \sim
\ell_{str}$, this description breaks down and black holes are more
appropriately described by string theory.

\begin{figure}
\epsfig{file=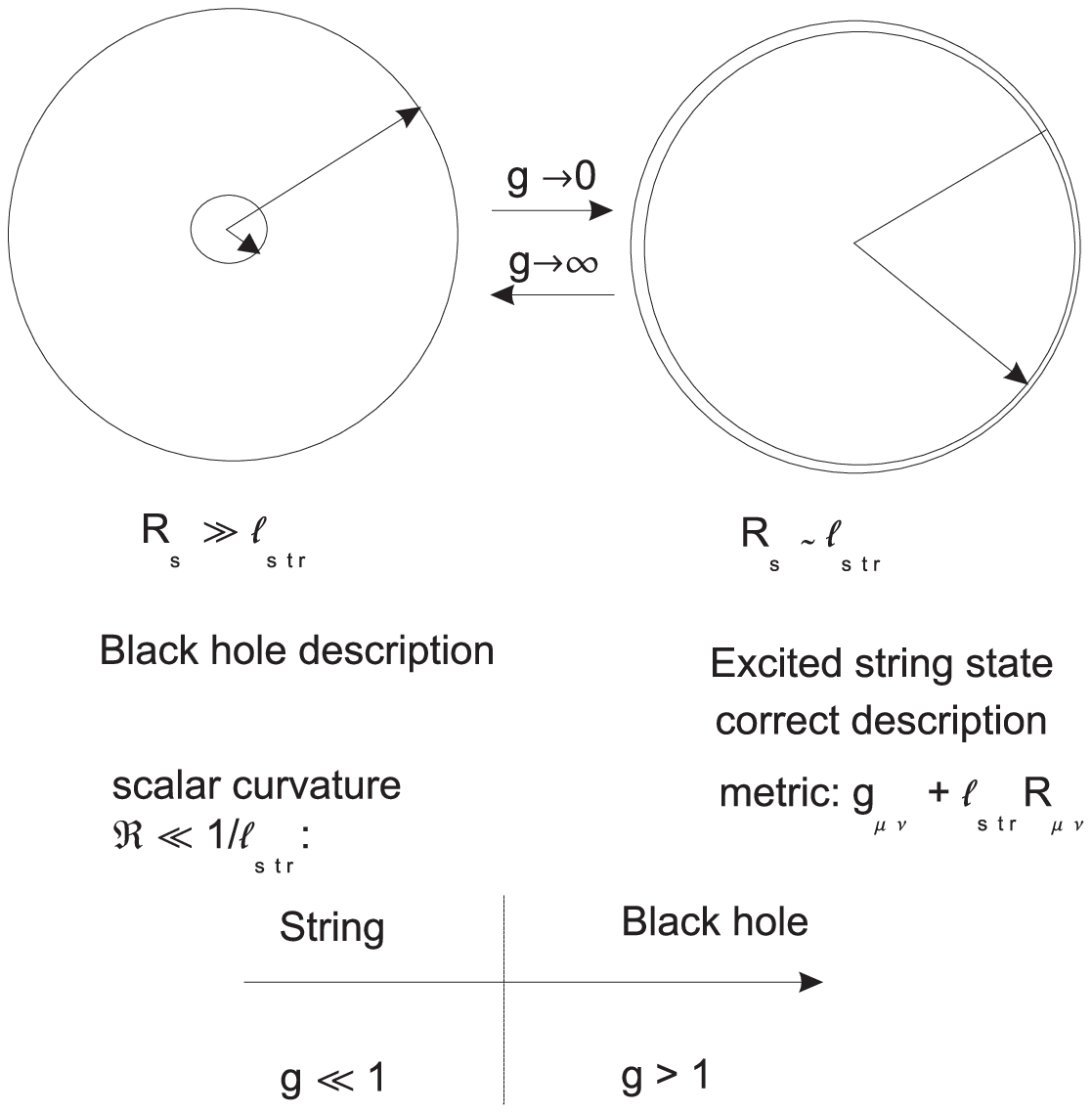}
\end{figure}

We are interested in extremal black holes with $M = Q$. We need to
consider supersymmetric black holes since BPS-states are then
independent of coupling $g$.  For such black holes, $M$ does not
change with $g^2$ and hence $M_{BH} = M_{str} = M$ for all
couplings.  We immediately run into a problem.  Most
supersymmetric black holes\footnote{Supersymmetric black holes
exist only for four and five dimensions} have zero horizon.  To
obtain supersymmetric black holes with a finite horizon, we need
several charges, namely three different charges in five dimension
and four charges in four dimensions.  These charges are carried by
fundamental strings and $D$-branes which couple to ten dimensional
$NS$-fields and $RR$-fields. Hence it was only with the
introduction of open strings and D-branes into the theory that
black holes could be addressed. We are interested in a
perturbative description of string theory as $g \rightarrow 0$.
Note that
\begin{enumerate}
\item p-branes carrying electric NS charge have mass $\sim 1$
\item p-branes carrying magnetic NS charge have mass $\sim
\frac{1}{g^2}$
\item D-branes carrying RR charge have mass $\sim \frac{1}{g}$
\end{enumerate}

Since $G M$  determines the gravitational field of mass M, and $G
= g^2 \ell_S^2$, we see that as $g \rightarrow 0$ only (1) and (3)
have $GM \rightarrow 0$.

In five spacetime dimensions, we need a three form field coupling
to one of the required charges.  A magnetic NS charge carrying
p-brane has no weak coupling description.  Fortunately, in Type
IIB, we have a two form gauge field $A^{(2)}$ which couples to a
D-1 brane. This two form gauge field is dual to $\tilde{A}^{(6)}$
which couples to a D5-brane. Since these D-branes have an exact
description as $g \rightarrow 0$, we will be able to count all the
microstates of these D-brane configurations.

\section{Five Dimensional Charged Black Holes}

We begin with Type IIB string theory in $d= 10$.  The low energy
effective background theory is 10-dimensional supergravity.  One
can find classical solutions for this theory in which the
$D$-branes appear as solitons carrying RR-charge.  One first
obtains in $d=10$ classical solutions with non-zero: $Q_5$, D5
brane charge; $Q_1$, D1-brane charge; $N$, momentum of D1-brane.

In Type IIB, consider compactification\cite{taylor2, ganor} on
$M^5 \times T^5$ with  $T^5 = T^4 \times S^1$. On compactifying
from $d= 10$ to $d = 5$, one obtains a 5-dimensional black hole.
Taking the extremal limit yields $T_H \rightarrow 0$ and entropy
as $$ S_{BH} = 2 \pi \sqrt{N Q_1 Q_5} $$ Let the volume of $T^4$
be $(2 \pi)^4 V$, $R$ be the radius of the circle $S^{1}$ and
$d\Omega_3$ be the line element on a unit three-sphere.  The
canonical black hole metric in the non-compact dimensions is given
by
\begin{equation}
ds^2 = -\frac{1}{\lambda^{3/2}} dt^2 + \lambda^{1/3} (dr^2 + r^2
d\Omega_3^2)
\end{equation}
where $\dis \lambda = (1 + \frac{r_1^2}{r^2})(1 +
\frac{r_5^2}{r^2}) (1 + \frac{r_N^2}{r^2})
$
and \bea r_1^2 & = & \frac{(RV)^{2/3}}{V \sqrt{g}} Q_1 \\ 
r_5^2 &= & (R V)^{2/3} \sqrt{g} Q_5 \\ 
r_N^2 &=& \frac{(R V )^{2/3}}{R^2 V} N
 \eea

Why do we need $N$, $Q_1$ and $Q_5$? Essentially it is to obtain a
finite area for the horizon of the black hole.  As one approaches
the D5-brane horizon, the volume parallel to the brane shrinks due
to brane tension and the volume perpendicular to the brane
expands.  By superposing a D1-brane along one of the $T^5$
directions, the remaining directions are stabilized.  The volume
along the D1-brane due to tension also tends to shrink to zero and
to balance this tension momentum $N$ is given to the D1-brane.
Note that all the branes appear as point-like particles in $M^5$,
and it is really the microstates of this `point' particle that we
are computing.

\subsection{D-Brane Description of supersymmetric Black Holes}

We wrap a D5-brane $Q_5$ times on $T^5$ and a D1-brane $Q_1$ times
on $S^1$.  Since the theory is boost invariant, the momentum $N$
can only come from the massless excitations of this system. Recall
since D5- and D1- brane form a BPS system of bound states, we have
$$ M = \frac{RV}{g} Q_5 + \frac{R}{g} Q_1 + \frac{N}{R} $$ where
$g$ is the string coupling.

The massless modes must move along $S^1$ in one direction to
maintain the BPS condition.  A massive excitation would violate
the BPS condition.  Excitations of the branes are described by
open strings and the ones which contribute are those that go from
1-brane to 1-brane, namely (1,1), as well as (1,5), (5,1) and
(5,5). One can do an explicit calculation to compute the open
string contributions.

\begin{center}
\epsfig{file=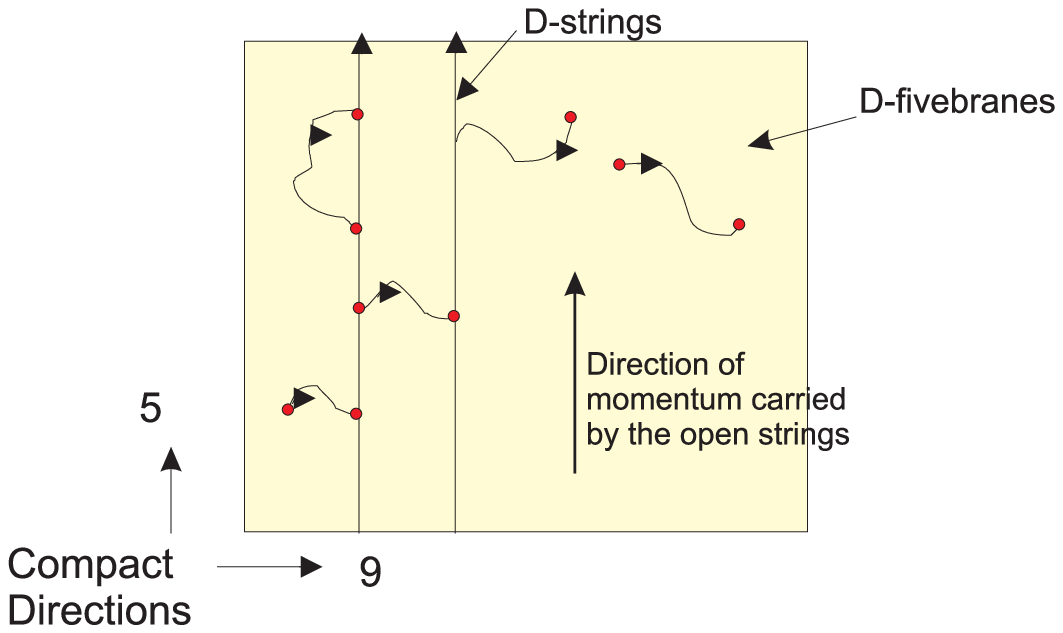}
\end{center}

To compute the entropy of the charged black hole, we need to count all
the microstates which can yield momentum $N$ for a $Q_5$, $Q_1$ BPS
bound state of a D5 with a D1-brane. Since only massless excitations can
contribute to the microstates, we can use the Yang-Mills description of
D-branes to do this counting.

Start with $U(Q_5)$ ${\cal N}=1$ super Yang-Mills in $d=10$ and
dimensionally reduce to $d=5 + 1$ $$ A_\mu  \rightarrow \left\{
\begin{array}{ll} A_\alpha, & \alpha = 0,5,6,7,8,9 \\ A_I \equiv
X_I & I = 1,2,3,4
\end{array}
\right.
$$
$$
S_{YM} = \frac{1}{g_{YM}^2} \int d^{5+1} \xi (\mbox{\rm tr ~} F_{\alpha \beta}
F^{\alpha \beta} + \frac{1}{2} \mbox{\rm tr~} D_\alpha X_I D^\alpha X^I +
\frac{1}{4} \mbox{\rm tr ~} [X_I, X_J]^2 + {\rm fermions} )
$$
where $g_{YM}^2= g$ is the closed string coupling.

To introduce D1-branes we consider topologically non-trivial gauge
field configurations for $A_\alpha$.  Recall the interaction term
for brane coupling is given by $$ \int_{\Sigma_{5 + 1}}  c \wedge
\mbox{\rm tr} \expon^F $$ where $c$ is the sum of the the
$RR$-fields. Consider the term $\dis \int_{\Sigma_{5 + 1}} A^{(2)}
\mbox{\rm tr } F \wedge F
$
where $A^{(2)}$ is the $RR$ 2-form gauge field, with an instanton
configuration in the compact directions 5,6,7,8, i.e. $$ A_9 = 0,
\, \, A_\alpha(x^5, \cdots, x^8) = {\rm instanton} $$ having
instanton number $Q_1$.  We then obtain a D1-brane (since
instanton is a soliton which is independent of $x^9$) coupled to
the $RR$-field, namely, $$ \int_{\Sigma_{5+1}}  c \wedge \mbox{\rm
tr } \expon^F = Q_1 \int_{\Sigma_2} d^{1+1} \xi A^{(2)} $$

In effect, we have created $Q_1$ number of D1-branes fused together with
the D5-brane.

To give momentum $N$ to the $Q_1$ D1-brane along $\hat{9}$, note
that the instanton configuration depends on the moduli $\zeta^a$,
namely, $$ A_\alpha (x^5, \cdots, x^8, \zeta^a) $$ For $Q_1$
instantons in $U(Q_5)$ gauge theory, the number of moduli
parameters is $4Q_1Q_5$, and moduli space is $\dis {\cal
M}=\frac{(T^4)^{Q_1 Q_5}}{S(Q_1Q_5)} $. The massless excitations
of the D1-branes can be realized as small oscillations of the
moduli, i.e. $\zeta^a = \zeta^a(t + x^9)$, the moduli are all left
movers so as to maintain the BPS condition.  Hence \bea S_{YM} & =
& \frac{1}{g^2_{YM}} \int d^{1+1} \xi \int d^4 \xi \mbox{\rm tr ~}
F^2 \\ & = & \frac{Q_1}{g^2_{YM}} \int d^{1+1} \xi G_{ab}(\zeta)
\partial^\alpha \zeta^a \partial_\alpha \zeta^a
\eea This action is an ${\cal N}=4$ nonlinear sigma model, and
together with its fermionic partner forms a (4,4) superconformal
field theory on a space of length $2 \pi R$.  This theory is
described by $4Q_1Q_5$ free bosons and free fermions, i.e. $\dis
B_B=N_F= 4 Q_1 Q_5$.

The condition of only left movers $\zeta^a(t + x^9)$ yields $L_0 =
N$ and $\bar{L_0}=0$.  From conformal field theory, \bea d(N) & =
& \mbox{{\rm number of microstates for } N} \\ & = & \expon^{2\pi
\sqrt{\frac{c}{6} L_0}} = \expon^{S_{BH}} \eea Since \bea c & = &
N_B + \frac{1}{2} N_F \\ & = & 6 Q_1 Q_5 \eea we finally obtain
the result of Strominger and Vafa \cite{strom1}, namely $$ S_{BH}
= 2 \pi \sqrt{Q_1 Q_5 N} $$ Limitations of the calculation is that
$gN, gQ_1, gQ_5
>> 1$.  For physically interesting cases, we need to understand
non-susy black holes in four dimension.  In these cases $N \sim 1$
and the formalism has to be extended.

In conclusion,a massive BPS $D$-brane  state is identical to an
extremal black hole.
 This seems paradoxical since string states are defined on a flat
background spacetime.  The magic of BPS is the answer.  We start
with weak coupling $g$ where we can count the number of
microstates of the D-brane system.  As we increase $g$, the
counting is unchanged as the number of BPS states do not change.
For large $g$, the $D$-brane system undergoes gravitational
collapse and becomes a black hole with a horizon. Thus, a quantum
black hole is a strongly coupled highly charged bound state of
$D$-branes.

\bigskip
\section{AdS/CFT Correspondence}\label{ads}

In section \ref{sectbackgrd}, we saw that a system of $N$
coincident D $p$-branes is  a classical solution of the low energy
effective string action in which only the metric, the dilaton and
the RR $(p + 1)$-form potential are non-vanishing.  The metric is
given by eq(\ref{branemetric}). For large values of $r$, the
metric becomes flat. Since the curvature is small, the classical
supergravity theory provides a good description of
D-brane\cite{ads1,ads2}.

We specialize eq(\ref{branemetric}) to a $D$3-brane. For a
$D$3-brane, the metric is given by \beq 
ds^2 = f^{-1/2} (-dt^2 + dx_{\rho}^2) + f^{1/2}(r^2 + r^2
d\sigma_5^2) \label{d3brane}\eeq where $x_{\rho}=(x_0, x_1, \cdots
x_p )$ are the coordinates
along the $D$3-brane worldvolume and \beq 
f = 1 + \frac{4 \pi g_s N \alpha^{\prime 2}}{r^4} \eeq with
$d\sigma_5^2$ as the metric on $S^5$. We can rewrite
eq(\ref{d3brane}) as \beq \label{lim}
ds^2 = \left\{ f^{-1/2} (-dt^2 + dx_{\rho}^2) + f^{1/2}r^2
\right\} + f^{1/2} r^2 d\sigma_5^2. \eeq Note that the coordinate
$r$ is the distance transverse to the $D$3-world brane. The terms
within the bracket will ultimately yield in an appropriate limit
the metric for $AdS_5$. Consider \beq 
r \rightarrow 0; ~ ~ f^{1/2} r^2 \rightarrow \alpha^\prime \sqrt{4
\pi g_s N}, \mbox{{\rm constant.}} \eeq This means that $S^5$
which would become infinitesimally small to form the transverse
space, due to the $r^2$ prefactor in the transverse direction, now
forms a neck due to eq(\ref{lim}) with $S^5$ held fixed at some
constant size as $r \rightarrow 0$. Also, note that the $AdS_5$
space is formed by combining the $D$3-brane world volume with an
extra dimension from the transverse direction.  The $D$3-brane
world volume resides at ``the end of the neck" (see figure below).
In the limit $r \rightarrow 0$, $S^5$ carries a non-trivial
RR-charge and provides the source for the $A^{(3)}$ RR-field
emanating from the $D$3-brane.

\begin{center}
\epsfig{file=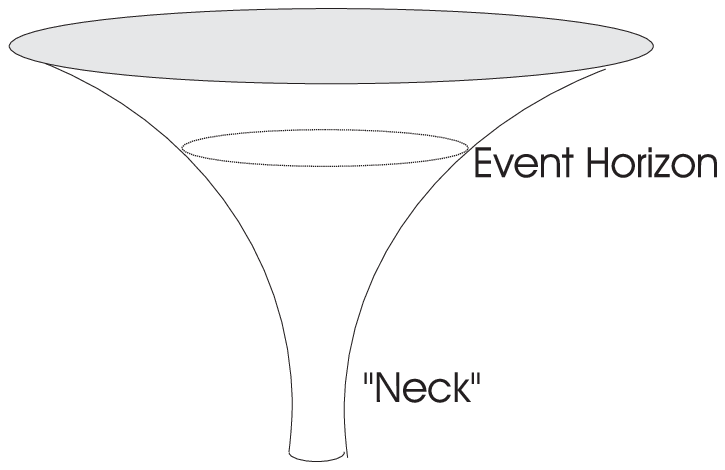}
\end{center}

Thus if we consider the near-horizon Maldacena limit in which \beq 
r \rightarrow 0, ~ ~ \alpha^\prime \rightarrow 0 \eeq 
with $U \equiv \frac{r}{\alpha^\prime}$ held fixed\footnote{Note
that the Maldacena limit is a weak string coupling limit in
contrast with the strong 't Hooft coupling.}, the metric in
eq(\ref{branemetric}) becomes \beq 
\frac{ds^2}{\alpha^\prime} \rightarrow \frac{U^2}{\sqrt{4 \pi N
g_s}} (-dt^2 + dx_1^2  + dx_{\rho}^2) + \frac{\sqrt{4 \pi
N g_s }}{U^2} dU^2 + \sqrt{4 \pi N g_s} d\Omega_5^2 \label{limitmetric}\eeq 
Eq(\ref{limitmetric}) is the metric of the manifold $AdS_5 \times
S^5$ in which the two radii of $AdS_5$ and $S^5$ are equal and
given by \beq 
R_{AdS_5}^2 = R_{S^5}^2 \equiv b^2 = \alpha^\prime \sqrt{4 \pi N
g_s}. \eeq 
It is also interesting to note that in this limit, the Yang-Mills
coupling constant, \beq g_{YM}^2 = 2 g_s (2 \pi)^{p-2}
(\alpha^\prime)^{(p-3)/2} \rightarrow 4 \pi g_s \eeq becomes
dimensionless, so that \beq 
\lambda \equiv \frac{b^2}{\alpha^\prime} = \sqrt{N g_{YM}^2} >>
1\eeq This also implies that the four dimensional world volume is
conformally invariant.

We have seen that a system of $N$ coincident D 3-branes possesses
${\cal N}= 4$ super Yang-Mills gauge theory in 3+1 dimensions with
$U(N)$ gauge group. We see that the classical solution in
eq(\ref{limitmetric}) is a good approximation in the large $N$
limit in which the radii of $AdS_5$ and $S^5$ are huge. Thus,
Maldacena \cite{ads3} conjectures that the strongly interacting
${\cal N}=4$ super Yang-Mills with gauge group $U(N)$ is
equivalent in the large $N$ limit to the ten dimensional Type IIB
superstring theory compactified on $AdS_5 \times S^5$. Indeed,
since supergravity is not a consistent quantum theory, one can
extend the conjecture to any value of $\lambda$ and say that
${\cal N}=4$ super Yang-Mills is equivalent to type IIB string
theory compactified on the special background of $AdS_5 \times
S^5$.

We can also compare the global symmetries. Type IIB string theory
on $AdS_5 \times S^5$ has isometry group $SO(4,2) \times SO(6)$
since $S^5$ has $SO(6)$ symmetry. It turns out that these
symmetries are also the relevant symmetries for ${\cal N}=4$ super
Yang-Mills with gauge group $U(N)$ in 3+1 dimensions. Indeed the
$SO(4,2)$ or $SU(2,2)$ is realized as a conformal invariance in
super Yang-Mills theory. Furthermore, in 10 dimensions, ${\cal
N}=1$ pure super Yang-Mills contains gauge field potentials
$A_\mu,$, $\mu = 0, 1, \cdots , 9$ giving $10-2 =8$ degrees of
freedom in the adjoint representation of $U(N)$\cite{ads1}.
Together with the 8-dimensional Majorana-Weyl ``gluinos"
$\lambda_\alpha$, $\alpha=1,2, \cdots 8$, the theory has 16
Majorana supercharges $Q_\alpha$, $\alpha=1, 2, \cdots 16$. Under
dimensional reduction, the 16 supercharges becomes 4 sets of
complex Majoranas $Q_{\alpha}^A, \bar Q_{\dot\alpha}^A$, $\alpha =
1,2$, $A=1, \cdots 4$ which transform as the $\{ {\bf 4}\}$ and
$\{ {\bf \bar 4}\}$ rep of the $R$-symmetry group of $SU(4)$ and
the scalar fields $\phi_i$ transform as $\{ {\bf 6}\}$ under
$SU(4)$. so that the theory is invariant under $SU(4)$. However,
Type IIB theory has 32 supercharges and super Yang-Mills has only
16.  From the perspectives of the $N$ coincident BPS D3-branes,
half the Type IIB supersymmetries are broken.  One also notes that
the remaining 16 fermionic generators  can arise from the
extension of the conformal group under supersymmetries.

\end{document}